\documentclass[lettersize,journal]{IEEEtran}
\usepackage[dvipdfmx]{graphicx}
\usepackage{amsmath,amsfonts}
\usepackage{algorithm}
\usepackage{algpseudocode}
\usepackage{array}
\usepackage{xcolor} 
\usepackage{textcomp}
\usepackage{stfloats}
\usepackage{url}
\usepackage{verbatim}
\usepackage{graphicx}
\usepackage{cite}
\usepackage{amsmath,amssymb,amsfonts}
\usepackage{bbm}
\usepackage[acronym,shortcuts]{glossaries}
\usepackage{subfigure}
\usepackage{hyperref}
\newacronym{mmWave}{mmWave}{millimeter-wave}
\newacronym{OFDM}{OFDM}{orthogonal frequency division multiplexing}
\newacronym{CoMP}{CoMP}{coordinated multi-point}
\newacronym{SRM}{SRM}{sum rate maximization}
\newacronym{CSI}{CSI}{channel state information}
\newacronym{5G}{5G}{fifth-generation mobile communication systems}
\newacronym{6G}{6G}{sixth-generation mobile communication systems}
\newacronym{LOS}{LOS}{line-of-sight}
\newacronym{NLOS}{NLOS}{non-line-of-sight}
\newacronym{QoS}{QoS}{quality of service}
\newacronym{ERM}{ERM}{empirical risk minimization}
\newacronym{MSGD}{MSGD}{mini-batch stochastic gradient descent}
\newacronym{BCD}{BCD}{block coordinate descent}
\newacronym{RIS}{RIS}{reflected intelligence surface}
\newacronym{BMSGD}{BMSGD}{block mini-batch stochastic gradient descent}
\newacronym{BSGD}{BSGD}{block stochastic gradient descent}
\newacronym{OutMin}{OutMin}{outage minimization}
\newacronym{BS}{BS}{base station}
\newacronym{UPA}{UPA}{uniform planar array}
\newacronym{ULA}{ULA}{uniform linear array}
\newacronym{UE}{UE}{user equipment}
\newacronym{CPU}{CPU}{central processing unit}
\newacronym{SDMA}{SDMA}{space division multiple access}
\newacronym{TDD}{TDD}{time-division duplexing}
\newacronym{AoD}{AoD}{angle of departure}
\newacronym{NR}{NR}{new radio}
\newacronym{AWGN}{AWGN}{additive white Gaussian noise}
\newacronym{SINR}{SINR}{signal-to-noise interference ratio}
\newacronym{SNR}{SNR}{signal-to-noise ratio}
\newacronym{SGD}{SGD}{stochastic gradient descent}
\newacronym{EM}{EM}{Edmundson--Madansky}
\newacronym{QT}{QT}{quadratic transform}
\newacronym{MRT}{MRT}{maximum ratio transmission}
\newacronym{MMSE}{MMSE}{minimum mean square error}
\newacronym{CDF}{CDF}{cumulative distribution function}
\newacronym{MIMO}{MIMO}{multiple-input multiple-output}
\newacronym{RF}{RF}{radio frequency}
\newacronym{FP}{FP}{fractional programming}
\newacronym{SotA}{SotA}{state-of-the-art}
\newacronym{DFT}{DFT}{discrete Fourier transform}
\newacronym{AltMin}{AltMin}{alternating minimization}
\newacronym{AltMax}{AltMax}{alternating maximization}
\newacronym{eMBB}{eMBB}{enhanced mobile broadband}
\newacronym{mMTC}{mMTC}{massive machine type communication}
\newacronym{URLLC}{URLLC}{ultra-reliable and low latency communications}
\newacronym{THz}{THz}{terahertz}
\newacronym{SC}{SC}{single-carrier}
\newacronym{PAPR}{PAPR}{peak-to-average power ratio}
\newacronym{CP}{CP}{cyclic prefix}
\newacronym{FBMC}{FBMC}{filter bank multi-carrier}
\newacronym{AoA}{AoA}{angle of arrival}
\newacronym{QAM}{QAM}{quadrature amplitude modulation}
\newacronym{BER}{BER}{bit error rate}
\newacronym{PSK}{PSK}{phase shift keying}
\newacronym{SVD}{SVD}{singular value decomposition}
\newacronym{SC-FDE}{SC-FDE}{single-carrier frequency-domain equalization}
\newacronym{PA}{PA}{power amplifier}
\newacronym{OTFS}{OTFS}{orthogonal time frequency space}
\newacronym{SC-TDE}{SC-TDE}{single-carrier time-domain equalization}
\newacronym{MSE}{MSE}{mean square error}
\newacronym{CCDF}{CCDF}{complementary cumulative distribution function}
\newacronym{OQAM}{OQAM}{offset quadrature amplitude modulation}
\newacronym{KKT}{KKT}{Karush–Kuhn–Tucker}
\newacronym{UTD}{UTD}{uniform theory of diffraction}

\definecolor{carolinablue}{rgb}{0.6, 0.73, 0.89}

\begin{document}

% \title{Closed-Form Near-Field Beam Generation \\Using Uniform Linear Array}
%\title{Near-Field Beams Generation Via Analog Beamforming with Uniform Linear Array}
% \title{Optimization and Characterization of Near-Field Beams via One-dimensional Array Signal Processing}
% \title{Optimization and Characterization of Near-Field Beams from Perspectives of Array Signal Processing}
\title{Optimization and Characterization of Near-Field Beams with Uniform Linear Arrays}

\author{Sota Uchimura,~\IEEEmembership{Graduate Student Member, IEEE}, Josep Miquel Jornet,~\IEEEmembership{Fellow, IEEE}, \\ and Koji Ishibashi,~\IEEEmembership{Senior Member, IEEE}\vspace{-2ex}
%
% <-this % stops a space
\thanks{This work has been submitted to the IEEE for possible publication.  Copyright may be transferred without notice, after which this version may no longer be accessible.}
\thanks{S. Uchimura and K. Ishibashi are with the Advanced Wireless and Communication Research Center (AWCC), The University of Electro-Communications, Tokyo 182-8285, Japan (e-mail: uchimura@awcc.uec.ac.jp, koji@ieee.org), (\emph{Corresponding author: Sota Uchimura})}
\thanks{J. M. Jornet is with the Department of Electrical and Computer Engineering \& institute for the Wireless Internet of Things, Northeastern University, Boston, MA 02115 USA (e-mail: j.jornet@northeastern.edu)}}

\markboth{Journal of \LaTeX\ Class Files,~Vol.~14, No.~8, August~2021}%
{Shell \MakeLowercase{\textit{et al.}}: A Sample Article Using IEEEtran.cls for IEEE Journals}


\maketitle
\begin{abstract}

% In this paper, we consider near-field beams with wavefronts that can mitigate signal attenuation and blockage effects using a \ac{ULA}. % in wireless communications in the near-field region.
In this paper, we consider near-field beams that can mitigate signal attenuation and blockage effects using a \ac{ULA}. % in wireless communications in the near-field region.
In particular, closed-form expressions for phase distributions in a \ac{ULA} are derived to generate Bessel beams and curving beams based on the desired propagation directions and trajectories.
Based on the phase distributions, the maximum steering angle and propagation distance of Bessel beams with a \ac{ULA} are revealed.
In addition, from the sampling theorem in the spatial domain, the requirements for \acp{ULA} to properly generate Bessel beams are clarified.
For curving beams, trajectories to reach a user while avoiding one obstacle are designed via the Lagrangian method.
Numerical results obtained by electromagnetic wave simulations confirm the effectiveness of the analyses for Bessel beams and the curving beam designs.
Furthermore, the characteristics of Gaussian beams, beamfocusing, Bessel beams, and curving beams are summarized in terms of the statistical behavior of their intensity and signal processing.
\end{abstract}

\glsresetall


\glsresetall

\section{Introduction}

\IEEEPARstart{T}{he} full utilization of high-frequency bands, such as the \ac{mmWave} and \ac{THz} bands, has been envisaged as a promising solution to address the further complicated demands in future wireless communications systems \cite{Wang2023, Jiang2024_Suv}.
One of the hurdles that hides the benefits provided by such bands is the significant spreading losses due to higher carrier frequencies \cite{Jornet2011}.
%
% Considering the compensation of the received power to detect signals, therefore, high-frequency bands require signal propagation with high directivity achieved by large aperture antennas, which mean directional antennas, large antenna arrays, or large metasurfaces \cite{Andrews2017,Liu2021,Bodet2023_OJ}.
%
Therefore, to compensate for the received signal power, high-frequency bands require signal propagation with high directivity achieved by large aperture antennas, namely directional antennas, large antenna arrays, or large metasurfaces \cite{Andrews2017,Liu2021,Bodet2023_OJ}.
%
% However, high directivity at transceivers, together with less diffraction with shorter wavelengths, imposes random path blockages caused by small objects, such as human bodies, which rapidly and suddenly decrease the received signal power\cite{Petrov2017,Uchimura2023}.
%
However, the combination of \emph{i)} the high directivity of \ac{THz} transmitters, resulting in narrow beams, \emph{ii)} the higher photon energy of \ac{THz} waves, which leads to strong reflection and/or absorption by common materials, and \emph{iii)} the short wavelength of \ac{THz} signals, which makes diffraction even from small objects significant, results in partial and even full blockage by common obstacles, resulting in transient, rapidly and suddenly drops in the received signal power \cite{Petrov2017,Uchimura2023}.

To mitigate such channel impairments, transmission strategies exploiting \ac{CSI} have been conceived \cite{Ayach2014,Uchimura2024_TWC,Uchimura2025_TWC}.
%
% On the one hand, in this context, \ac{CSI} modeling and estimation methods have been established for various systems, antenna architectures, and modulation schemes \cite{Molisch2005,Heath2016,Sarieddeen2021,Uchimura2024_TVT,Yoshida2024,Ueda2025}.
% %
% Their accuracy and effectiveness have been verified theoretically, numerically, and experimentally, especially in wireless communications in the far-field region, where the planar wave assumption is valid.
%
% On the one hand, in this context, channel modeling and estimation methods have been established for various systems, antenna architectures, and modulation schemes.
% %
% Their accuracy and effectiveness have been verified theoretically, numerically and experimentally, especially in wireless communications in the far-field region \cite{Uchimura2024_TVT,Heath2016,Sarieddeen2021,MacCartney2016,Mukherjee2022,Fieramosca2024,Poddar2024}, where the planar wave assumption is valid.
%
On the one hand, channel modeling and estimation methods have been established for various systems, and their effectiveness has been verified theoretically, numerically, and experimentally, especially in the far-field region \cite{Uchimura2024_TVT,Heath2016,Sarieddeen2021,MacCartney2016,Mukherjee2022,Fieramosca2024,Poddar2024}, where the planar wavefront assumption is valid.
%
% On the one hand, channel modeling and estimation methods have been established for various systems, and their accuracy and effectiveness have been verified theoretically, numerically, and experimentally, especially in the far-field region \cite{Uchimura2024_TVT,Heath2016,Sarieddeen2021,MacCartney2016,Mukherjee2022,Fieramosca2024,Poddar2024}, where the planar wavefront assumption is valid.
%
On the other hand, a large antenna aperture enlarges the near-field region, where signal propagation is modeled by the spherical wavefront \cite{1Balanis2005, Headland2018}.
%
% Especially in the \ac{THz} bands, data transmission must be operated in the near-field to achieve the anticipated higher data rates with practical transmit power in the presence of the mobility of users \cite{Petrov2023}.
Especially in the \ac{THz} bands, data transmission must be operated in the near-field to achieve the anticipated higher data rates with practical transmit power in the presence of the user mobility \cite{Petrov2023}.
%
% Thus, channel modeling and estimation methods should be rebuilt for wireless communications in the near-field region \cite{Liu2023,Bodet2024,Chen2024,Castellanos2024,Monemi2024,You2025}.
% Thus, channel modeling and estimation methods should be rebuilt for {\color{blue}the near-field region}\cite{Liu2023,Bodet2024,Chen2024,Castellanos2024,Monemi2024,You2025}.
%
% Thus, channel modeling and estimation methods should be rebuilt to address the non-linear phase variation of electromagnetic waves in transceivers due to the sperical wavefront\cite{Liu2023,Bodet2024,Chen2024,Castellanos2024,Monemi2024,You2025}.
Thus, channel modeling and estimation methods should be rebuilt to address the non-linear phase variation of spherical waves in transceivers \cite{Liu2023,Bodet2024,Chen2024,Castellanos2024,Monemi2024,You2025}.
%
% However, channel models to accurately capture near-field effects and the characterization of a reference point of a boundary between the far- and near-field regions have still been pursued \cite{Bodet2024,Chen2024,Castellanos2024,Monemi2024,You2025}.
However, channel models to accurately capture near-field effects and the characterization of a boundary between the far- and near-field regions have still been pursued \cite{Bodet2024,Chen2024,Castellanos2024,Monemi2024,You2025}.
%
% Moreover, random blockage effects make channel modeling more complicated not only in the far-field, but also in the near-field \cite{MacCartney2016,Mukherjee2022,Fieramosca2024,Poddar2024}. 
Moreover, random blockage effects make channel modeling more complicated in both the far- and near-fields \cite{MacCartney2016,Mukherjee2022,Fieramosca2024,Poddar2024}. 
%
% {\color{red}Thus, schemes for acquiring full \ac{CSI} are still insufficient to characterize the near-field and high-frequency bands, even from the perspectives of theoretical modeling and estimation methods.
% %
% From the above, transmission technologies that entirely depend on the full \ac{CSI} are impractical in the near-field region.}
%
% In other words, the absence of accurate channel models for the near-field makes acquiring full \ac{CSI} impossible. Thus, transmission technologies that entirely depend on the full \ac{CSI} are impractical in the near-field region.
In other words, the absence of accurate near-field channel models makes acquiring full \ac{CSI} impossible. Thus, transmission technologies that entirely depend on the full \ac{CSI} are impractical in the near-field region.

Fortunately, the near-field region brings unique wavefronts that generate novel beams, which lead to non-diffraction, self-healing, and self-accelerating properties \cite{Durnin1987, Siviloglou2007}.
From the perspective of signal processing, manipulating the phase and/or amplitude of signals to form the wavefronts only with limited information, such as positions, can mitigate channel impairments.
%
% To name a few, beamfocusing, Bessel beams, and curving beams have been studied for designs of wireless communications systems \cite{Zhang2022, Reddy2023, Guerboukha2024}.
%
To name a few, beamfocusing, Bessel beams, and curving beams have been studied for wireless communications systems \cite{Zhang2022, Reddy2023, Guerboukha2024}.
Since beamfocusing is easy to handle in traditional phased-arrays due to signal propagation with the spherical wavefront, its theoretical designs to deal with the randomness caused by the user mobility have been well investigated \cite{Zhang2022,You2025}.
%
% Since beamfocusing is easy to handle in phased-arrays due to signal propagation with the spherical wavefront, its theoretical designs to deal with the randomness caused by user mobility have been well investigated \cite{Zhang2022,You2025}.
%
In contrast, Bessel and curving beams should be adjusted according to the locations of users and obstacles while maintaining a conical wavefront and a curved wavefront, respectively.
%
% As one of the approaches sparked to generate those beams based on the desired propagation directions and trajectories, hardware designs for lenses, antennas, and metasurfaces have been developed \cite{Yang2023,Zhongsheng2024,Lee2025,Gabriel2022}.
%
As one of the approaches sparked to generate those beams based on the desired directions and trajectories, hardware designs for lenses, antennas, and metasurfaces have been developed \cite{Yang2023,Zhongsheng2024,Lee2025,Gabriel2022}.
However, optimizing the hardware structure to generate one specific beam prevents the generation of other types of beam with the same signal source.
%
% In fact, it has been demonstrated that no beam can achieve the best performance in all situations in wireless communications, which means that beams should be switched according to the communication environment and scenarios \cite{Petrov2024,Petrov2024_TC}.
%
In fact, it has been demonstrated that no beam can achieve the best performance in all wireless channels, which means that beams should be switched according to the communication environment and scenarios \cite{Petrov2024,Petrov2024_TC}.
%
% Therefore, algorithmic approaches are also required, which can generate any type of near-field beam with the same signal source and without additional hardware costs while dealing with the randomness caused by the movement of users and obstacles.
% Therefore, algorithmic approaches are also required, which can generate any type of near-field beam with the same signal source while dealing with the randomness caused by the movement of users and obstacles.
%
Therefore, algorithmic approaches are also required, which can generate any type of beam with the same signal source while dealing with the randomness caused by the movement of users and obstacles.

% From the above, designs of Bessel and curving beams based on the desired directions and trajectories only through the manipulation of phase distributions in transmitters have been proposed \cite{Simon2024,Droulias2024}, both of which can be achieved by adjustment of phases of phase shifters in phased-array architectures \cite{Heath2016}.
%
From the above, designs of Bessel and curving beams based on the desired directions and trajectories only through the manipulation of phase distributions in transmitters have been proposed \cite{Simon2024,Droulias2024}, both of which can be achieved by adjustment of phases of phase shifters in phased-arrays \cite{Heath2016}.
In \cite{Simon2024}, phase distributions in antenna arrays to steer Bessel beams toward the desired azimuth and elevation angles are obtained by solving non-linear simultaneous equations.
%
% In \cite{Simon2024}, phase distributions in antenna arrays to steer Bessel beams toward the desired azimuth and elevation angles are obtained by solving non-linear equations.
%
However, the limitations for steering angles and propagation distances and robustness against blockages of Bessel beams with antenna arrays are still veiled due to the lack of closed-form expressions of the phase distributions.
%
% The work in \cite{Droulias2024} comprehensively analyzed curving beams in terms of design methods and propagation properties, such as trajectory designs to connect users, the maximum propagation distance, and requirements for antenna arrays to generate curving beams.
%
In \cite{Droulias2024}, curving beams are comprehensively analyzed in terms of propagation properties and design methods, such as the maximum propagation distance and trajectory designs to reach users.
However, optimization for trajectories to avoid obstacles based on the relationship between the positions of users and obstacles is not considered.
Although trajectory optimization was proposed in \cite{Gabriel2022}, this approach considered only free-space propagation in the algorithm.
Furthermore, comparisons between different types of beams have not been completed.
%
% In other words, even with conventional works \cite{Durnin1987,Siviloglou2007,Zhang2022, Reddy2023, Guerboukha2024,Yang2023,Zhongsheng2024,Lee2025,Gabriel2022,Simon2024,Droulias2024}, insights into the proper choice of beams to meet the requirements in wireless communications systems with one-dimensional or two-dimensional antenna arrays are insufficient, which should be examined in more detail for the further development of wireless communications \cite{Sarieddeen2021,Petrov2024}.
%
In other words, even with conventional works \cite{Durnin1987,Siviloglou2007,Zhang2022, Reddy2023, Guerboukha2024,Yang2023,Zhongsheng2024,Lee2025,Gabriel2022,Simon2024,Droulias2024}, insights into the proper choice of beams to meet the requirements in systems with one-dimensional or two-dimensional antenna arrays are insufficient, which should be examined in more detail for the further development of wireless communications \cite{Sarieddeen2021,Petrov2024}.

In this paper, therefore, near-field beams optimized based on the relationship between the positions of users and obstacles are compared through theoretical analyses and electromagnetic wave simulations in wireless communications systems with a \ac{ULA}.
Specifically, closed-form expressions for phase distributions in a \ac{ULA} are derived to generate Bessel beams toward the desired azimuth angles.
%
% Based on the phase distributions, the necessary and sufficient conditions are clarified to ensure diffraction-free propagation along the desired directions using the \ac{ULA}.
%
Based on the phase distributions, the necessary and sufficient conditions are clarified to ensure diffraction-free propagation using the \ac{ULA}.
The relationship between the maximum propagation distance, robustness against blockages, and the size of the \ac{ULA} is also revealed.
%
% In addition, closed-form expressions for phase distributions for curving beams, which lead to trajectories that can reach a user while avoiding one obstacle, are derived via the Lagrangian method.
%
In addition, closed-form expressions for phase distributions for curving beams, which lead to trajectories that can reach a user while avoiding one obstacle, are derived.
%
% Numerical results confirm the effectiveness of the mathematical analyses for Bessel beams and the proposed curving beam designs.
% Numerical results confirm the effectiveness of the analyses for Bessel beams and the proposed curving beam designs.
Numerical results confirm the effectiveness of the analyses for Bessel beams and the optimized curving beams.
Through comparisons with Gaussian beams, beamfocusing, Bessel beams, and curving beams, the contributions of each beam to performance improvements are summarized from their intensity and requirements in the beam generation.

\emph{Notation}:
The following notation is used throughout the article. 
Vectors is denoted by lower-case bold letters, as in $\mathbf{x}$. 
%
% The sets of natural and real numbers are represented by $\mathbb{N}$ and $\mathbb{R}$, respectively.
% The set of and real numbers is represented by $\mathbb{R}$.
%
The set of and real numbers, the imaginary unit, the transpose operator, the $\ell_2$-norm, and the absolute value are denoted by $\mathbb{R}$, $j$,  $(\cdot)^\mathrm{T}$, $\|\cdot\|_{2}$, and $|\cdot|$ respectively.
% The transpose operator, the $\ell_2$-norm, and the absolute value are denoted by $(\cdot)^\mathrm{T}$, $\|\cdot\|_{2}$, and $|\cdot|$ respectively.
% %
The open interval is denoted by $(a,b)$ with two arbitrary real numbers $a$ and $b$.
% The imaginary unit is denoted by $j$.
% %
% The transpose operator, the $\ell_2$-norm, and the absolute value are denoted by $(\cdot)^\mathrm{T}$, $\|\cdot\|_{2}$, and $|\cdot|$ respectively.
%
% The closed and open intervals are denoted by $[a,b]$ and $(a,b)$, respectively, with two arbitrary real numbers $a$ and $b$.

%
%----
%

\section{System Model}

% \begin{figure}[t]
% \centering
% %\colorbox{carolinablue}
% \includegraphics[width=0.85\columnwidth]{fig/Nt=Nr=64_Ns=1_16QAM_fin.eps}
% \caption{Comparison of the proposed \ac{SC}, conventional \ac{SC}, and \ac{OFDM} schemes under single-stream transmission in a massive \ac{MIMO} setting.}
% \label{fig:BER_RF1}
% \end{figure}

\subsection{Coordinate System}

Consider signal propagation in the $xy$-plane in the Cartesian coordinate system, where a \ac{ULA} is equipped with $N$ antenna elements and is located on the $x$-axis.
Let $x_{\mathrm{t},n}\in\mathcal{X}_\mathrm{arr}$ denote the position of the $n$-th antenna element in the $x$-coordinate with $\mathcal{X}_\mathrm{arr}\triangleq\{x_{\mathrm{t},n}\}_{n=1}^{n=N}$ denoting the set of the $x$-coordinates of the positions of the antenna elements.
The \ac{ULA} is assumed to be symmetric about the $y$-axis, where the position of the $n$-th antenna element is given by $\mathbf{p}_{\mathrm{t},n}\triangleq[x_{\mathrm{t},n},y_{\mathrm{t},n}]^\mathrm{T}=[\tfrac{-N+2n-1}{2}\Delta,0]^\mathrm{T}\in\mathbb{R}^2$ with $\Delta\in\mathbb{R}^{+}$ denoting the antenna spacing.
Let $R\triangleq x_{\mathrm{t},N}=\tfrac{N-1}{2}\Delta$ and $-R\triangleq x_{\mathrm{t},1}=\tfrac{-N+1}{2}\Delta$ denote the $x$-coordinates of the positions of the $N$-th and first antenna elements, respectively.
Consequently, the aperture size of the \ac{ULA} is denoted by $2R$.

Signal propagation from the \ac{ULA} to the positive region of the $y$-axis is considered, such that the domain of the azimuth angle $\theta_\mathrm{A}$ is defined as $\theta_\mathrm{A}\in(-\tfrac{\pi}{2}, \tfrac{\pi}{2})$, whose value is zero $\theta_\mathrm{A}=0$ on the $y$-axis.
%
% Note that the following analyses and results can be applied to signal propagation toward the inverse direction straightforwardly.
Note that the following analyses can be applied to the reverse propagation straightforwardly.

% The propagation region is in the near-field.

%
%

\subsection{Electric Field and Phase Manipulation}

% Let $\mathbf{p}_\mathrm{u}=[x_\mathrm{u},y_\mathrm{u}]^\mathrm{T}\in\mathbb{R}^2$ denote the position of the user in the $xy$-plane.
%
% Following the literature on electromagnetic channel modeling \cite{Chen2024,Castellanos2024}, the electric field at the point $\mathbf{p}_\mathrm{u}$ from the $n$-th antenna element is described by
%
Let $\mathbf{p}_\mathrm{u}=[x_\mathrm{u},y_\mathrm{u}]^\mathrm{T}\in\mathbb{R}^2$ denote the position of the user.
Following the literature \cite{Chen2024,Castellanos2024}, the electric field at the position $\mathbf{p}_\mathrm{u}$ from the $n$-th antenna element is described by
\begin{equation}
    \mathbf{e}_n(\mathbf{p}_\mathrm{u}) \propto  \tfrac{\Omega}{r_n}\exp(-jkr_n)I_n \mathbf{u}_{n}(\mathbf{p}_\mathrm{u})
\end{equation}
where $\mathbf{u}_n(\mathbf{p}_n)$ and $r_n=\|\mathbf{p}_\mathrm{u}-\mathbf{p}_{\mathrm{t},n}\|_2\in\mathbb{R}^+$ denote the unit vector for the polarization of the $n$-th antenna element at the position $\mathbf{p}_{\mathrm{u}}$ and the distance between the $n$-th antenna element and position $\mathbf{p}_\mathrm{u}$, respectively.
The scalar $\Omega$ [V/A] is a constant related to the impedance.
The wavenumber $k\in\mathbb{R}^{+}$ is given by $k=\tfrac{2\pi}{\lambda}$, where $\lambda=\tfrac{c}{f_\mathrm{c}}$ denotes the wavelength calculated with the speed of light $c$ and the central frequency $f_\mathrm{c}\in\mathbb{R}^{+}$.

The scalar $I_n=\gamma_n\exp(j\phi_n)$ denotes the excited current of the $n$-th antenna element in the complex form, where $\gamma_n\in\mathbb{R}^{+}$ and $\phi_n$ denote the magnitude and phase for the $n$-th antenna element, respectively. 
%
% Then, the phase $\phi_n\triangleq kd_n$ can be decomposed into the wavenumber $k$ and the minimum distance $d_n\in\mathbb{R}^+$ between the position of the $n$-th antenna element and the wavefront function, which describes the wavefront of the whole antenna array \cite{Simon2024,Headland2018}.
Then, the phase $\phi_n\triangleq kd_n$ can be decomposed into the wavenumber and the minimum distance $d_n\in\mathbb{R}^+$ between the $n$-th antenna element and the wavefront function, which describes the wavefront of the whole antenna array \cite{Simon2024,Headland2018}.
%
% Thus, near-field beams with specific wavefronts can be generated via phase-shifter designs that manipulate the phases for the antenna elements $\phi_n,\forall n$, which is considered in what follows.
Thus, beams with specific wavefronts can be generated via phase-shifter designs for manipulating the phases $\phi_n,\forall n$, which is considered in what follows.

% in what follows, given the wavefront function, analog beamforming, which designs the phases for the antenna elements $\phi_n,\forall n$ to generate near-field beams, is considered.
% %
% The phases of the current are manipulated based on wavefront functions, thereby forming the corresponding wavefront of the antenna array.
% %
% In other words, analog precording, which designs the phases of the antenna elements defined by $\phi_n\triangleq kd_n, (n=1,\ldots,N)$, enables near-field beam generation in hybrid antenna array architectures.

%
%---
%

\section{Bessel Beam Generation with a ULA}

In this section, the limitations of Bessel beams are considered in terms of the maximum steering angle and propagation distance, as well as their robustness against blockages.
Closed-form expressions for phase distributions to generate Bessel beams reveal that the steering angle is restricted to less than 45 [deg], and a \ac{ULA} with half-wavelength spacing can generate any Bessel beam.
The tradeoff between the propagation distance and robustness against blockages is also clarified based on the analysis for the maximum propagation distance.

% \begin{figure}[t]
% \centering
% 	\begin{minipage}{0.40\columnwidth}
% 	\subfigure[{Wavefront function}]
% 	{
%     \includegraphics[width=\linewidth]{fig/systems.pdf}
% 	\label{fig:wavefront}
% 	}
% 	\end{minipage}
% 	%
% 	\begin{minipage}{0.40\columnwidth}
% 	\subfigure[{Rotating coordinate system}]
% 	{
%   \includegraphics[width=\linewidth]{fig/rot_systems.pdf}
% 	\label{fig:rotating}
% 	}
% 	\end{minipage}
%  \\
%  \begin{minipage}{0.40\columnwidth}
% 	\subfigure[{Geometrical optics}]
% 	{
%     \includegraphics[width=\linewidth]{fig/GO.pdf}
% 	\label{fig:GO}
% 	}
% 	\end{minipage}
% 	%
% 	\begin{minipage}{0.40\columnwidth}
% 	\subfigure[Obstacle]
% 	{
% 	\includegraphics[width=\linewidth]{fig/obs.pdf}
% 	\label{fig:obs}
% 	}
% 	\end{minipage}
% 	% \caption{Received power and phase distributions}
%         \caption{Cartesian coordinate system and wavefront function}
% 	\label{fig:B_coordinate}
% \end{figure}

\begin{figure}[t]
\centering
	\begin{minipage}{0.40\columnwidth}
	\subfigure[{Wavefront function}]
	{
    \includegraphics[width=\linewidth]{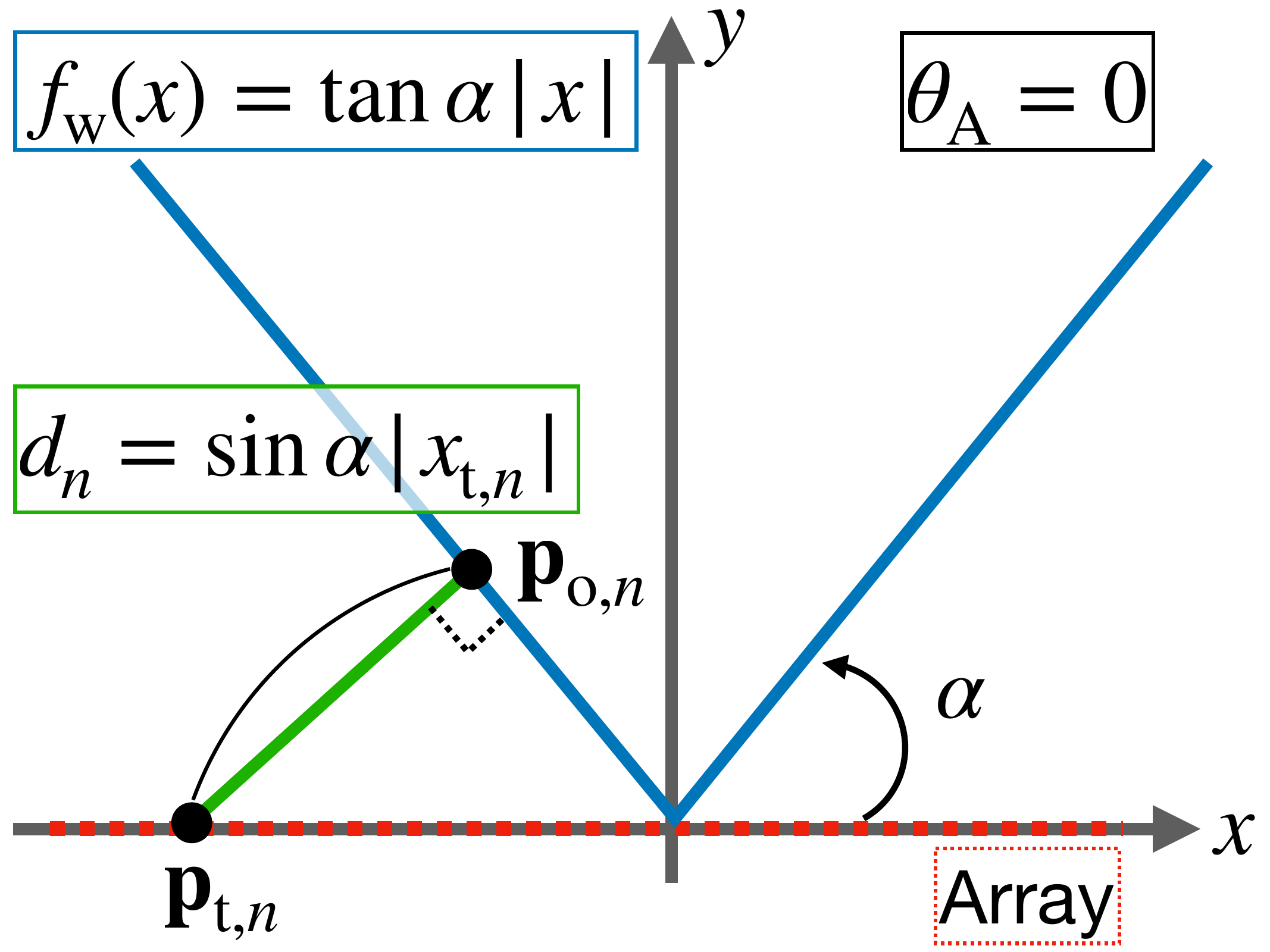}
	\label{fig:wavefront}
	}
	\end{minipage}
	\begin{minipage}{0.40\columnwidth}
	\subfigure[{Rotating coordinate system}]
	{
  \includegraphics[width=\linewidth]{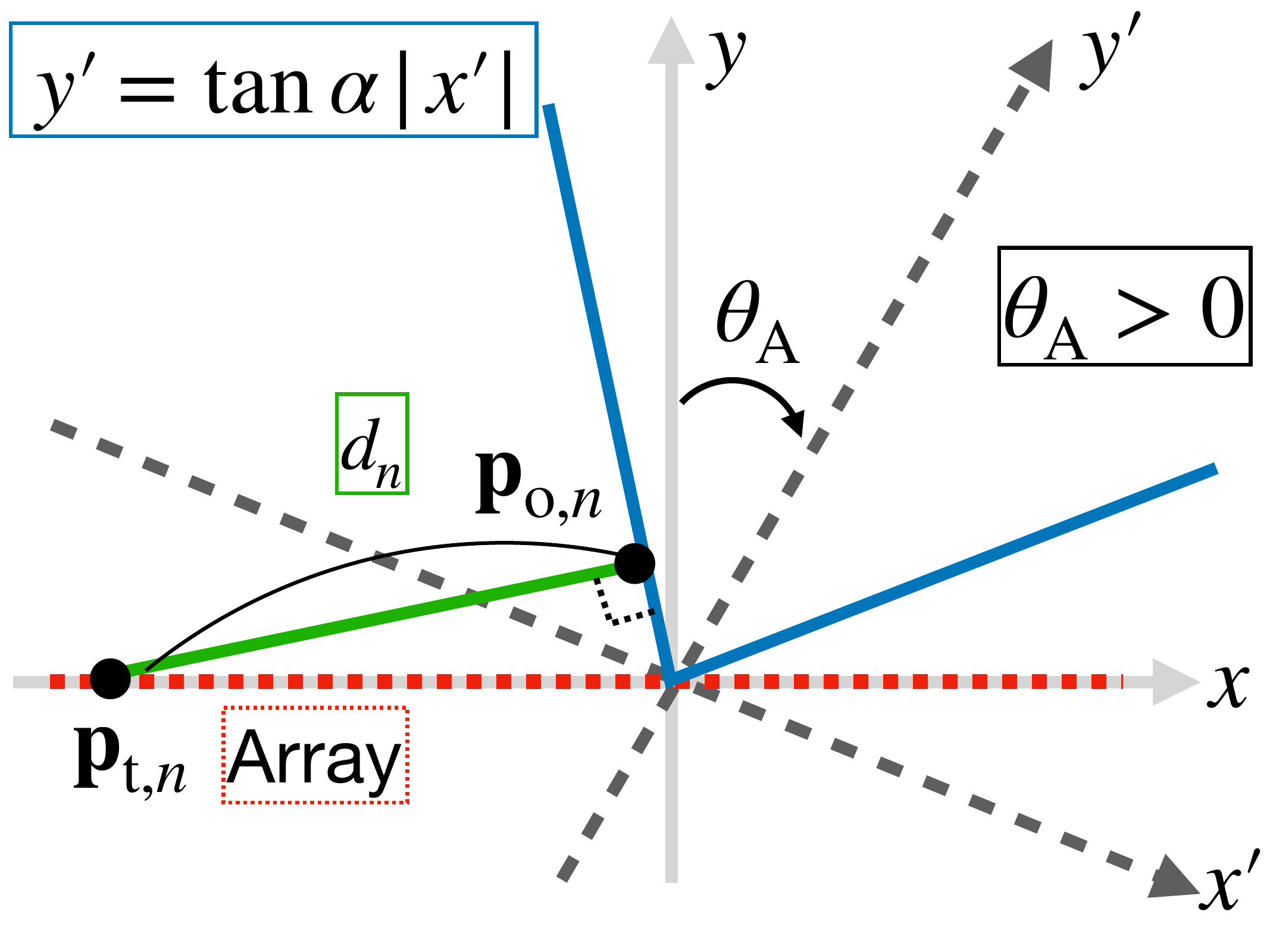}
	\label{fig:rotating}
	}
	\end{minipage}
    \caption{Cartesian coordinate system and wavefront function}
% 	\label{fig:B_coordinate}
\end{figure}

%
%----
%

\subsection{Closed-Form Expression for Phase Distribution}
\label{sec:Bessel_closed}

A Bessel beam has a conical wavefront \cite{Durnin1987}, which leads to a wavefront function described by an absolute function in \ac{ULA} systems.
As shown in Fig. \ref{fig:wavefront}, in the non-steering case $\theta_\mathrm{A}=0$, the wavefront function is given by $f_\mathrm{w}(x)=\tan\alpha|{x}|$, where $\alpha\in(0,\tfrac{\pi}{2})$ is the controllable parameter.
Then, the minimum distance between the $n$-th antenna element and the wavefront function can be calculated as $d_n=\sin\alpha|{x_{\mathrm{t},n}}|$.
Thus, the phases for generating the non-steering Bessel beam are given by $\phi_n = k\sin\alpha|{x_{\mathrm{t},n}}|,\forall n$ \cite{Headland2018}.

The basic idea of steering the Bessel beam with the \ac{ULA} is to define the wavefront function in the rotating coordinate system based on the desired angle \cite{Simon2024}, as shown in Fig. \ref{fig:rotating}.
In the $xy$-plane, the rotating coordinates can be defined by
\begin{equation}
\begin{bmatrix}
x^\prime
\\
y^\prime
\end{bmatrix}
=
\begin{bmatrix}
\cos\theta_\mathrm{A}&-\sin\theta_\mathrm{A}
\\
\sin\theta_\mathrm{A}&\cos\theta_\mathrm{A}
\end{bmatrix}
\begin{bmatrix}
x
\\
y
\end{bmatrix}. \label{eq:rotation}
\end{equation}
%
% which yields the relationship $y^\prime = \tan\alpha|{x}^\prime|$ to steer the Bessel beam toward the desired angle $\theta_\mathrm{A}$.

% Thus, the wavefront function to steer the Bessel beam toward the desired angle $\theta_\mathrm{A}$ is obtained from the solutions with respect to $y$ of the following equation
%
Thus, the wavefront function to steer the Bessel beam toward the desired angle is obtained from the solutions concerning $y$ of the equation $y^\prime = \tan\alpha|{x}^\prime|$, which equals
\begin{equation}
    x\sin\theta_\mathrm{A}+y\cos\theta_\mathrm{A}
    =
    \tan\alpha|{x\cos\theta_\mathrm{A}-y\sin\theta_\mathrm{A}|}. \label{eq:wavefront_steering}
\end{equation}

Consequently, the wavefront function is given by
\begin{equation}
    f_\mathrm{w}(x)
    =
    \begin{cases}
    \tan(\alpha-\theta_\mathrm{A}) x,~~~~\mathrm{if}~x\ge 0
    \\
    -\tan(\alpha+\theta_\mathrm{A}) x,~\mathrm{otherwise}
    \end{cases},
    \label{eq:wavefront_Bessel}
\end{equation}
under the necessary condition $0<\alpha+|\theta_\mathrm{A}|<\tfrac{\pi}{2}$ to generate beams.
The details of the derivation are given in Appendix A.

Based on the minimum distance between the $n$-th antenna elements and the wavefront function $f_\mathrm{w}(x)$ in \eqref{eq:wavefront_Bessel}, the phase of the $n$-th antenna element is calculated as
\begin{equation}
    \phi_n
    =
    \begin{cases}
k|{\sin(\alpha-\theta_\mathrm{A})}|x_{\mathrm{t},n},~~~\mathrm{if}~{x_{\mathrm{t},n}}\ge 0
\\
-k|{\sin(\alpha+\theta_\mathrm{A})}|x_{\mathrm{t},n},~\mathrm{otherwise}
\end{cases},
\label{eq:closed_Bessel}
\end{equation}
where the derivation is given in Appendix B.

The phase distribution determined by the phases given by \eqref{eq:closed_Bessel} means that the steering Bessel beam comprises two beams with linear phase distributions with respect to $x_{\mathrm{t},n}$, which are generated by the antenna elements located in the positive and negative regions of the $x$-axis, respectively.
% The phase distribution determined by the phases given in \eqref{eq:closed_Bessel} means that the steering Bessel beam comprises two beams with \textcolor{blue}{linear phase distributions}, which are generated by the antenna elements located in the positive and negative regions of the $x$-axis, respectively.

%
%-------------------------------------
%

\subsection{Steering Capability}
\label{sec:B_steer}

In this subsection, the maximum steering angle of the Bessel beam with the \ac{ULA} is derived based on the necessary and sufficient condition for the parameter $\alpha$ to generate the Bessel beam toward the desired angle.
% As described in Section \ref{sec:Bessel_closed}, the relationship $0<\alpha+|\theta_\mathrm{A}|<\tfrac{\pi}{2}$ shows the necessary condition to generate the Bessel beam toward the desired angle.
% %
% In this section, the maximum steering angle of the Bessel beam with the \ac{ULA} is derived based on the necessary and sufficient condition for the parameter $\alpha$.

% % From the property of the spherical wavefront of each antenna element, 
% In order to convey signals to the positive region of the $y$-axis, the wavefront function has to take non-negative values $f_\mathrm{w}(x)\ge 0$.
% %
% % Therefore, the necessary and sufficient condition for the parameter $\alpha$ to generate the Bessel beam toward the desired angle $\theta_\mathrm{A}$ is given by
% Therefore, the necessary and sufficient condition for the parameter $\alpha$ \textcolor{blue}{is given by}
%
In order to convey signals to the positive region of the $y$-axis, the wavefront function has to take non-negative values $f_\mathrm{w}(x)\ge 0$, leading to the the necessary and sufficient condition for the parameter $\alpha$ given by
\begin{equation}
    |\theta_\mathrm{A}| \le \alpha < \tfrac{\pi}{2}-|\theta_\mathrm{A}|, \label{eq:max_steering}
\end{equation}
where the derivation is given in Appendix C.

The condition in \eqref{eq:max_steering} indicates that only signal propagation toward azimuth angles within the open interval $\theta_\mathrm{A}\in(-\tfrac{\pi}{4},\tfrac{\pi}{4})$ can benefit from the Bessel beam in the \ac{ULA} systems.

%
% In the following sections, it is assumed that the parameter $\alpha$ and the angle $\theta_\mathrm{A}$ satisfy the condition in \eqref{eq:max_steering}.
% the necessity and sufficient condition in \eqref{eq:max_steering} is satisfied.

% Bessel beams and their properties can be granted only in

% Moreover, under the condition described in \eqref{eq:max_steering}, the inequality $\cos(\alpha-\theta_\mathrm{A})>0$ holds, which yields the simple minimum distance expression given by
% %
% \begin{equation}
%     d_{\mathrm{t},n}
%     \!\!=\!\!
%     \begin{cases}
% \!\tfrac{x_{\mathrm{t},n}\big|\tfrac{\tan\alpha-\tan\theta_\mathrm{A}}{1+\tan\alpha\tan\theta_\mathrm{A}}\big|}{1+\big(\tfrac{\tan\alpha-\tan\theta_\mathrm{A}}{1+\tan\alpha\tan\theta_\mathrm{A}}\big)^2}
% \sqrt{\!\big(\!\tfrac{\tan\alpha-\tan\theta_\mathrm{A}}{1+\tan\alpha\tan\theta_\mathrm{A}}\!\big)^2\!\!\!+\!1},~\mathrm{if}~{x_{\mathrm{t},n}}\ge 0
% \\
% \!\tfrac{-x_{\mathrm{t},n}\big|\tfrac{\tan\theta_\mathrm{A}+\tan\alpha}{\tan\alpha\tan\theta_\mathrm{A}-1}\big|}{1+\big(\tfrac{\tan\theta_\mathrm{A}+\tan\alpha}{\tan\alpha\tan\theta_\mathrm{A}-1}\big)^2}
% \sqrt{\!\big(\!\tfrac{\tan\theta_\mathrm{A}+\tan\alpha}{\tan\alpha\tan\theta_\mathrm{A}-1}\!\big)^2\!\!\!+\!1}, \mathrm{otherwise}
% \end{cases}\!\!\!\!\!\!,
% \label{eq:closed_min_dist}
% \end{equation}

% The derivation of the solution in \eqref{eq:closed_min_dist} is given in Appendix C.

%
%-------
%

\subsection{Maximum Propagation Distance}
\label{sec:prop_dist}

Bessel beams can achieve diffraction-free propagation only with infinite antenna aperture and infinite transmit power.
In practice, therefore, the quasi-non-diffraction range can be estimated with the parameter $\alpha$ and the aperture size of an antenna.
%
% Mathematically, under the assumption of signal propagation along the $y$-axis, the maximum propagation distance, which is a reference point for the beginning of the degradation of the intensity of the Bessel beam, is given by $\tfrac{R}{\tan\alpha}$ \cite{Durnin1987}.
Mathematically, under the assumption of signal propagation along the $y$-axis, the maximum propagation distance, which is a reference point for the beginning of the degradation of the intensity, is given by $\tfrac{R}{\tan\alpha}$ \cite{Durnin1987}.
%
% Mathematically, the maximum propagation distance, which is a reference point for the beginning of the degradation of the intensity of the Bessel beam, is derived under the assumption of signal propagation along the $y$-axis in \cite{Durnin1987}, and is given by $\tfrac{R}{\tan\alpha}$.
%
In this subsection, that analysis is extended to the steering version of the Bessel beam. %, resulting in the maximum propagation distance as a function of the parameter $\alpha$, aperture size, and angle $\theta_\mathrm{A}$.

% As mentioned in Section \ref{sec:Bessel_closed}, the Bessel beam comprises the two beams generated by the antenna elements located in the positive and negative regions of the $x$-axis, respectively.
As mentioned in Section \ref{sec:Bessel_closed}, the Bessel beam comprises the two beams.
Thus, two reference points for the beginning of the degradation of the intensity along the desired propagation direction are considered, which are determined by the antenna positions in the positive and negative regions of the $x$-axis, respectively.
Note that the $y$-axis in the rotating coordinate system corresponds to the desired propagation direction.

\begin{figure}[t]
\centering
 \begin{minipage}{0.40\columnwidth}
	\subfigure[{Propagation distance}]
	{
    \includegraphics[width=\linewidth]{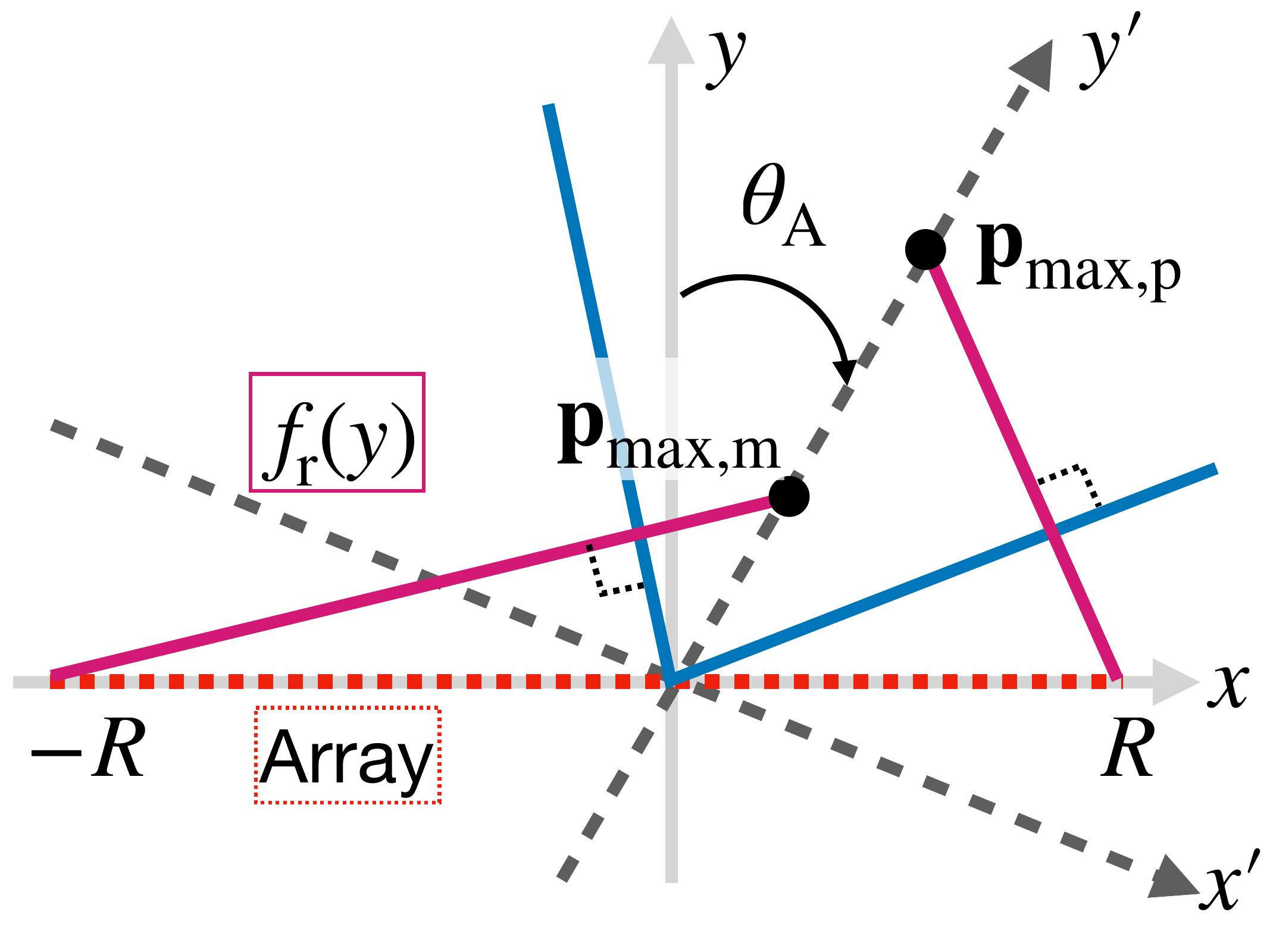}
	\label{fig:GO}
	}
	\end{minipage}
	\begin{minipage}{0.40\columnwidth}
	\subfigure[Self-healing capability]
	{
	\includegraphics[width=\linewidth]{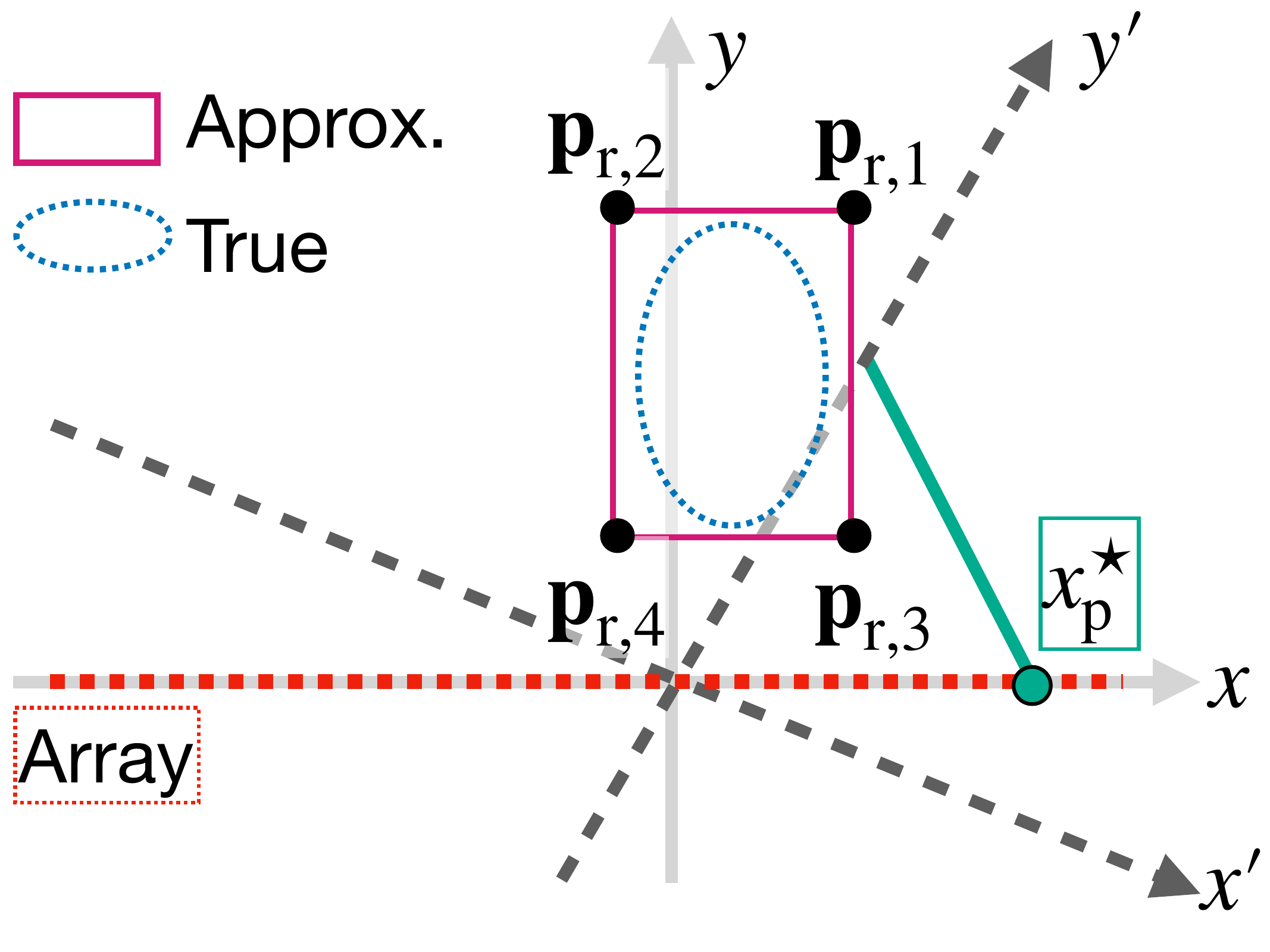}
	\label{fig:obs}
	}
	\end{minipage}
        \caption{Analyses based on geometric optics}
	\label{fig:B_GO}
\end{figure}

Let $\mathbf{p}_{\mathrm{max,p}}\in\mathbb{R}^2$ and $\mathbf{p}_{\mathrm{max,m}}\in\mathbb{R}^2$ denote the reference points corresponding to the beams generated by the positive and negative regions of the $x$-axis, respectively, as shown in Fig. \ref{fig:GO}.
%
% Based on geometrical optics \cite{Marchand1966}, the reference points $\mathbf{p}_{\mathrm{max,p}}$ and $\mathbf{p}_{\mathrm{max,m}}$ are determined by the intersections between the $y$-axis in the rotating coordinate system and the direct rays from each antenna element, which are described by the linear functions perpendicular to the wavefront function $f_\mathrm{w}(x)$ \cite{Durnin1987}.
% %
% From the orthogonality of the lines, the product of the slopes of the wavefront function $f_\mathrm{w}(x)$ and the linear functions for the direct rays is $-1$.
%
% Moreover, the intercepts of the linear functions are given by the $x$-coordinates of the positions of the antenna elements.
%
Based on geometrical optics \cite{Marchand1966}, the reference points $\mathbf{p}_{\mathrm{max,p}}$ and $\mathbf{p}_{\mathrm{max,m}}$ are determined by the intersections between the $y$-axis in the rotating coordinate system and the direct rays from each antenna element.
Those direct rays are described by the linear functions perpendicular to the wavefront function $f_\mathrm{w}(x)$ \cite{Durnin1987}, where the product of the slopes of $f_\mathrm{w}(x)$ and the linear functions is $-1$ from the orthogonality of the lines.
Moreover, the intercepts of the linear functions are given by the $x$-coordinates of the positions of the antenna elements.
Consequently, as shown in Fig. \ref{fig:GO}, the direct ray from the $n$-th antenna element is described by
\begin{equation}
    f_\mathrm{r}(y)
    =
    \begin{cases}
        -\tan(\alpha-\theta_\mathrm{A})y + x_{\mathrm{t},n},~\mathrm{if}~x_{\mathrm{t},n} \ge0
        \\
        \tan(\alpha+\theta_\mathrm{A})y + x_{\mathrm{t},n},~~~\mathrm{otherwise}
    \end{cases}, \label{eq:linear}
\end{equation}

% Geometrically and mathematically, the intersections between the $y$-axis in the rotating coordinate system and the linear functions $f_\mathrm{r}(y)$ with the intercepts $x_{\mathrm{t},N}=R$ or $x_{\mathrm{t},1}=-R$ are the farthest from the origin of the coordinate system among all the intersections.
%
The intersection between the $y$-axis in the rotating coordinate system and the linear function $f_\mathrm{r}(y)$ with the intercept $x_{\mathrm{t},N}=R$ or $x_{\mathrm{t},1}=-R$ is the farthest from the origin of the coordinate system among all the intersections.
Moreover, the relationship between the $x$- and $y$-coordinates of the intersections is described by $(x,y)=(y\tan\theta_\mathrm{A},y)$.
Thus, the reference points $\mathbf{p}_{\mathrm{max,p}}$ and $\mathbf{p}_{\mathrm{max,m}}$ can be estimated as
\begin{align}
    \mathbf{p}_{\mathrm{max,p}} &= \Big[\tfrac{R\tan\theta_\mathrm{A}}{\tan\theta_\mathrm{A}+\tan(\alpha-\theta_\mathrm{A})},\tfrac{R}{\tan\theta_\mathrm{A}+{\tan(\alpha-\theta_\mathrm{A})}}\Big]^\mathrm{T}, \label{eq:point_p}
    \\
    \mathbf{p}_{\mathrm{max,m}} &= \Big[\tfrac{-R\tan\theta_\mathrm{A}}{\tan\theta_\mathrm{A}-\tan(\alpha+\theta_\mathrm{A})},\tfrac{-R}{\tan\theta_\mathrm{A}-\tan(\alpha+\theta_\mathrm{A})}\Big]^\mathrm{T}, \label{eq:point_m}
\end{align}
%
% where the derivations are given in Appendix D.

Following the definition in \cite{Durnin1987,Zhongsheng2024}, both the beams generated by the positive and negative regions of the $x$-axis should contribute to signal propagation within the quasi-non-diffraction range.
Thus, the maximum propagation distance $d_\mathrm{max}$ along the desired direction $\theta_\mathrm{A}$ is estimated by $d_\mathrm{max}=\min(\|\mathbf{p}_{\mathrm{max,p}}\|_2, \|\mathbf{p}_{\mathrm{max,m}}\|_2)$, resulting in
\begin{equation}
    d_{\mathrm{max}}
    =
    R\tfrac{\cos(\alpha+|\theta_\mathrm{A}|)}{\sin\alpha}
    =
    \tfrac{R}{\tan\alpha}\tfrac{\cos(\alpha+|\theta_\mathrm{A}|)}{\cos\alpha}.
    \label{eq:dist1}
\end{equation}
where the derivation is given in Appendix D.

% Under the necessary and sufficient condition in \eqref{eq:max_steering}, the function $\tfrac{\cos(\alpha+|\theta_\mathrm{A}|)}{\cos\alpha}$ is monotonically decreasing with respect to both $\alpha$ and $|\theta_\mathrm{A}|$, where the parameter $\alpha$ should be greater than or equal to $|\theta_\mathrm{A}|$.
Under the condition in \eqref{eq:max_steering}, the function $\tfrac{\cos(\alpha+|\theta_\mathrm{A}|)}{\cos\alpha}$ returns positive values less than or equal to 1 and is monotonically decreasing with respect to both $\alpha$ and $|\theta_\mathrm{A}|$, where $\alpha$ should be greater than or equal to $|\theta_\mathrm{A}|$.
%
% From the above, the steering of the Bessel beam shrinks the maximum propagation \textcolor{blue}{distance} drastically compared to that of the non-steering version, which is given by $\tfrac{R}{\tan\alpha}$.
%
% \textcolor{blue}{Thus}, the steering of the Bessel beam shrinks the maximum propagation \textcolor{blue}{distance} drastically compared to that of the non-steering version, which is given by $\tfrac{R}{\tan\alpha}$.
Thus, the steering of the Bessel beam shrinks the maximum propagation distance drastically compared to that of the non-steering version (\emph{i.e.}, $\tfrac{R}{\tan\alpha}$).
%
% Note that the distance $d_\mathrm{max}$ in \eqref{eq:dist1} is $\tfrac{R}{\tan\alpha}$ when $\theta_\mathrm{A}=0$.

Owing to the steering nature caused by the \ac{ULA}, however, part of the beam can contribute to signal propagation along the desired direction beyond the maximum propagation distance $d_\mathrm{max}$.
Then, the limitation of the propagation distance is estimated by $d_\mathrm{lim}=\max(\|\mathbf{p}_{\mathrm{max,p}}\|_2, \|\mathbf{p}_{\mathrm{max,m}}\|_2)$ given by
\begin{equation}
    d_{\mathrm{lim}}
    =
    R\tfrac{\cos(\alpha-|\theta_\mathrm{A}|)}{\sin\alpha}
    = 
    \tfrac{R}{\tan\alpha}\tfrac{\cos(\alpha-|\theta_\mathrm{A}|)}{\cos\alpha},
    \label{eq:dist2}
\end{equation}
where $\tfrac{\cos(\alpha-|\theta_\mathrm{A}|)}{\cos\alpha}$ is greater than or equal to 1 under the condition in \eqref{eq:max_steering}, resulting in a longer propagation distance if lower intensity is acceptable compared to the region within the maximum distance $d_\mathrm{max}$.
%
% Thus, if lower intensity than that in the region within the maximum distance $d_\mathrm{max}$ is acceptable, the propagation distance is longer compared to the non-steering cases.
% Thus, if lower intensity than that in the region within the maximum distance $d_\mathrm{max}$ is acceptable, the propagation distance is longer compared to the non-steering cases.

% Since the function 
% which is the monotonically increasing function with respect to the azimuth angle $\theta_\mathrm{A}$ under the necessity and sufficient condition in \eqref{eq:max_steering}.

% From the discussions about the distances $d_\mathrm{max,1}$ and $d_\mathrm{max,2}$, the propagation region, which is contributed by only one side of the antenna array, is enlarged with the larger azimuth angle $\theta_\mathrm{A}$.
%
% The derivations for $d_{\mathrm{max}}$ and $d_{\mathrm{lim}}$ are given in Appendix E.

%
%------
%

%
%-------------
%

\subsection{Adjustment for Antenna Arrays}
\label{sec:sampling}

% In Section \ref{sec:prop_dist}, the maximum propagation distance is derived as a function of the parameter $\alpha$, the aperture size, and the azimuth angle $\theta_\mathrm{A}$.
% %
% % Although reducing $\alpha$ can increase the maximum propagation distance, it is limited by the necessary and sufficient condition in \eqref{eq:max_steering}.
% Although reducing $\alpha$ can increase the maximum propagation distance, it is limited by the \textcolor{blue}{condition} in \eqref{eq:max_steering}.
%
From Sections \ref{sec:B_steer} and \ref{sec:prop_dist}, although reducing $\alpha$ can increase the maximum propagation distance, it is limited by the condition in \eqref{eq:max_steering}.
%
% Thus, the number of antenna elements $N$ and the antenna spacing $\Delta$ are critical indicators in system designs to achieve the desired maximum propagation distance $d_\mathrm{max,d}$, which are discussed in this subsection.
Thus, the number of antenna elements $N$ and the antenna spacing $\Delta$ are critical indicators in system designs to achieve the desired propagation distance $d_\mathrm{max,d}$, which are discussed in this subsection.

In the \ac{ULA} systems, the antenna aperture $2R$ is calculated as $2R=(N-1)\Delta$.
%
% Thus, given the azimuth angle $\theta_\mathrm{A}$, parameter $\alpha$, and fixed antenna spacing $\Delta$, the minimum number of antenna elements to achieve the desired maximum propagation distance $d_\mathrm{max,d}$ is given by
%
From the inequality $d_\mathrm{max,d}\ge d_\mathrm{max}$, given the azimuth angle $\theta_\mathrm{A}$, parameter $\alpha$, and fixed antenna spacing $\Delta$, the minimum number of antenna elements to achieve the desired maximum propagation distance $d_\mathrm{max,d}$ is given by
%
% \begin{align}
%     d_\mathrm{max,d}&\ge d_\mathrm{max}=\tfrac{(N-1)\Delta}{2}\tfrac{\cos(\alpha+|\theta_\mathrm{A}|)}{\sin\alpha}, \nonumber
%     \\
%     N &= \Big\lceil 2d_\mathrm{max,d} \tfrac{\sin\alpha}{\Delta\cos(\alpha+|\theta_\mathrm{A}|)}+1\Big\rceil, \label{eq:max_dist_num_antenna}
% \end{align}
\begin{equation}
    N = \Big\lceil 2d_\mathrm{max,d} \tfrac{\sin\alpha}{\Delta\cos(\alpha+|\theta_\mathrm{A}|)}+1\Big\rceil, \label{eq:max_dist_num_antenna}
\end{equation}
where $\lceil\cdot\rceil$ denotes the cell function.

In turn, given the azimuth angle $\theta_\mathrm{A}$, parameter $\alpha$, and fixed number of antenna elements $N$, the minimum antenna spacing to achieve the desired propagation distance $d_\mathrm{max,d}$ is given by $\Delta=2d_\mathrm{max,d}\tfrac{\sin\alpha}{(N-1)\cos(\alpha+|\theta_\mathrm{A}|)}$.
%
% \begin{equation}
%     \Delta=2d_\mathrm{max,d}\tfrac{\sin\alpha}{(N-1)\cos(\alpha+|\theta_\mathrm{A}|)}.
% \end{equation}
%
% \begin{align}
%     \Delta
%     &=2d_\mathrm{desired,1}\tfrac{\sin\alpha}{(N-1)(\cos(\alpha+|\theta_\mathrm{A}|)},
%     \\
%     \Delta
%     &=2d_\mathrm{desired,2}\tfrac{\sin\alpha}{(N-1)(\cos(\alpha-|\theta_\mathrm{A}|)}.
% \end{align}
%
Note that phase distributions are discretized by the positions of antenna elements in antenna array systems.
Thus, antenna spacing should satisfy the sampling theorem in the spatial domain to properly generate beams characterized by continuous phase distributions. %, which is originally based on the continuous linear phase distribution \cite{Durnin1987}, exactly.

% In this model, since the \ac{ULA} is located on the $x$-axis, the sampling theorem is given by $\Delta<\tfrac{\pi}{k_\mathrm{x}}$ \cite{Pizzo2022}, where $k_\mathrm{x}\in\mathbb{R}^+$ denotes the magnitude of the $x$-component of the wavenumber $k$, which satisfies $k=\sqrt{k_\mathrm{x}^2+k_\mathrm{y}^2}$ with $k_\mathrm{y}\in\mathbb{R}^+$ denoting the magnitude of the $y$-component of the wavenumber.
In this model, since the \ac{ULA} is located on the $x$-axis, the sampling theorem is given by $\Delta<\tfrac{\pi}{k_\mathrm{x}}$ \cite{Pizzo2022}, where $k_\mathrm{x}\in\mathbb{R}^+$ denotes the magnitude of the transverse component of the wave vector, which satisfies $k=\sqrt{k_\mathrm{x}^2+k_\mathrm{y}^2}$ with $k_\mathrm{y}\in\mathbb{R}^+$ denoting the magnitude of the longitudinal component of the wave vector.
From the sampling theorem, the upper bound of the antenna spacing to generate the Bessel beam with the \ac{ULA} is given by
\begin{equation}
    \Delta < \tfrac{\lambda}{2}\tfrac{1}{\sin(\alpha+|\theta_\mathrm{A}|)},\label{eq:sampling}
\end{equation}
where the derivation is given in Appendix E.

% Under the necessary and sufficient condition in \eqref{eq:max_steering}, the function $\tfrac{1}{\sin(\alpha+|\theta_\mathrm{A}|)}$ returns only values greater than 1.
Under the condition in \eqref{eq:max_steering}, the function $\tfrac{1}{\sin(\alpha+|\theta_\mathrm{A}|)}$ returns only values greater than 1.
To elaborate further, the \ac{ULA} with half-wavelength spacing $\Delta=\tfrac{\lambda}{2}$ can generate any Bessel beam under the condition in \eqref{eq:max_steering}.
% \textcolor{blue}{Thus}, the \ac{ULA} with \textcolor{blue}{spacing} $\Delta=\tfrac{\lambda}{2}$ can generate any Bessel beam under the condition in \eqref{eq:max_steering}.

As described in Section \ref{sec:prop_dist}, a smaller $\alpha$ leads to a longer maximum propagation distance $d_\mathrm{max}$.
%
% In other words, under the necessary and sufficient condition in \eqref{eq:max_steering}, setting the parameter to $\alpha=|\theta_\mathrm{A}|$ leads to the minimum number of antenna elements and the minimum antenna spacing to achieve the desired propagation distance.
In other words, the parameter given by $\alpha=|\theta_\mathrm{A}|$ leads to the minimum number of antenna elements and the minimum antenna spacing to achieve the desired propagation distance while ensuring the desired azimuth angle.
However, a smaller parameter $\alpha$ also leads to less robustness against blockages, which is described in detail in the next subsection.

%

%

%
%------
%

\subsection{Self-Healing Capability}
\label{sec:healing}

Bessel beams are capable of reconstructing themselves even if they are partially blocked by obstacles.
%
% This property is called self-healing \cite{Zhongsheng2024} and brings robustness against random blockages in wireless communications systems.
This self-healing capability brings robustness against random blockages. % in wireless communications systems.
In this subsection, reference points for the beginning of self-healing are considered in signal propagation with an obstacle.
In particular, communication links are assumed to be blocked by cuboid or cylinder obstacles, which are used to model human bodies \cite{Mukherjee2022}.
%
% Following those modeling, in this section, signal propagation with an obstacle modeled as a rectangular or a circle in the $xy$-plane in this system is considered.
%
% Note that this assumption does not limit the following discussion to the cases of blockages caused by cuboid and cylinder obstacles.
Note that this assumption does not limit the following discussion to the blockages caused by cuboid or cylinder obstacles.
%
% Rather, this assumption indicates that blockage effects and robustness against blockages for near-field beams can be estimated simply by approximating complicated shapes of obstacles as a rectangular or a circle in the $xy$-plane.
Rather, this assumption indicates that robustness against blockages for near-field beams can be estimated simply by approximating complicated shapes of obstacles as a rectangular or a circle in the $xy$-plane.

\subsubsection{Rectangular obstacles}
\label{sec:healing_r}

First, the rectangular obstacle located in the $xy$-plane is considered, whose vertices are denoted by the positions $\mathbf{p}_{\mathrm{r},1}=[x_{\mathrm{r},1},y_\mathrm{f}]^\mathrm{T}\in\mathbb{R}^2$, $\mathbf{p}_{\mathrm{r},2}=[x_{\mathrm{r},2},y_\mathrm{f}]^\mathrm{T}\in\mathbb{R}^2$, $\mathbf{p}_{\mathrm{r},3}=[x_{\mathrm{r},1},y_\mathrm{n}]^\mathrm{T}\in\mathbb{R}^2$, and $\mathbf{p}_{\mathrm{r},4}=[x_{\mathrm{r},2},y_\mathrm{n}]^\mathrm{T}\in\mathbb{R}^2$.
%
% In order to consider practical scenarios, the positions of the vertices and user satisfy the following relationship $0<y_\mathrm{n}<y_\mathrm{u}$, $0<y_\mathrm{n}<y_\mathrm{f}$, and $x_\mathrm{r,1}>x_\mathrm{r,2}$.
The positions of the vertices and the user satisfy the following relationship $0<y_\mathrm{n}<y_\mathrm{f}$, $0<y_\mathrm{n}<y_\mathrm{u}$, and $x_\mathrm{r,1}>x_\mathrm{r,2}$, as shown in Fig. \ref{fig:obs}.
% %
% Then, let $\mathbf{p}_{\mathrm{r},1}=[x_{\mathrm{r},1},y_\mathrm{f}]^\mathrm{T}\in\mathbb{R}^2$ and $\mathbf{p}_{\mathrm{r},2}=[x_{\mathrm{r},2},y_\mathrm{f}]^\mathrm{T}\in\mathbb{R}^2$ denote the position of vertices far from the $x$-axis, where $x_\mathrm{r,1}>x_\mathrm{r,2}$.
% %
% Based on geometric optics and the analysis in Section \ref{sec:prop_dist}, 
%
% the rays of the $n$-th antenna element perpendicular to the conical wavefront is can be described by the linear function given by 
% %
% \begin{equation}
%     f_\mathrm{l}(y)
%     =
%     \begin{cases}
%         -\tfrac{\tan\alpha-\tan\theta_\mathrm{A}}{1+\tan\alpha\tan\theta_\mathrm{A}}y + x_{\mathrm{t},n},~\mathrm{if}~x_{\mathrm{t},n} \ge0
%         \\
%         -\tfrac{\tan\alpha+\tan\theta_\mathrm{A}}{\tan\alpha\tan\theta_\mathrm{A}-1}y + x_{\mathrm{t},n},~\mathrm{otherwise}
%     \end{cases}
% \end{equation}
%
%
% Based on geometric optics and the analysis in Section \ref{sec:prop_dist}, the antenna elements, which are not affected by partial blockage effects, can be estimated by the conditions $f_\mathrm{r}(y_\mathrm{f})>x_{\mathrm{r},1}$ and $f_\mathrm{r}(y_\mathrm{f})<x_{\mathrm{r},2}$ for the antenna elements located in the positive and negative regions of the $x$-axis, respectively.
Based on the analysis in Section \ref{sec:prop_dist}, the antenna elements that are not affected by blockage effects can be estimated by the conditions $f_\mathrm{r}(y_\mathrm{f})>x_{\mathrm{r},1}$ and $f_\mathrm{r}(y_\mathrm{f})<x_{\mathrm{r},2}$ in the positive and negative regions of the $x$-axis, respectively.
% %
From \eqref{eq:linear}, those conditions can be rewritten as
\begin{equation}
    {\begin{cases}
        x_{\mathrm{t},n}>x_\mathrm{r,1}+\tan(\alpha-\theta_\mathrm{A}) y_\mathrm{f},~\mathrm{if}~x_{\mathrm{t},n}\ge 0
        \\
         x_{\mathrm{t},n}<x_\mathrm{r,2}-\tan(\alpha+\theta_\mathrm{A})y_\mathrm{f},~\mathrm{otherwise}
    \end{cases}}.
\end{equation}

Then, the reference points for the beginning of self-healing along the desired propagation direction are estimated by
%
% \begin{align}
%     d_\mathrm{h,p} &= \tfrac{x_\mathrm{p}^\star}{\tan\alpha}\tfrac{\cos(\alpha-\theta_\mathrm{A})}{\cos\alpha}, \label{eq:obs_min_dist_p}
%     \\
%     d_\mathrm{h,m} &= \tfrac{x_\mathrm{m}^\star}{\tan\alpha}\tfrac{\cos(\alpha+\theta_\mathrm{A})}{\cos\alpha}, \label{eq:obs_min_dist_m}
% \end{align}
\begin{equation}
    d_\mathrm{h,p} = {|x_\mathrm{p}^\star|}\tfrac{\cos(\alpha-\theta_\mathrm{A})}{\sin\alpha},~~
    d_\mathrm{h,m} = {|x_\mathrm{m}^\star|}\tfrac{\cos(\alpha+\theta_\mathrm{A})}{\sin\alpha}, \label{eq:obs_min_dist}
\end{equation}
where the derivation is the same as in Section \ref{sec:prop_dist} by replacing $R$ and $-R$ with $x_\mathrm{p}^\star$ and $x_\mathrm{m}^\star$, respectively, and the antenna positions $x_\mathrm{p}^\star$ and $x_\mathrm{m}^\star$ are determined by
\begin{align}
    x_\mathrm{p}^\star&\!\!=\!\!\underset{x\in\mathcal{X}_\mathrm{p}}{\min} x,~ \mathcal{X}_\mathrm{p}\!\!\triangleq\!\big\{x\!\in\!\mathcal{X}_\mathrm{arr}\!\mid\! x\!>\!x_\mathrm{r,1}\!\!+\tan(\alpha\!-\!\theta_\mathrm{A}) y_\mathrm{f}\big\},
    \\
    x_\mathrm{m}^\star&\!\!=\!\!\underset{x\in\mathcal{X}_\mathrm{m}}{\max} x,~\mathcal{X}_\mathrm{m}\!\!\triangleq\!\big\{\!x\!\in\!\mathcal{X}_\mathrm{arr}\!\!\mid\! x\!<\!x_\mathrm{r,2}\!-\tan(\alpha\!+\!\theta_\mathrm{A})y_\mathrm{f}\!\big\},
\end{align}
where the result $x_\mathrm{p}^\star<0$ means that the beam generated by the antenna elements located in the positive region of the $x$-axis is not blocked by the obstacle, and vice versa.

Let $d_\mathrm{UE}=\|\mathbf{p}_\mathrm{u}\|_2$ denote the distance between the origin of the coordinate system and the user.
From the above, the self-healing property of the Bessel beam is effective for the user located at the positions satisfying the conditions $d_\mathrm{h,p}<d_\mathrm{UE}$ or $d_\mathrm{h,m}<d_\mathrm{UE}$.
%
% From the above, \textcolor{blue}{the self-healing} is effective for the user located at the positions satisfying the conditions $d_\mathrm{h,p}<d_\mathrm{UE}$ or $d_\mathrm{h,m}<d_\mathrm{UE}$.
%
To elaborate further, shorter distances of $d_\mathrm{h,p}$ and $d_\mathrm{h,m}$ lead to greater robustness against blockages, which can be achieved by increasing $\alpha$.
%
% However, as mentioned in Section \ref{sec:prop_dist}, a larger $\alpha$ also leads to a shorter propagation distance, which may lead to the situations $d_\mathrm{max}<d_\mathrm{UE}$ or $d_\mathrm{lim}<d_\mathrm{UE}$.
%
However, as mentioned in Section \ref{sec:prop_dist}, a larger $\alpha$ also leads to a shorter propagation distance, which may lead to the situations $d_\mathrm{max}<d_\mathrm{UE}$ or $d_\mathrm{lim}<d_\mathrm{UE}$.
% However, \textcolor{blue}{a} larger $\alpha$ also leads to a shorter propagation distance, which may lead to the situations $d_\mathrm{max}<d_\mathrm{UE}$ or $d_\mathrm{lim}<d_\mathrm{UE}$.
%
Therefore, the parameter $\alpha$ should be carefully chosen based on the distances given by \eqref{eq:dist1}, \eqref{eq:dist2}, and \eqref{eq:obs_min_dist}.
% Thus, in the presence of obstacles, the parameter $\alpha$ should be carefully chosen based on the distances given by \eqref{eq:dist1}, \eqref{eq:dist2}, and \eqref{eq:obs_min_dist}.
%
% Furthermore, the result $x_\mathrm{+,\mathrm{arr}}^\star<0$ states that the beam generated by the antenna elements located in the region $x\ge0$ is not affected by the blockage effect.
% %
% Naturally, the vise verse is also true.

%
%
%

\subsubsection{Cylinder obstacles}
\label{sec:heal_cylinder}

% In the cases of the circle obstacle in the $xy$-plane, the information required to analyze the self-healing capability is the radius and position of the center of the obstacle, which are denoted by $r\in\mathbb{R}^+$ and $\mathbf{p}_\mathrm{c}\in[x_\mathrm{c},y_\mathrm{c}]^\mathrm{T}\in\mathbb{R}^2$, respectively.
In the cases of the circle obstacle in the $xy$-plane, the self-healing capability is analyzed based on the the radius and position of the center of the obstacle, which are denoted by $r\in\mathbb{R}^+$ and $\mathbf{p}_\mathrm{c}\in[x_\mathrm{c},y_\mathrm{c}]^\mathrm{T}\in\mathbb{R}^2$, respectively.

The tangent points between the circle and the rays perpendicular to the wavefront function are given by $(x_\mathrm{c,1},y_\mathrm{c,1})=(x_\mathrm{c}+r\cos(\alpha-\theta_\mathrm{A}), y_\mathrm{c}+r\sin(\alpha-\theta_\mathrm{A}))$ and $(x_\mathrm{c,2},y_\mathrm{c,2})=(x_\mathrm{c}-r\cos(\alpha+\theta_\mathrm{A}),y_\mathrm{c}+r\sin(\alpha+\theta_\mathrm{A}))$ for the regions $x\ge0$ and $x<0$, respectively.
% The tangent points between the circle and the \textcolor{blue}{direct rays from each antenna element} are given by $(x_\mathrm{c,1},y_\mathrm{c,1})=(x_\mathrm{c}+r\cos(\alpha-\theta_\mathrm{A}), y_\mathrm{c}+r\sin(\alpha-\theta_\mathrm{A}))$ and $(x_\mathrm{c,2},y_\mathrm{c,2})=(x_\mathrm{c}-r\cos(\alpha+\theta_\mathrm{A}),y_\mathrm{c}+r\sin(\alpha+\theta_\mathrm{A}))$ for the regions $x\ge0$ and $x<0$, respectively.
%
Since the relationship $x_\mathrm{c,1}>x_\mathrm{c,2}$ always holds under the necessary and sufficient condition in \eqref{eq:max_steering}, the conditions to avoid the obstacle can be described by
\begin{equation}
    {\begin{cases}
        x_{\mathrm{t},n}>x_\mathrm{c,1}+\tan(\alpha-\theta_\mathrm{A})y_\mathrm{c,1},~\mathrm{if}~x_{\mathrm{t},n}\ge 0
        \\
         x_{\mathrm{t},n}<x_\mathrm{c,2}-\tan(\alpha+\theta_\mathrm{A})y_\mathrm{c,2},~\mathrm{otherwise}
    \end{cases}}.
\end{equation}
The same analysis as in the rectangular case is established by replacing $(x_\mathrm{r,1}, y_\mathrm{f})$ with $(x_\mathrm{c,1},y_\mathrm{c,1})$ and $(x_\mathrm{r,2},y_\mathrm{f})$ with $(x_\mathrm{c,2},y_\mathrm{c,2})$ in the regions $x_{\mathrm{t},n}\ge0$ and $x_{\mathrm{t},n}<0$, respectively.
%

%
%-----
%

\section{Curving Beam Generation with a ULA}
\label{sec:curving}

% In this section, curving beams, whose wavefront is described by convex or concave functions, are discussed.
% %
% In particular, the design of the parameters for the curving beams based on the parabolic trajectories is proposed to avoid an obstacle.
%
In this section, designs of curving beams are proposed to reach a user while avoiding one obstacle.

%
%------
%

\subsection{Parabolic Trajectory and Phase Distribution}
\label{sec:trajectory}

In curving beam designs, their trajectories are defined first according to the movement of users or locations of users and obstacles to simplify the designs.
%
% Following the existing work \cite{Droulias2024}, the parabolic trajectory given by $ f_\mathrm{t}(y) = \beta(y-p)^2+q$ is considered, where the parameters to control the curvature $\beta$ and the vertex $(p,q)$ should be designed to achieve the desired trajectory.
Following \cite{Droulias2024}, the parabolic trajectory given by $ f_\mathrm{t}(y) = \beta(y-p)^2+q$ is considered, where the parameters to control the curvature $\beta$ and the vertex $(p,q)$ are designed to achieve the desired trajectory.

In particular, the parameters that satisfy the equation $x_\mathrm{u} = \beta(y_\mathrm{u}-p)^2 + q$ are required to connect the trajectory and the user located at $\mathbf{p}_\mathrm{u}$ \cite{Droulias2024}.
Thus, the parameter given by $q=x_\mathrm{u}-\beta(y_\mathrm{u}-p)^2$ transforms the trajectory function into
\begin{equation}
    f_\mathrm{t}(y)=\beta(y^2-y_\mathrm{u}^2)-2\beta p(y-y_\mathrm{u})+x_\mathrm{u}. \label{eq:c_traj}
\end{equation}

The envelope, created by tangent lines to a trajectory function, forms a curving beam, where the tangent lines describe rays from antenna elements \cite{Droulias2024,Gabriel2022,Guerboukha2024}.
%
% The phase distribution of the curving beam should be are designed, such that the ray of each antenna element has a tangent point to the trajectory function in \eqref{eq:c_traj} [].
%
% As a result, the set of tangent points can be seen as the desired trajectory.
%
Under the assumption on non-zero curvature in the parabolic trajectory $\beta\neq 0$, for the $n$-th antenna element, the $y$-coordinate of the tangent point to the trajectory function in \eqref{eq:c_traj} is given by \cite{Droulias2024}
\begin{equation}
     y_{\mathrm{t}}(x_{\mathrm{t},n})=\sqrt{\tfrac{\beta p^2 + q - x_{\mathrm{t},n}}{\beta}}=\sqrt{\tfrac{-\beta y_\mathrm{u}^2+2\beta p y_\mathrm{u}+x_\mathrm{u}- x_{\mathrm{t},n}}{\beta}}.
\end{equation}
where $y(x_{\mathrm{t},1})=y(-R)$ or $y(x_{\mathrm{t},N})=y(R)$ denotes the limitation of the propagation distance for the curving beam with the trajectory characterized by the function in \eqref{eq:c_traj}.

% \begin{align}
%     y_{\mathrm{t}}(x_{\mathrm{t},n})&=\sqrt{\tfrac{\beta p^2 + q - x_{\mathrm{t},n}}{\beta}},\nonumber
%     \\
%     &=\sqrt{\tfrac{-\beta y_\mathrm{u}^2+2\beta p y_\mathrm{u}+x_\mathrm{u}}{\beta}}.
% \end{align}

% Then, the continuous phase distribution to generate curving beams based on the parabolic trajectory $f_\mathrm{p}(y)$ is given by []
% %
% \begin{equation}
%     \phi(x)
%     =
%     k\int\tfrac{2\beta_\mathrm{}\Big(\sqrt{\tfrac{\beta_\mathrm{} p_\mathrm{}^2 + q - x}{\beta_\mathrm{}}}-p_\mathrm{}\Big)}{\sqrt{1+4\beta_\mathrm{}^2\Big(\sqrt{\tfrac{\beta_\mathrm{} p_\mathrm{}^2 + q - x}{\beta_\mathrm{}}}-p_\mathrm{}\Big)^2}} dx, \label{eq:C_phase}
% \end{equation}
% %
% which is the generalized version of the phase distribution given in \cite{Droulias2024} considering both positive and negative curvatures $\beta >0$ and $\beta<0$.

% The integral in \eqref{eq:C_phase} has the closed-form solutions [].
%
Consequently, the phase of the $n$-th antenna element to generate the curving beam is given by \cite{Droulias2024}
%
% \begin{equation}
%     \phi_n
%     =k\Big\{\tfrac{\log(\sqrt{c_1}-c_2)}{4|{\beta}|}-\beta(\tfrac{p}{2\beta} + \tfrac{1}{2\beta}\sqrt{\tfrac{(\beta p^2 + q - x_{\mathrm{t},n})}{\beta}})\sqrt{c_1}\Big\},\label{eq:c_phase}
% \end{equation}
\begin{equation}
    \phi_n
    =k\Big\{\tfrac{\log(\sqrt{c_1}-c_2)}{4|{\beta}|}-({p} + \sqrt{\tfrac{(\beta p^2 + q - x_{\mathrm{t},n})}{\beta}})\tfrac{\sqrt{c_1}}{2}\Big\},\label{eq:c_phase}
\end{equation}
where each constant is given by
\begin{align}
    c_1&\!=\!{4\beta^2p^2 \!+\! 4\beta(\beta p^2 \!+\! q \!-\! x_{\mathrm{t},n}) \!-\! 8\beta^2p\sqrt{\tfrac{\beta p^2 + q - x_{\mathrm{t},n}}{\beta}} + 1},\nonumber
    \\
    c_2&=\big({2\beta^2 p - 2 \beta^2\sqrt{\tfrac{\beta p^2 + q - x_{\mathrm{t},n}}{\beta}}}\big) \tfrac{1}{|{\beta}|}.\nonumber
\end{align}

Note that the phase in \eqref{eq:c_phase} is the generalized version of the phase given in \cite{Droulias2024} and is applicable to both positive and negative curvatures $\beta >0$ and $\beta<0$.
%
% The phase given by \eqref{eq:c_phase} and the parameter $q$ given by $q=x_\mathrm{u}-\beta(y_\mathrm{u}-p)^2$ mean that the design of the curving beam with the parabolic trajectory in \eqref{eq:c_traj} is equivalent to the design of the parameters $\beta$ and $p$.
From \eqref{eq:c_phase} and the parameter $q=x_\mathrm{u}-\beta(y_\mathrm{u}-p)^2$, the design of the curving beam with the parabolic trajectory in \eqref{eq:c_traj} is equivalent to the design of the parameters $\beta$ and $p$.
In the following subsections, the design of the parameters $\beta$ and $p$ in \eqref{eq:c_traj} to avoid one obstacle while maintaining user connectivity is considered.

% In the following sections, the design of the parameters $\beta$ and $p$ in \eqref{eq:wavefront_curving_re} to avoid an obstacle is considered based on the relationship between the tangent points and the positions of a user and an obstacle.
% %
% In the following subsection, it is assumed that the positions of the user and obstacle are available in the parameter design.

%
%------
%

\subsection{Problem formulation}
\label{sec:avoid_rec}

\begin{figure}[t]
\centering
 \begin{minipage}{0.40\columnwidth}
	\subfigure[{Positive curvature $\beta>0$}]
	{
    \includegraphics[width=\linewidth]{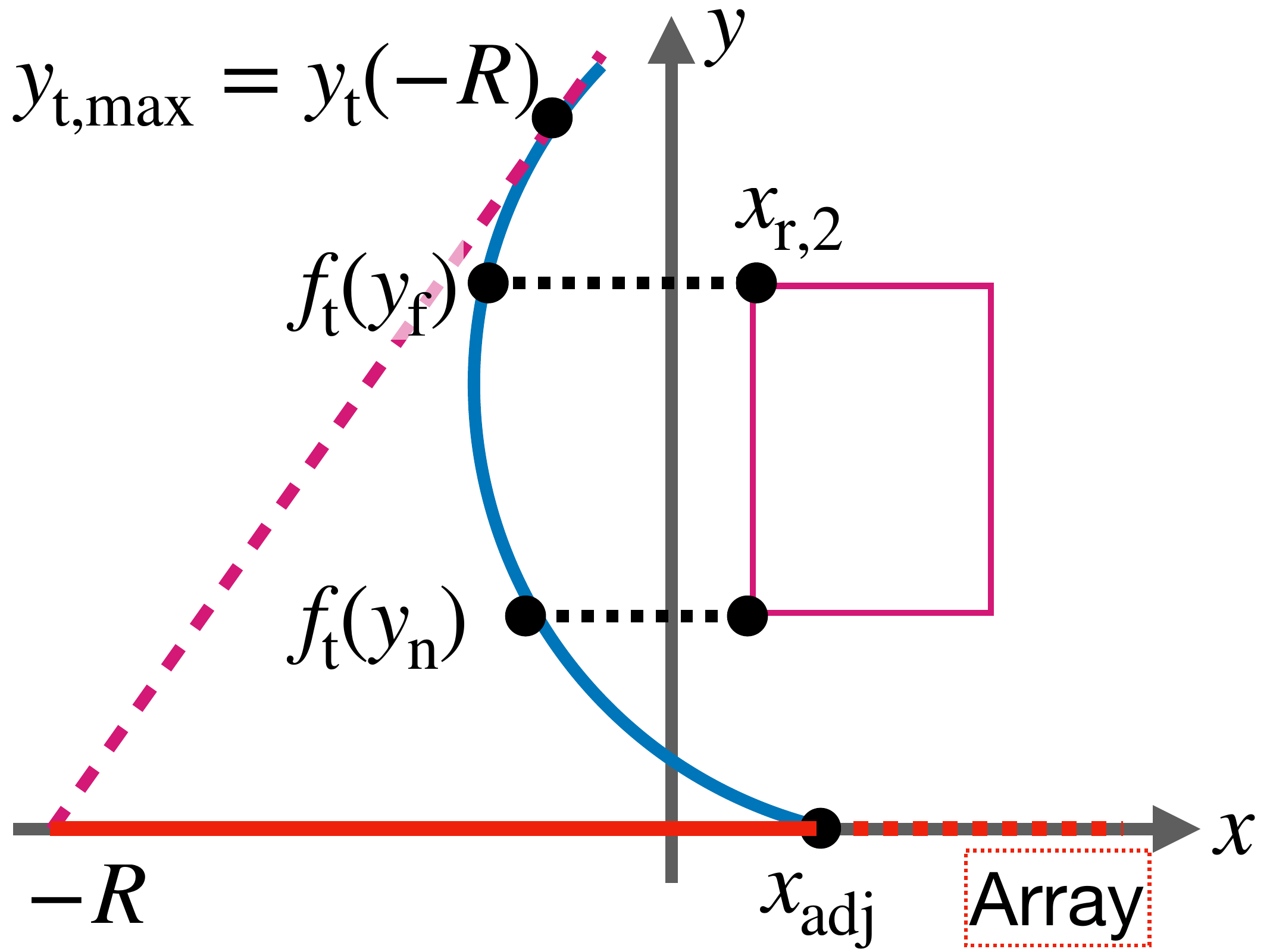}
	\label{fig:positive}
	}
	\end{minipage}
	\begin{minipage}{0.40\columnwidth}
	\subfigure[{Negative curvature $\beta<0$}]
	{
	\includegraphics[width=\linewidth]{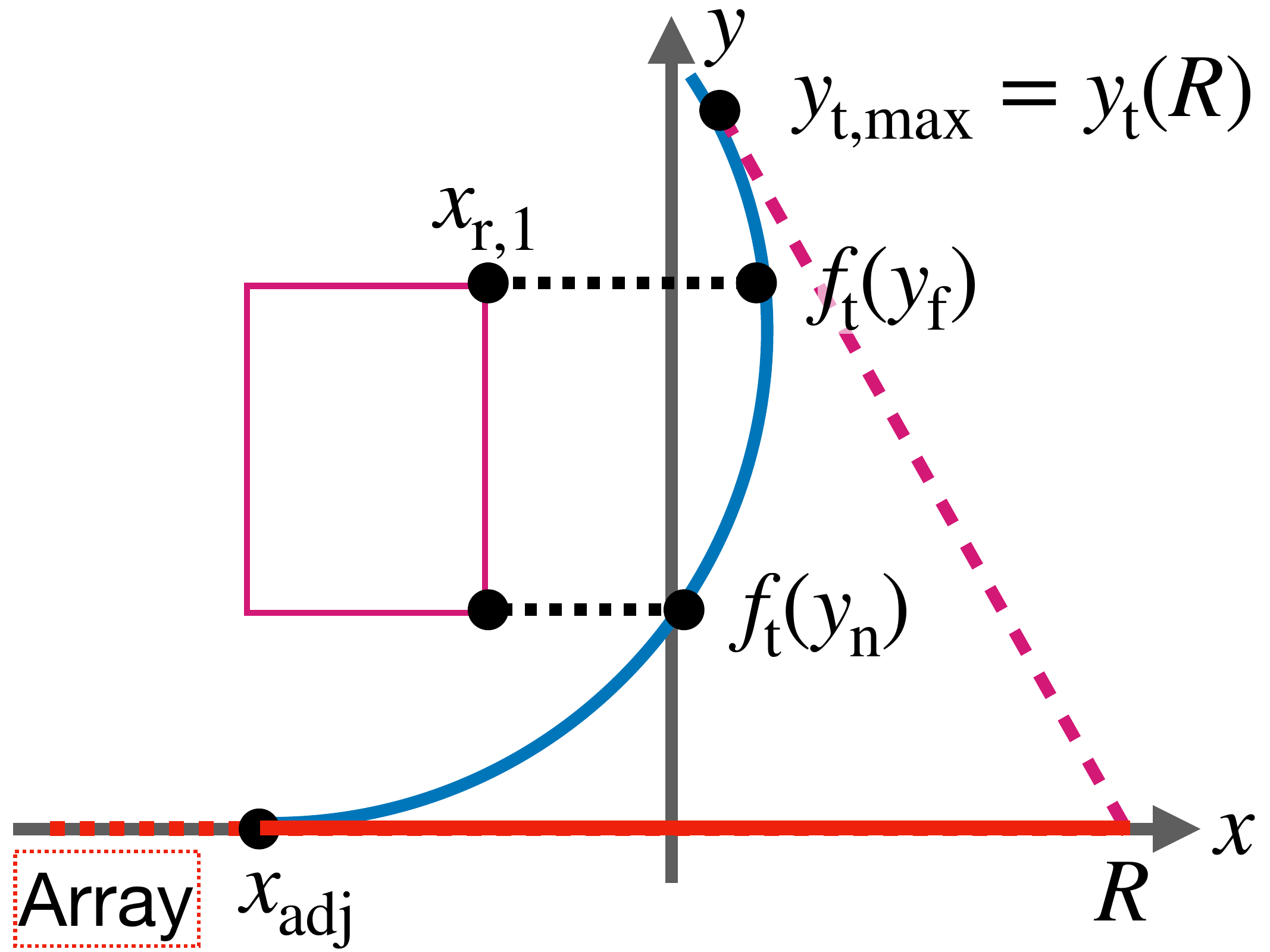}
	\label{fig:negative}
	}
	\end{minipage}
        \caption{Trajectory design for curving beams to avoid one obstacle}
	\label{fig:curving}
\end{figure}

First, the rectangular obstacle defined in Section \ref{sec:healing_r} and the parabolic trajectory given by \eqref{eq:c_traj} with positive curvature $\beta>0$ are considered.
%
% The rectangular obstacle defined in Section \ref{sec:healing_r} is considered.
% %
% First, the trajectory in \eqref{eq:c_traj} with positive curvature $\beta>0$ is considered.
%
In this case, as shown in Fig. \ref{fig:positive}, the parameter design to avoid the obstacle is formulated as
\begin{subequations}
  \begin{align}
  \label{opt:find}
      \mathrm{Find}&~\beta>0,~p
      \\
      \mathrm{subject~to}~&f_\mathrm{t}(y_\mathrm{n})\le x_\mathrm{r,2},
      \\
      &f_\mathrm{t}(y_\mathrm{f})\le x_\mathrm{r,2},
      \\
      &y_\mathrm{t,\mathrm{min}}\le y_\mathrm{u},
      \\
      &y_\mathrm{u}\le y_\mathrm{t,\mathrm{max}},
  \end{align}
\end{subequations}
% {\color{blue}
% \begin{subequations}
%   \begin{align}
%   \label{opt:find}
%       \mathrm{Find}&~\beta>0,~p
%       \\
%       \mathrm{subject~to}~&f_\mathrm{t}(y_\mathrm{n})\le x_\mathrm{r,2},~f_\mathrm{t}(y_\mathrm{f})\le x_\mathrm{r,2},
%       \\
%       &y_\mathrm{t,\mathrm{min}}\le y_\mathrm{u},
%       \\
%       &y_\mathrm{u}\le y_\mathrm{t,\mathrm{max}},
%   \end{align}
% \end{subequations}
%
where $y_\mathrm{t,\mathrm{min}}\in\mathbb{R}^+$ and $y_\mathrm{t,\mathrm{max}}\in\mathbb{R}^+$ denote the closest and farthest $y$-coordinates of the tangent points from the origin of the coordinate system, respectively.

% The definitions of the $y$-coordinates given by $y_\mathrm{t,\mathrm{min}}\triangleq y_{\mathrm{t}}(R)$ and $y_\mathrm{t,\mathrm{max}}\triangleq y_{\mathrm{t}}(-R)$ lead to the curving beam with all the antenna elements.
The $y$-coordinates defined by $y_\mathrm{t,\mathrm{min}}\triangleq y_{\mathrm{t}}(R)$ and $y_\mathrm{t,\mathrm{max}}\triangleq y_{\mathrm{t}}(-R)$ lead to the curving beam with all the antenna elements.
%
% However, as described in Section \ref{sec:healing}, the positions of the antenna elements affect robustness against blockages.
However, the positions of the antenna elements affect robustness against blockages.
Moreover, exploiting all antenna elements may be impossible depending on the relationship between the positions of the user and the obstacle.
From those perspectives, the curving beam with all antenna elements is not always optimal to maximize robustness against blockages, which is the motivation to jointly optimize $\beta$, $p$, and the number of antenna elements to exploit for beam generation.

Geometrically, the propagation distance is maximized only by $y_\mathrm{t,\mathrm{max}}=y_\mathrm{t}(-R)$ under the assumption $\beta>0$, regardless of $\beta$ and $p$.
%
% To elaborate further, if the definition of $y_\mathrm{t,\mathrm{max}}=y_\mathrm{t}(-R)$ is impossible in the problem in \eqref{opt:find}, the curvature of the trajectory should be set to negative $\beta<0$.
%
% Meanwhile, any antenna element can achieve the minimum tangent point given by $y_{\mathrm{t},\mathrm{min}}=0$ by adjusting the parameters $\beta$ and $p$.
Meanwhile, any antenna element can achieve the minimum tangent point $y_{\mathrm{t},\mathrm{min}}=0$ by adjusting the parameters $\beta$ and $p$.
%
% To elaborate further, the number of antenna elements to exploit for beam generation is optimized based on $y_\mathrm{t,\mathrm{min}}$ with fixed $y_\mathrm{t,\mathrm{max}}=y_\mathrm{t}(-R)$.
% %
% From the above, the parameter design is reformulated as % problem in \eqref{opt:find} is transformed into
%
From the above, the number of antenna elements to exploit for beam generation is optimized based on $y_\mathrm{t,\mathrm{min}}$ with fixed $y_\mathrm{t,\mathrm{max}}=y_\mathrm{t}(-R)$, which yields the following problem
%
% \begin{subequations}
% \begin{align}
% \label{opt:find2}
%     \mathrm{Find}&~\beta>0,~p,~x_\mathrm{adj},
%     \\
%     \mathrm{subject~to}~&x_\mathrm{adj} \in \mathcal{X}_\mathrm{arr},\label{const:array}
%     \\
%     &\beta(y_\mathrm{n}^2-y_\mathrm{u}^2)-2\beta p(y_\mathrm{n}-y_\mathrm{u})+x_\mathrm{u}\le x_\mathrm{r,2},\label{const:avoid_n}
%     \\
%     &\beta(y_\mathrm{f}^2-y_\mathrm{u}^2)-2\beta p(y_\mathrm{f}-y_\mathrm{u})+x_\mathrm{u}\le x_\mathrm{r,2},\label{const:avoid_f}
%     \\
%     &\sqrt{\tfrac{-\beta y_\mathrm{u}^2+2\beta p y_\mathrm{u}+x_\mathrm{u}- x_{\mathrm{adj}}}{\beta}} \le y_\mathrm{u},\label{const:min_tangent}
%     \\
%     &y_\mathrm{u}\le \sqrt{\tfrac{-\beta y_\mathrm{u}^2+2\beta p y_\mathrm{u}+x_\mathrm{u}+ R}{\beta}}, \label{const:max_tangent}
%     \\
%     &
%     {-\beta y_\mathrm{u}^2+2\beta p y_\mathrm{u}+x_\mathrm{u}- x_{\mathrm{adj}}}\ge 0, \label{const:sqrt}
% \end{align}
% \end{subequations}
%
\begin{subequations}
\begin{align}
\label{opt:find2}
    \mathrm{Find}&~\beta>0,~p,~x_\mathrm{adj},
    \\
    \mathrm{subject~to}~&x_\mathrm{adj} \in \mathcal{X}_\mathrm{arr},\label{const:array}
    \\
    &\beta(y_\mathrm{n}^2-y_\mathrm{u}^2)-2\beta p(y_\mathrm{n}-y_\mathrm{u})+x_\mathrm{u}\le x_\mathrm{r,2},\label{const:avoid_n}
    % \\
    % &\beta(y_\mathrm{f}^2-y_\mathrm{u}^2)-2\beta p(y_\mathrm{f}-y_\mathrm{u})+x_\mathrm{u}\le x_\mathrm{r,2},\label{const:avoid_f}
    % \\
    % &\sqrt{\tfrac{-\beta y_\mathrm{u}^2+2\beta p y_\mathrm{u}+x_\mathrm{u}- x_{\mathrm{adj}}}{\beta}} \le y_\mathrm{u},\label{const:min_tangent}
    % \\
    % &y_\mathrm{u}\le \sqrt{\tfrac{-\beta y_\mathrm{u}^2+2\beta p y_\mathrm{u}+x_\mathrm{u}+ R}{\beta}}, \label{const:max_tangent}
    % \\
    % &
    % {-\beta y_\mathrm{su}^2+2\beta p y_\mathrm{u}+x_\mathrm{u}- x_{\mathrm{adj}}}\ge 0, \label{const:sqrt}
\end{align}
\begin{align}
    % \mathrm{Find}&~\beta>0,~p,~x_\mathrm{adj},
    % \\
    % \mathrm{subject~to}~&x_\mathrm{adj} \in \mathcal{X}_\mathrm{arr},\label{const:array}
    % \\
    % &\beta(y_\mathrm{n}^2-y_\mathrm{u}^2)-2\beta p(y_\mathrm{n}-y_\mathrm{u})+x_\mathrm{u}\le x_\mathrm{r,2},\label{const:avoid_n}
    % \\
    &\beta(y_\mathrm{f}^2-y_\mathrm{u}^2)-2\beta p(y_\mathrm{f}-y_\mathrm{u})+x_\mathrm{u}\le x_\mathrm{r,2},\label{const:avoid_f}
    \\
    &\sqrt{\tfrac{-\beta y_\mathrm{u}^2+2\beta p y_\mathrm{u}+x_\mathrm{u}- x_{\mathrm{adj}}}{\beta}} \le y_\mathrm{u},\label{const:min_tangent}
    \\
    &y_\mathrm{u}\le \sqrt{\tfrac{-\beta y_\mathrm{u}^2+2\beta p y_\mathrm{u}+x_\mathrm{u}+ R}{\beta}}, \label{const:max_tangent}
    \\
    &
    {-\beta y_\mathrm{u}^2+2\beta p y_\mathrm{u}+x_\mathrm{u}- x_{\mathrm{adj}}}\ge 0, \label{const:sqrt}
\end{align}
\end{subequations}
where the constraint in \eqref{const:sqrt} is the sufficient condition for $\tfrac{-\beta y_\mathrm{u}^2+2\beta{p} y_\mathrm{u}+x_\mathrm{u}+ R}{\beta}\ge 0$ and $\tfrac{-\beta y_\mathrm{u}^2+2\beta{p} y_\mathrm{u}+x_\mathrm{u}- x_{\mathrm{adj}}}{\beta}\ge 0$ owing to $R\ge x_\mathrm{adj}$ and $\beta > 0$.
To improve robustness against blockage, the differences between the trajectory and the obstacle $f_\mathrm{t}(y_\mathrm{n})-x_\mathrm{r,2}$ and $f_\mathrm{t}(y_\mathrm{f})-x_\mathrm{r,2}$ should be minimized, which is equivalent to the maximization of $x_\mathrm{r,2}-f_\mathrm{t}(y_\mathrm{n})$ and $x_\mathrm{r,2}-f_\mathrm{t}(y_\mathrm{f})$.
In addition, to ensure near-field communications, the number of antenna elements to exploit for beam generation should be maximized, as shown in Fig. \ref{fig:positive}.
%
% In this context, the objective function is given by the combination of the minimization of the differences and the maximization of $x_\mathrm{adj}$, which yields
%
Thus, the objective for the parameter design is given by the linear combination of the minimization of the differences and the position $-x_\mathrm{adj}$, which yields
\begin{subequations}
\label{opt:minimization_r}
\begin{align}
    \underset{\beta>0,p,x_\mathrm{adj}}{\mathrm{minimize}}&~f_\mathrm{para}
    \\
    \mathrm{subject~to}&~\eqref{const:array}, \eqref{const:avoid_n}, \eqref{const:avoid_f}, \eqref{const:min_tangent}, \eqref{const:max_tangent}, \eqref{const:sqrt},
    % \eqref{const:array}-\eqref{const:sqrt},
\end{align}
\end{subequations}
where the objective function is given by
\begin{equation}
    f_\mathrm{para} = \beta(y_\mathrm{n}^2\!+\!y_\mathrm{f}^2\!-\!2y_\mathrm{u}^2)+2\beta{p}(2y_\mathrm{u}\!-\!y_\mathrm{n}\!-\!y_\mathrm{f})-wx_\mathrm{adj},\label{eq:c_obj}
\end{equation}
where $w\in\mathbb{R}^{+}$ is the hyperparameter to balance blockage avoidance and full exploitation of the \ac{ULA}.
%
% The tuning of $w$ is described in Section \ref{sec:closed_curving}.

The optimization problem in \eqref{opt:minimization_r} cannot be solved directly due to the coupling between $\beta$ and $p$, discrete constraint in \eqref{const:array}, fractional expression inside the square root function in \eqref{const:min_tangent} and \eqref{const:max_tangent}, and open set $\beta>0$.
The new variable $\tilde{p}=\beta p$ is introduced and optimized instead of $p$ to overcome the coupling issue.
The discrete constraint in \eqref{const:array} is relaxed into the continuous constraint $-R\le x_\mathrm{adj}\le R$.
Based on the fact that both sides in the inequalities in \eqref{const:min_tangent} and \eqref{const:max_tangent} are positive, the inequalities ${{-\beta y_\mathrm{u}^2+2\beta p y_\mathrm{u}+x_\mathrm{u}- x_{\mathrm{max,arr}}}} \le {\beta}y_\mathrm{u}^2$ and ${\beta}y_\mathrm{u}^2\le {{-\beta y_\mathrm{u}^2+2\beta p y_\mathrm{u}+x_\mathrm{u}+ R}}$ are equivalent to \eqref{const:min_tangent} and \eqref{const:max_tangent}, respectively.
Finally, since the optimal solution of $\beta=0$ indicates that the curving beam is unnecessary to avoid the obstacle, the open set $\beta>0$ is replaced by the set of non-negative values $\beta\ge0$.
As a result, the optimization problem to design the curving beam is reformulated as
%
% \begin{subequations}
% \label{opt:final_r_p}
% \begin{align}
%     \underset{\beta\ge0,\tilde{p},x_\mathrm{adj}}{\mathrm{minimize}}&~f_\mathrm{para}, \label{obj:r_p}
%     \\
%     \mathrm{subject~to}&~-R\le x_\mathrm{adj} \le R,\label{const:array_final}
%     \\
%     &\beta(y_\mathrm{n}^2-y_\mathrm{u}^2)-2\tilde{p}(y_\mathrm{n}-y_\mathrm{u})+x_\mathrm{u}\le x_\mathrm{r,2},\label{const:avoid_n_finala}
%     \\
%     &\beta(y_\mathrm{f}^2-y_\mathrm{u}^2)-2\tilde{p}(y_\mathrm{f}-y_\mathrm{u})+x_\mathrm{u}\le x_\mathrm{r,2},\label{const:avoid_f_final}
%     \\
%     &{-2\beta y_\mathrm{u}^2+2\tilde{p} y_\mathrm{u}+x_\mathrm{u}- x_{\mathrm{adj}}} \le 0,\label{const:min_tangent_final}
%     \\
%     &{-2\beta y_\mathrm{u}^2+2\tilde{p} y_\mathrm{u}+x_\mathrm{u}+R} \ge 0, \label{const:max_tangent_final}
%     \\
%     &
%     {-\beta y_\mathrm{u}^2+2\tilde{p} y_\mathrm{u}+x_\mathrm{u}- x_{\mathrm{adj}}}\ge 0. \label{const:sqrt_final}
% \end{align}
% \end{subequations}
\begin{subequations}
\label{opt:final_r_p}
\begin{align}
    \underset{\beta\ge0,\tilde{p},x_\mathrm{adj}}{\mathrm{minimize}}&~f_\mathrm{para}, \label{obj:r_p}
    \\
    \mathrm{subject~to}&~-R\le x_\mathrm{adj} \le R,\label{const:array_final}
    \\
    &\beta(y_\mathrm{n}^2-y_\mathrm{u}^2)-2\tilde{p}(y_\mathrm{n}-y_\mathrm{u})+x_\mathrm{u}\le x_\mathrm{r,2},\label{const:avoid_n_finala}
    \\
    &\beta(y_\mathrm{f}^2-y_\mathrm{u}^2)-2\tilde{p}(y_\mathrm{f}-y_\mathrm{u})+x_\mathrm{u}\le x_\mathrm{r,2},\label{const:avoid_f_final}
    \\
    &{-2\beta y_\mathrm{u}^2+2\tilde{p} y_\mathrm{u}+x_\mathrm{u}- x_{\mathrm{adj}}} \le 0,\label{const:min_tangent_final}
    \\
    &{-2\beta y_\mathrm{u}^2+2\tilde{p} y_\mathrm{u}+x_\mathrm{u}+R} \ge 0, \label{const:max_tangent_final}
    \\
    &
    {-\beta y_\mathrm{u}^2+2\tilde{p} y_\mathrm{u}+x_\mathrm{u}- x_{\mathrm{adj}}}\ge 0. \label{const:sqrt_final}
\end{align}
\end{subequations}
which can be solved via linear programming solvers.

Let $x_{\mathrm{adj}}^\star$ denote the optimal solution of $x_{\mathrm{adj}}$.
In the cases of $\beta>0$, the relationship $y_\mathrm{t}(x_{\mathrm{t},n_1}) \ge y_\mathrm{t}(x_{\mathrm{t},n_2}),(x_{\mathrm{t},n_1}\le x_{\mathrm{t},n_2})$ holds.
Thus, the full exploitation of the \ac{ULA} under the constraints in \eqref{const:array_final}, \eqref{const:min_tangent_final}, and \eqref{const:sqrt_final}  is achieved by the following projection of $x_{\mathrm{adj}}^\star$ into the domain $\mathcal{X}_\mathrm{arr}$
\begin{equation}
    x^\star_{\mathrm{t}}\!=\!\underset{x\in x_\mathrm{cand}}{\mathrm{argmin}}~|x_\mathrm{adj}^\star-x|,~ \mathcal{X}_\mathrm{cand}\!\triangleq\! \{x\in\mathcal{X}_\mathrm{arr}\mid x_\mathrm{adj}^\star\ge x\}. \label{eq:max_array}
\end{equation}

By replacing $x_\mathrm{adj}$ with fixed $x^\star_{\mathrm{t}}$, the optimization problem in \eqref{opt:final_r_p} is solved again, resulting in the optimal solutions $\beta^\star$ and $\tilde{p}^\star$.
Since the parameter $p$ can take any real number, the optimal solution of $p$ is calculated as $p^\star = \tfrac{\tilde{p}^\star}{\beta^\star}$.

The infeasible region of the problem in \eqref{opt:final_r_p} shows the relationship between the positions of the user, the \ac{ULA}, and the obstacle, where any curving beam based on the trajectory in \eqref{eq:c_traj} with positive curvature cannot avoid the obstacle.

In such a situation, one possible option to generate the curving beam is to design trajectories with negative curvature $\beta<0$.
The basic idea of the design for trajectories with negative curvature is the same as those with positive curvature.
However, the objective function and the constraints in \eqref{const:avoid_n_finala} through \eqref{const:sqrt_final} should be modified based on negative curvature and the positions of the obstacle, as shown in Fig. \ref{fig:negative}.
Specifically, the optimization problem to design the curving beam with the parabolic trajectory in \eqref{eq:c_traj} with negative curvature is formulated as
\begin{subequations}
\label{opt:final_r_m}
\begin{align}
    \underset{\beta\le0,\tilde{p},x_\mathrm{adj}}{\mathrm{maximize}}&~f_\mathrm{para}
    \\
    \mathrm{subject~to}~&-R\le x_\mathrm{adj} \le R,\label{const:array_final_m}
    \\
    &~\beta(y_\mathrm{n}^2-y_\mathrm{u}^2)-2\tilde{p}(y_\mathrm{n}-y_\mathrm{u})+x_\mathrm{u}\ge x_\mathrm{r,1},\label{const:avoid_n_final_m}
    \\
    &\beta(y_\mathrm{f}^2-y_\mathrm{u}^2)-2\tilde{p}(y_\mathrm{f}-y_\mathrm{u})+x_\mathrm{u}\ge x_\mathrm{r,1},\label{const:avoid_f_final_m}
    \\
    &{-2\beta y_\mathrm{u}^2+2\tilde{p} y_\mathrm{u}+x_\mathrm{u}- R} \le 0,\label{const:min_tangent_final_m}
    \\
    &{-2\beta y_\mathrm{u}^2+2\tilde{p} y_\mathrm{u}+x_\mathrm{u}- x_{\mathrm{adj}}} \ge 0, \label{const:max_tangent_final_m}
    \\
    &
    {-\beta y_\mathrm{u}^2+2\tilde{p} y_\mathrm{u}+x_\mathrm{u}- x_{\mathrm{adj}}}\le 0, \label{const:sqrt_final_m}
\end{align}
\end{subequations}
where solutions can be obtained in the same manner as in the design for positive curvature $\beta>0$.
Note that the projection of the optimal solution $x_{\mathrm{adj}}^\star$ into the domain $\mathcal{X}_\mathrm{arr}$ is based on the following criterion
\begin{equation}
    x^\star_{\mathrm{t}}\!=\!\underset{x\in x_\mathrm{cand}}{\mathrm{argmin}}~|x_\mathrm{adj}^\star-x|,~\mathcal{X}_\mathrm{cand}\!\triangleq\! \{x\in\mathcal{X}_\mathrm{arr}\mid x_\mathrm{adj}^\star\le x\}. \label{eq:min_array}
\end{equation}

%
%----
%

The above parameter designs are also effective in the cases of circle obstacles. 
%
% In those cases, the same optimization problems in \eqref{opt:final_r_p} and \eqref{opt:final_r_m} can be used by approximating the circle obstacle as the square obstacle, whose vertices are given by $\mathbf{p}_{\mathrm{r,1}}=[x_\mathrm{r,1},y_\mathrm{f}]^\mathrm{T}=[x_\mathrm{c}+\tfrac{r}{2},y_\mathrm{c}+\tfrac{r}{2}]^\mathrm{T}\in\mathbb{R}^2$, $\mathbf{p}_{\mathrm{r,2}}=[x_\mathrm{r,2},y_\mathrm{f}]^\mathrm{T}=[x_\mathrm{c}-\tfrac{r}{2},y_\mathrm{c}+\tfrac{r}{2}]^\mathrm{T}\in\mathbb{R}^2$, $\mathbf{p}_{\mathrm{r,3}}=[x_\mathrm{r,1},y_\mathrm{n}]^\mathrm{T}=[x_\mathrm{c}+\tfrac{r}{2},y_\mathrm{c}-\tfrac{r}{2}]^\mathrm{T}\in\mathbb{R}^2$, and $\mathbf{p}_{\mathrm{r,4}}=[x_\mathrm{r,2},y_\mathrm{n}]^\mathrm{T}=[x_\mathrm{c}-\tfrac{r}{2},y_\mathrm{c}-\tfrac{r}{2}]^\mathrm{T}\in\mathbb{R}^2$, respectively.
%
In those cases, the same optimization problems in \eqref{opt:final_r_p} and \eqref{opt:final_r_m} can be used by approximating the circle obstacle as the square obstacle, whose vertices are given by $\mathbf{p}_{\mathrm{r,1}}=[x_\mathrm{r,1},y_\mathrm{f}]^\mathrm{T}=[x_\mathrm{c}+r,y_\mathrm{c}+{r}]^\mathrm{T}\in\mathbb{R}^2$, $\mathbf{p}_{\mathrm{r,2}}=[x_\mathrm{r,2},y_\mathrm{f}]^\mathrm{T}=[x_\mathrm{c}-{r},y_\mathrm{c}+{r}]^\mathrm{T}\in\mathbb{R}^2$, $\mathbf{p}_{\mathrm{r,3}}=[x_\mathrm{r,1},y_\mathrm{n}]^\mathrm{T}=[x_\mathrm{c}+{r},y_\mathrm{c}-{r}]^\mathrm{T}\in\mathbb{R}^2$, and $\mathbf{p}_{\mathrm{r,4}}=[x_\mathrm{r,2},y_\mathrm{n}]^\mathrm{T}=[x_\mathrm{c}-{r},y_\mathrm{c}-{r}]^\mathrm{T}\in\mathbb{R}^2$, respectively.

\begin{figure*}[t]
  \begin{align}
  \label{eq:closed_curving}
      (\beta,\tilde{p},x_\mathrm{adj})
      =
      \begin{cases}
          \Big(
          \tfrac{-(x_\mathrm{r,2} - x_\mathrm{u})}{(y_\mathrm{f} - y_\mathrm{u})(y_\mathrm{n} - y_\mathrm{u})},~
          \tfrac{-(x_\mathrm{r,2}y_\mathrm{f} - x_\mathrm{u}y_\mathrm{f} + x_\mathrm{r,2}y_\mathrm{n} - x_\mathrm{u}y_\mathrm{n})}{2(y_\mathrm{f} - y_\mathrm{u})(y_\mathrm{n} - y_\mathrm{u})},~
          \tfrac{x_\mathrm{r,2}y_\mathrm{u}^2 + x_\mathrm{u}y_\mathrm{f}y_\mathrm{n} - x_\mathrm{r,2}y_\mathrm{f}y_\mathrm{u} - x_\mathrm{r,2}y_\mathrm{n}y_\mathrm{u}}{(y_\mathrm{f} - y_\mathrm{u})(y_\mathrm{n} - y_\mathrm{u})}
          \Big)
          \\
          \Big(
          \tfrac{-(Ry_\mathrm{f} - Ry_\mathrm{u} + x_\mathrm{u}y_\mathrm{f} - x_\mathrm{r,2}y_\mathrm{u})}{y_\mathrm{u}(y_\mathrm{f} - y_\mathrm{u})^2},~
          \tfrac{-(Ry_\mathrm{f}^2 - Ry_\mathrm{u}^2 + x_\mathrm{u}y_\mathrm{f}^2 - 2x_\mathrm{r,2}y_\mathrm{u}^2 + x_\mathrm{u}y_\mathrm{u}^2)}{2y_\mathrm{u}(y_\mathrm{f} - y_\mathrm{u})^2},~
          \tfrac{-(Ry_\mathrm{f}^2 - x_\mathrm{r,2}y_\mathrm{u}^2 - Ry_\mathrm{f}y_\mathrm{u} + x_\mathrm{u}y_\mathrm{f}y_\mathrm{u})}{(y_\mathrm{f} - y_\mathrm{u})^2}
          \Big)
          \\
          \Big(
          \tfrac{-(Ry_\mathrm{n} - Ry_\mathrm{u} + x_\mathrm{u}y_\mathrm{n} - x_\mathrm{r,2}y_\mathrm{u})}{y_\mathrm{u}(y_\mathrm{n} - y_\mathrm{u})^2},~
          \tfrac{-(Ry_\mathrm{n}^2 - Ry_\mathrm{u}^2 + x_\mathrm{u}y_\mathrm{n}^2 - 2x_\mathrm{r,2}y_\mathrm{u}^2 + x_\mathrm{u}y_\mathrm{u}^2)}{2y_\mathrm{u}(y_\mathrm{n} - y_\mathrm{u})^2},~
          \tfrac{-(Ry_\mathrm{n}^2 - x_\mathrm{r,2}y_\mathrm{u}^2 - Ry_\mathrm{n}y_\mathrm{u} + x_\mathrm{u}y_\mathrm{n}y_\mathrm{u})}{(y_\mathrm{n} - y_\mathrm{u})^2}
          \Big)
          \\
          \Big(
          \tfrac{Ry_\mathrm{f} - Ry_\mathrm{u} - x_\mathrm{u}y_\mathrm{f} + x_\mathrm{r,2}y_\mathrm{u}}{y_\mathrm{f}y_\mathrm{u}(y_\mathrm{f} - y_\mathrm{u})},~
          \tfrac{Ry_\mathrm{f}^2 - Ry_\mathrm{u}^2 - x_\mathrm{u}y_\mathrm{f}^2 + x_\mathrm{r,2}y_\mathrm{u}^2}{2y_\mathrm{f}y_\mathrm{u}(y_\mathrm{f} - y_\mathrm{u})},~
          R
          \Big)
          \\
          \Big(
          \tfrac{Ry_\mathrm{n} - Ry_\mathrm{u} - x_\mathrm{u}y_\mathrm{n} + x_\mathrm{r,2}y_\mathrm{u}}{y_\mathrm{n}y_\mathrm{u}(y_\mathrm{n} - y_\mathrm{u})},~
          \tfrac{Ry_\mathrm{n}^2 - Ry_\mathrm{u}^2 - x_\mathrm{u}y_\mathrm{n}^2 + x_\mathrm{r,2}y_\mathrm{u}^2}{2y_\mathrm{n}y_\mathrm{u}(y_\mathrm{n} - y_\mathrm{u})},~
          R
          \Big)
          \\
          \Big(
          \tfrac{-(Ry_\mathrm{f} - Ry_\mathrm{u} + x_\mathrm{u}y_\mathrm{f} - x_\mathrm{r,2}y_\mathrm{u})}{y_\mathrm{u}(y_\mathrm{f} - y_\mathrm{u})^2},~
          \tfrac{-(Ry_\mathrm{f}^2 - Ry_\mathrm{u}^2 + x_\mathrm{u}y_\mathrm{f}^2 - 2x_\mathrm{r,2}y_\mathrm{u}^2 + x_\mathrm{u}y_\mathrm{u}^2)}{2y_\mathrm{u}(y_\mathrm{f} - y_\mathrm{u})^2},~
          R
          \Big)
          \\
          \Big(
          \tfrac{-(Ry_\mathrm{n} - Ry_\mathrm{u} + x_\mathrm{u}y_\mathrm{n} - x_\mathrm{r,2}y_\mathrm{u})}{y_\mathrm{u}(y_\mathrm{n} - y_\mathrm{u})^2},~
          \tfrac{-(Ry_\mathrm{n}^2 - Ry_\mathrm{u}^2 + x_\mathrm{u}y_\mathrm{n}^2 - 2x_\mathrm{r,2}y_\mathrm{u}^2 + x_\mathrm{u}y_\mathrm{u}^2)}{2y_\mathrm{u}(y_\mathrm{n} - y_\mathrm{u})^2},~
          R
          \Big)
          \\
          \Big(
          \tfrac{Ry_\mathrm{f} - Ry_\mathrm{u} - x_\mathrm{u}y_\mathrm{f} + x_\mathrm{r,2}y_\mathrm{u}}{y_\mathrm{u}(y_\mathrm{f} - y_\mathrm{u})^2},~
          \tfrac{-(Ry_\mathrm{u}^2 - Ry_\mathrm{f}^2 + x_\mathrm{u}y_\mathrm{f}^2 - 2x_\mathrm{r,2}y_\mathrm{u}^2 + x_\mathrm{u}y_\mathrm{u}^2)}{2y_\mathrm{u}(y_\mathrm{f} - y_\mathrm{u})^2},~
          R
          \Big)
          \\
          \Big(
          \tfrac{Ry_\mathrm{n} - Ry_\mathrm{u} - x_\mathrm{u}y_\mathrm{n} + x_\mathrm{r,2}y_\mathrm{u}}{y_\mathrm{u}(y_\mathrm{n} - y_\mathrm{u})^2},~
          \tfrac{-(Ry_\mathrm{u}^2 - Ry_\mathrm{n}^2 + x_\mathrm{u}y_\mathrm{n}^2 - 2x_\mathrm{r,2}y_\mathrm{u}^2 + x_\mathrm{u}y_\mathrm{u}^2)}{2y_\mathrm{u}(y_\mathrm{n} - y_\mathrm{u})^2},~
          R
          \Big)
      \end{cases},
  \end{align}
  \hrulefill
  \end{figure*}

\subsection{Closed-Form Solutions for Parameters}
\label{sec:closed_curving}

The optimization problems in \eqref{opt:final_r_p} and \eqref{opt:final_r_m} consist of a differentiable objective function, multiple real scalar variables, and linear inequalities, where solutions can be obtained from the \ac{KKT} conditions in closed form and depend on the hyperparameter $\omega$ in addition to the positions.
%
% are linear programming problems with a differentiable objective and
%
% The solutions of \eqref{opt:final_r_p} and \eqref{opt:final_r_m} depend on the hyperparameter $\omega$ and the positions of the user and the obstacle.
%
As an example, the closed-form solutions for the problem in \eqref{opt:final_r_p} are considered.

Some points that satisfy the \ac{KKT} conditions for the problem in \eqref{opt:final_r_p} include the value $x_\mathrm{adj}^\star = -R$, which means that exploiting only one antenna element is optimal in terms of the objective function in \eqref{obj:r_p}.
In other words, an improper hyperparameter $w$ prevents the generation of the curving beams.
Meanwhile, all the candidates for the points that satisfy the \ac{KKT} conditions can be obtained in advance via the Lagrangian method, thereby excluding the candidates including $x_\mathrm{adj}^\star = -R$.
% solution candidates for the problem in \eqref{opt:final_r_p} 
%
Note that such an operation does not imply the neglect of the global optimum solution in the problem in \eqref{opt:final_r_p}.
Rather, excluding the value $x_\mathrm{adj}^\star = -R$ is equivalent to balancing the two objective functions (\emph{i.e.}, improvement of the robustness and full exploitation of the \ac{ULA}) by tuning the hyperparameter $w$.
Moreover, the value $\beta=0$ means that the curving beam is not required to avoid the obstacle.

From the above, the possible combinations of the optimal solutions for the problem in \eqref{opt:final_r_p} are obtained, each of which is given at the top of this page, where the derivation is given in Appendix F.
The optimal solution for the problem in \eqref{opt:final_r_p} is one of the candidates in \eqref{eq:closed_curving}, which achieves the minimum value of the objective function in \eqref{obj:r_p} while satisfying the non-negative constraints for the Lagrangian multipliers and all the constraints in \eqref{const:array_final} through \eqref{const:sqrt_final}.
% , \eqref{const:avoid_f}, \eqref{const:sqrt}, \eqref{const:array_final}, \eqref{const:min_tangent_final},\eqref{const:max_tangent_final}, and \eqref{const:sqrt_final}.
%
% The closed-form solutions of the problem in \eqref{opt:final_r_m} can be obtained by the same process as in the problem \eqref{opt:final_r_p}.

From \eqref{eq:closed_curving}, it is obvious that part of the antenna elements cannot be exploited to generate the curving beam depending on the relationship between the positions of the user and the obstacle.
%
% In those cases, the other beams can be generated by the remaining antenna elements.
%
In such a case, in this paper, it is assumed that another curving beam based on the parabolic trajectory with the reverse curvature is generated by the remaining antenna elements.

\section{Performance Assessment}

% In this section, the effectiveness of the analysis for the Bessel beams and the parameter design for the curving beams are confirmed through the electromagnetic wave simulations.
In this section, the effectiveness of the analyses for the Bessel beams and the curving beams is confirmed through the electromagnetic wave simulations.
%
% Moreover, the characteristics of Gaussian beams, beamfocusing, Bessel beams, and curving beams are summarized in terms of the statistical behavior of their intensity and the perspectives of signal processing.
Moreover, the characteristics of Gaussian beams, beamfocusing, Bessel beams, and curving beams are summarized in terms of the statistical behavior of their intensity and signal processing.

%
%--
%

\subsection{Simulation Parameters and Setup}

In the simulations, the three-dimensional coordinate system is considered, where the \ac{ULA} located on the $x$-axis is modeled by the set of ideal dipole sources whose current density has only the $z$-component.
The central frequency is set to $f_\mathrm{c}=140$ [GHz].
%
% The obstacle is modeled by a cuboid or a cylinder in evaluations for signal propagation in the presence of the obstacle.
The obstacle is modeled by a cuboid or a cylinder in evaluations in the presence of the obstacle.
The transverse and longitudinal widths and height of the cuboid object are set to 0.28 [m], 0.47 [m], and 1.80[m], respectively, considering the human body \cite{MacCartney2016}.
The radius and height of the cylinder object are set to $0.28/2=0.14$ [m] and $1.8/10=0.18$  [m], respectively, considering the balance between the effectiveness, complexity, and memory in the simulations.
Both the obstacles are assumed to be symmetric about the $z$-axis and perfect electric conductors \cite{Fieramosca2024}.
%
% Note that the size of the obstacles is electrically extremely large, for example, $0.28\approx 130\lambda$, $0.47\approx 219 \lambda$, $1.8\approx 840\lambda$.
Note that the obstacles are electrically extremely large, for example, $0.28\approx 130\lambda$ [m].
%
% To efficiently evaluate the effects of such obstacles, the \ac{UTD} \cite{Paknys2016_2} and faceted \ac{UTD} \cite{Aguilar2022} are used to calculate the electrical field in the presence of the cuboid and cylinder obstacles, respectively.
Therefore, to efficiently evaluate the blockage effects, the \ac{UTD} \cite{Paknys2016_2} and faceted \ac{UTD} \cite{Aguilar2022} are used to calculate the electrical field in the presence of the cuboid and cylinder obstacles, respectively.

%
%-----
%

\begin{figure*}[t]
  \centering
    \begin{minipage}{1.0\columnwidth}
    \subfigure[{$\alpha=20$ [deg]}]
    {
      \includegraphics[width=\linewidth]{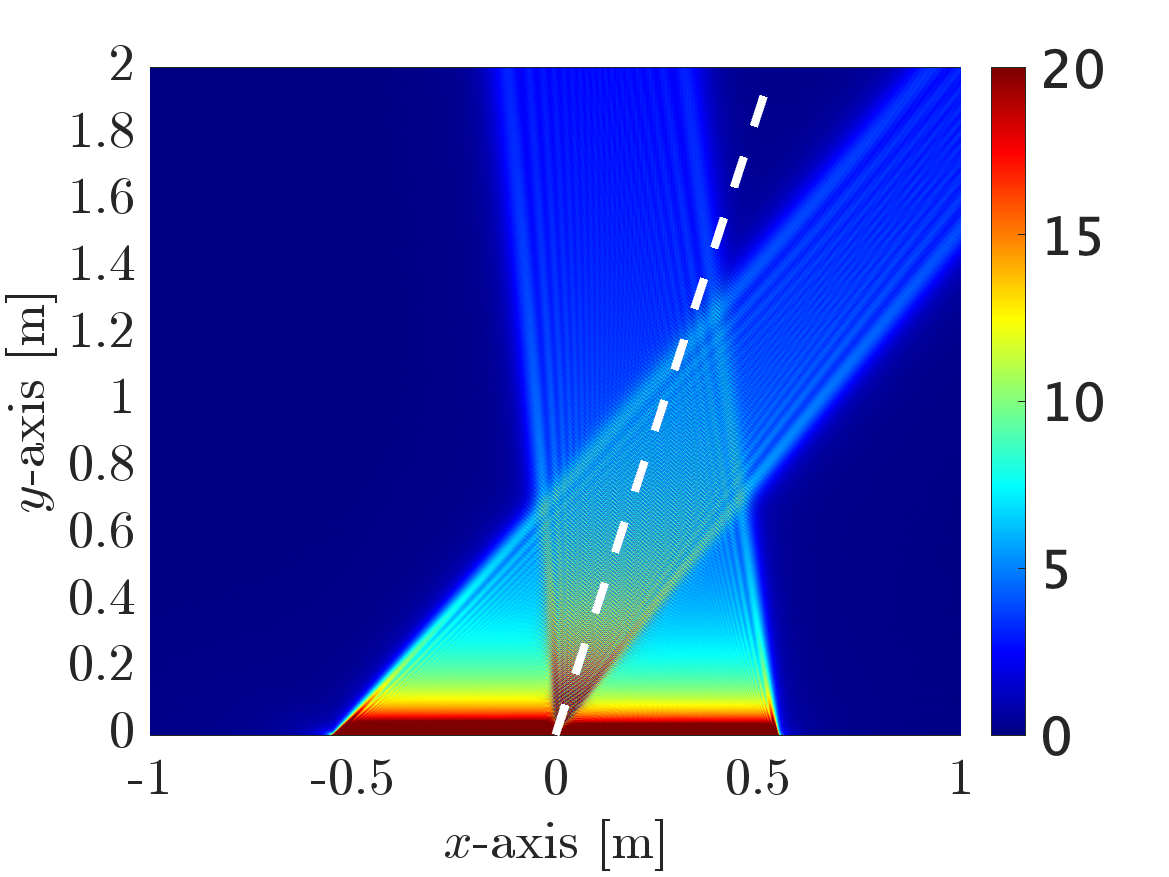}
    \label{fig:B_steer_15_alpha=20}
    }
    \end{minipage}
    \begin{minipage}{1.0\columnwidth}
    \subfigure[{$\alpha=15$ [deg]}]
    {
    \includegraphics[width=\linewidth]{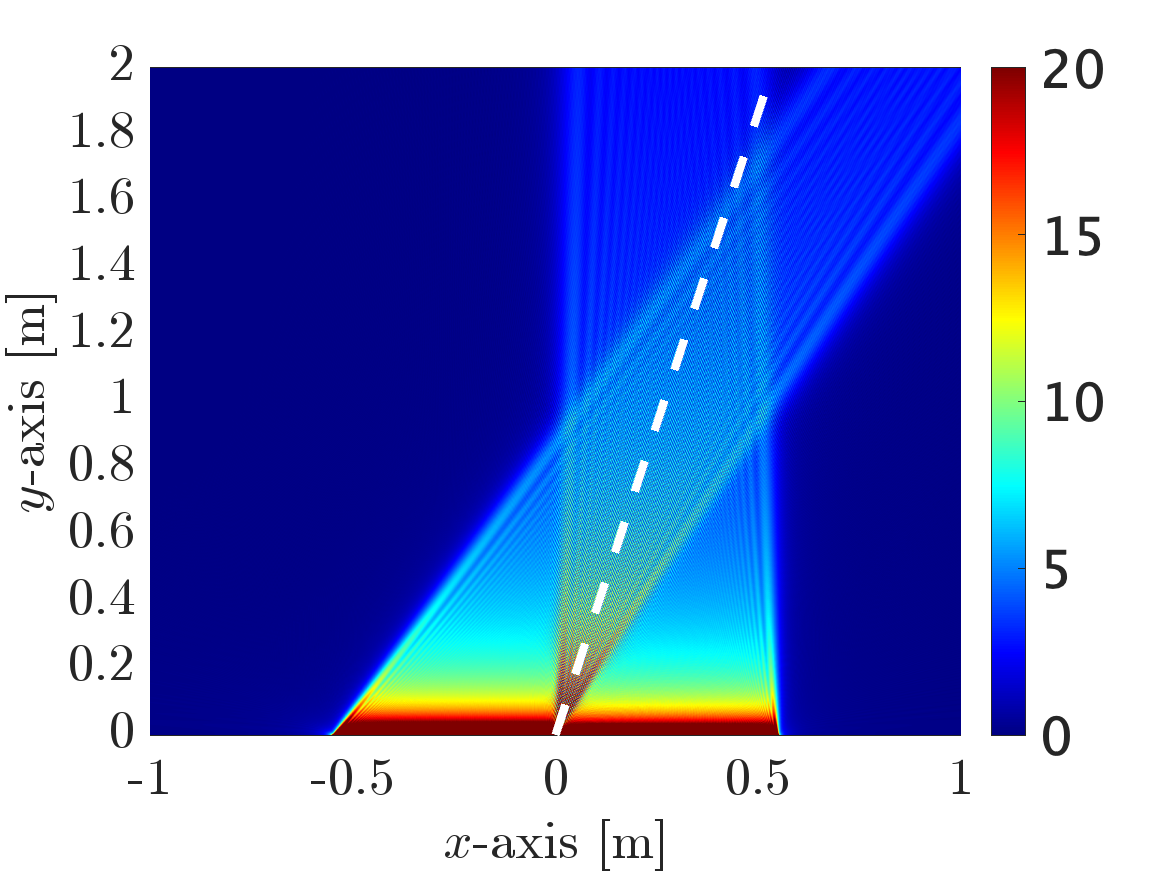}
    \label{fig:B_steer_15_alpha=15}
    }
    \end{minipage}
   \\
   \begin{minipage}{1.0\columnwidth}
    \subfigure[{$\alpha=10$ [deg]}]
    {
      \includegraphics[width=\linewidth]{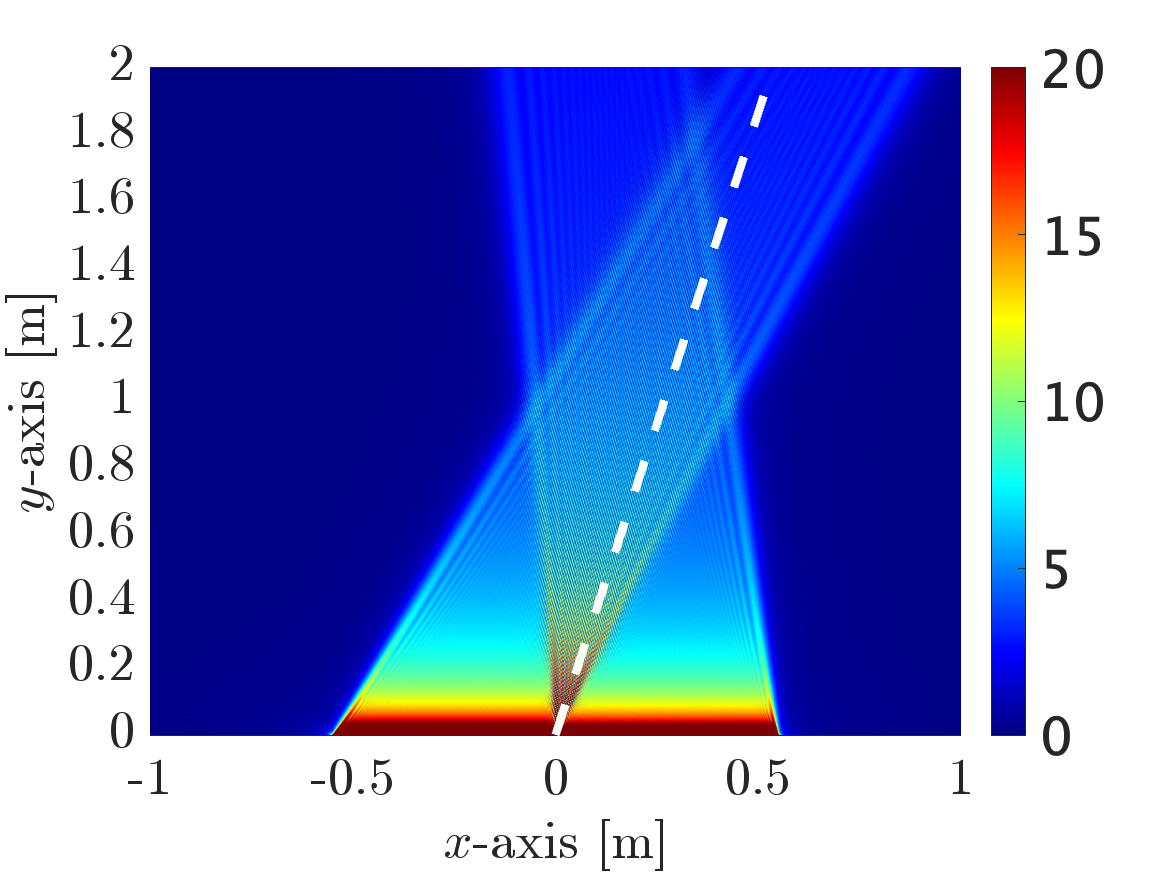}
    \label{fig:B_steer_15_alpha=10}
    }
    \end{minipage}
    \begin{minipage}{0.93\columnwidth}
    \subfigure[Along the desired direction]
    {
      \includegraphics[width=\linewidth]{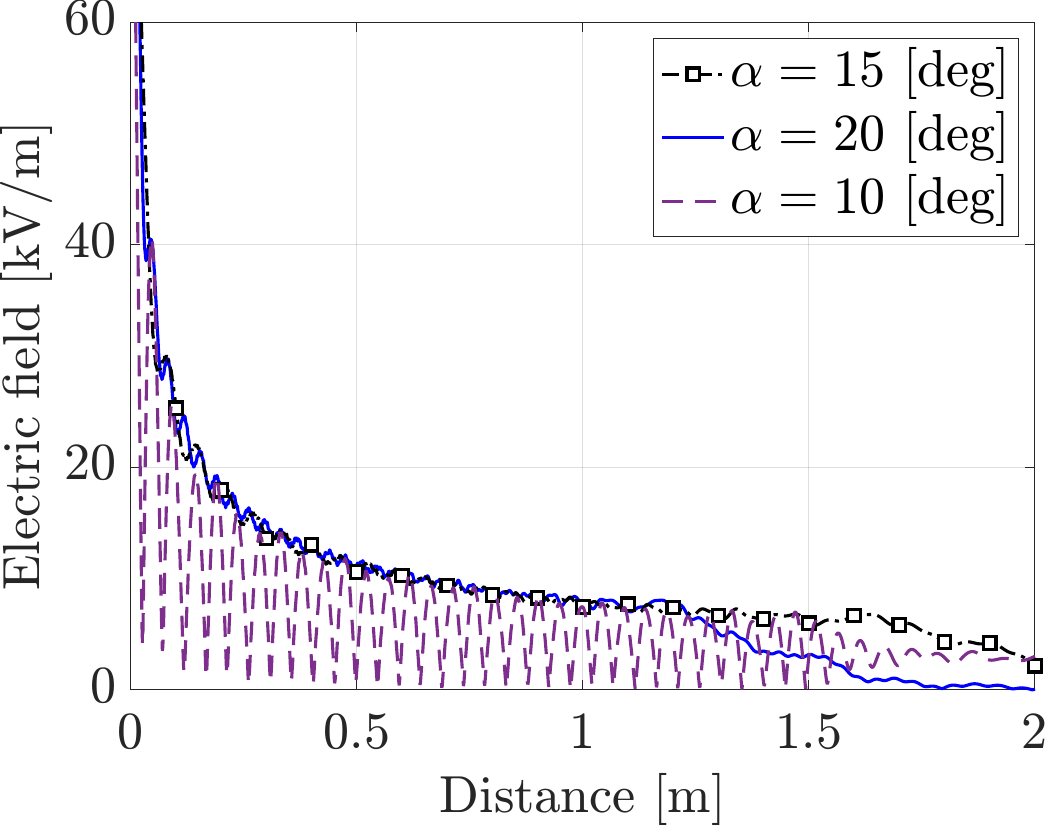}
    \label{fig:B_steer_15_field}
    }
    \end{minipage}
    % \caption{Received power and phase distributions}
          \caption{Amplitude of the electric field [kV/m] of the Bessel beams toward the angle $\theta_\mathrm{A}=15$ [deg] with various parameters $\alpha$ in the $xy$-plane}
    \label{fig:B_steer_15}
  \end{figure*}

\subsection{Steering Capability of Bessel beams}

In Fig. \ref{fig:B_steer_15}, the Bessel beams toward the azimuth angle $\theta_\mathrm{A}=15$ [deg] are evaluated, where the \ac{ULA} is equipped with $N=1024$ antenna elements with half-wavelength spacing $\Delta=\tfrac{\lambda}{2}$.
The amplitude of the electric field in the $xy$-plane is shown in Figs. \ref{fig:B_steer_15_alpha=20} through \ref{fig:B_steer_15_alpha=10}, where the parameter $\alpha$ is set to $\alpha={20,15,10}$ [deg], respectively.
In Fig. \ref{fig:B_steer_15_field}, that along the desired propagation direction is assessed.
%
% \begin{figure*}[t]
% \centering
% 	\begin{minipage}{1.0\columnwidth}
% 	\subfigure[{$\alpha=20$ [deg]}]
% 	{
%     % \includegraphics[width=\linewidth]{fig/steering/Steer=15_Alpha=20.pdf}
%     \includegraphics[width=\linewidth]{fig/steering/matlab/NearXY_alpha=20.eps}
% 	\label{fig:B_steer_15_alpha=20}
% 	}
% 	\end{minipage}
% 	%
% 	\begin{minipage}{1.0\columnwidth}
% 	\subfigure[{$\alpha=15$ [deg]}]
% 	{
%   % \includegraphics[width=\linewidth]{fig/steering/Steer=15_Alpha=15.pdf}
%   \includegraphics[width=\linewidth]{fig/steering/matlab/NearXY_alpha=15.eps}
% 	\label{fig:B_steer_15_alpha=15}
% 	}
% 	\end{minipage}
%  \\
%  \begin{minipage}{1.0\columnwidth}
% 	\subfigure[{$\alpha=10$ [deg]}]
% 	{
%     % \includegraphics[width=\linewidth]{fig/steering/Steer=15_Alpha=10.pdf}
%     \includegraphics[width=\linewidth]{fig/steering/matlab/NearXY_alpha=10.eps}
% 	\label{fig:B_steer_15_alpha=10}
% 	}
% 	\end{minipage}
% 	%
% 	\begin{minipage}{1.0\columnwidth}
% 	\subfigure[Along the desired direction]
% 	{
% 	% \includegraphics[width=\linewidth]{fig/steering/Steer=-15_15_Alpha=20.pdf}
%     \includegraphics[width=\linewidth]{fig/steering/matlab/Efield.pdf}
% 	\label{fig:B_steer_15_field}
% 	}
% 	\end{minipage}
% 	% \caption{Received power and phase distributions}
%         \caption{Amplitude of the electric field [kV/m] of the Bessel beams toward the angle $\theta_\mathrm{A}=15$ [deg] with various parameters $\alpha$ in the $xy$-plane}
% 	\label{fig:B_steer_15}
% \end{figure*}

As described in Section \ref{sec:B_steer}, with the parameters satisfying $\alpha\ge \theta_\mathrm{A}$, the Bessel beams achieve quasi-non-diffraction propagation along the desired propagation direction, denoted by the white dashed line in Fig. \ref{fig:B_steer_15}.
%
% Please note that the infinity array aperture and transmit power are necessary to realize diffraction-free propagation perfectly.
%
The comparison between the blue solid and black chained lines in Fig. \ref{fig:B_steer_15_field} confirms that a smaller $\alpha$ achieves a longer propagation distance.
In contrast, given the parameter $\alpha<\theta_\mathrm{A}$, as shown in \ref{fig:B_steer_15_alpha=10} and the purple dashed line with the square markers in Fig. \ref{fig:B_steer_15_field}, the Bessel beam cannot achieve quasi-non-diffraction propagation along the desired direction.
Thus, the performance gaps prove that the necessary and sufficient condition in \eqref{eq:max_steering} is valid.
Moreover, the shapes of the beams confirm that the two reference points $\mathbf{p}_{\mathrm{max,p}}$ and $\mathbf{p}_{\mathrm{max,m}}$ are not the same in the cases of $\theta_\mathrm{A}\neq 0$.

%
%-----
%

\subsection{Adjustment for Antenna Array}

\begin{figure*}[t]
\centering
	\begin{minipage}{1.0\columnwidth}
	\subfigure[{$\Delta=\tfrac{\lambda}{2}$, $N=3121$}]
	{
    \includegraphics[width=\linewidth]{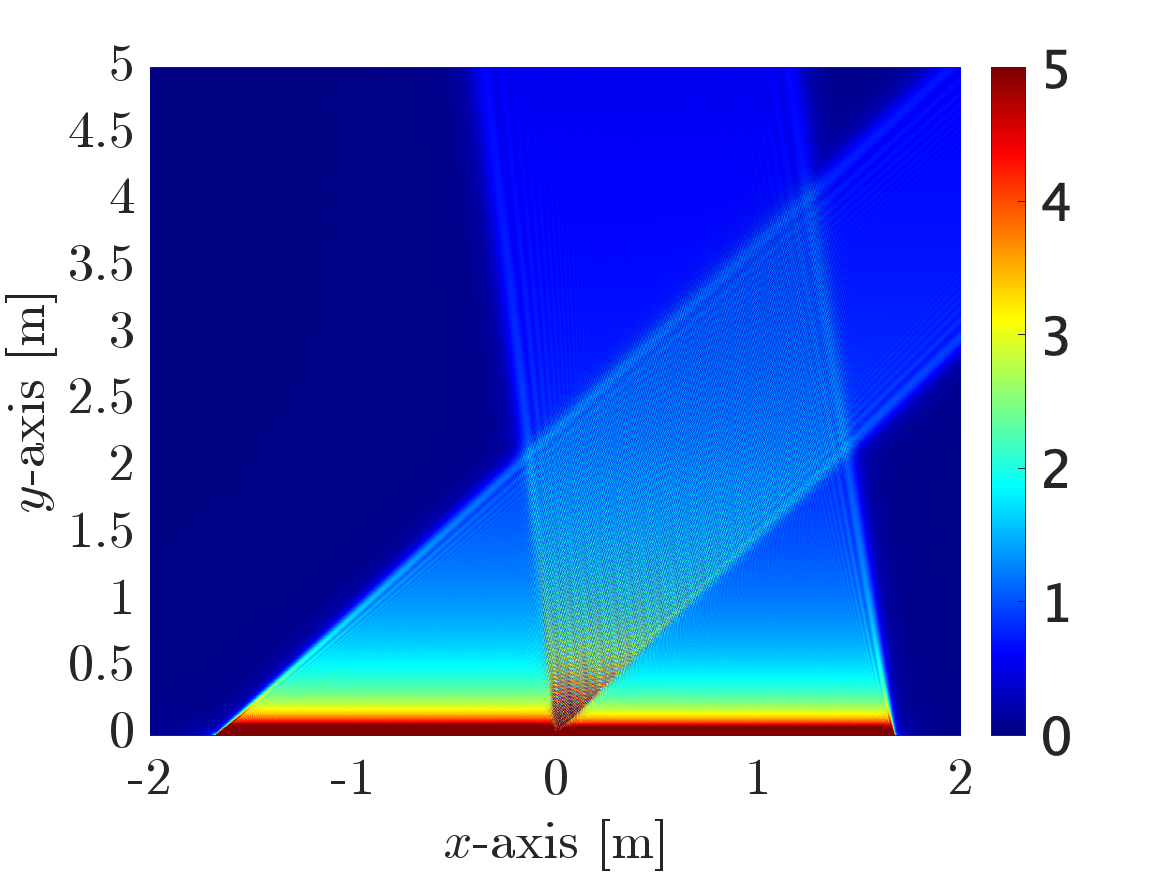}
	\label{fig:B_sampling_half}
	}
	\end{minipage}
	\begin{minipage}{1.0\columnwidth}
	\subfigure[{$\Delta=0.00186$, $N=1797$}]
	{
  \includegraphics[width=\linewidth]{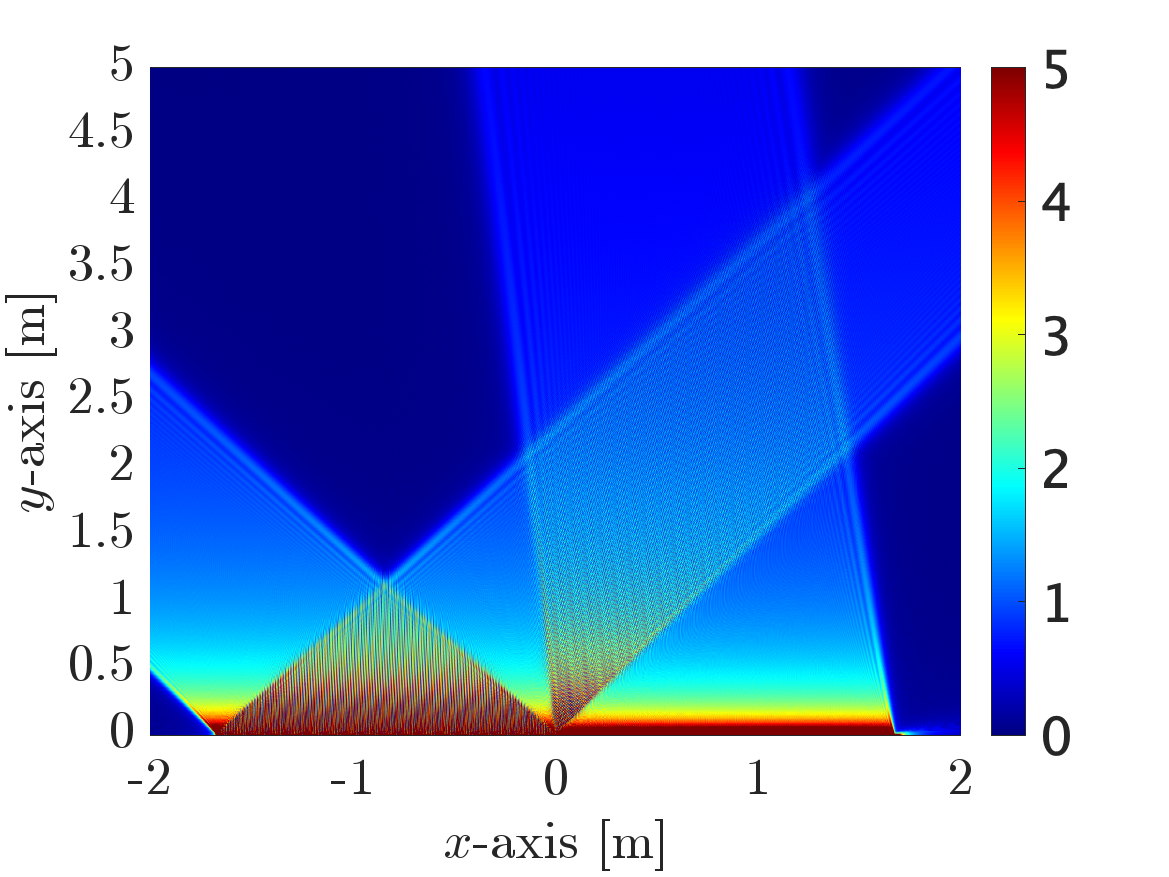}
	\label{fig:B_sampling_1}
	}
	\end{minipage}
 \\
 \begin{minipage}{1.0\columnwidth}
	\subfigure[{$\Delta=0.00372$, $N=899$}]
	{
    \includegraphics[width=\linewidth]{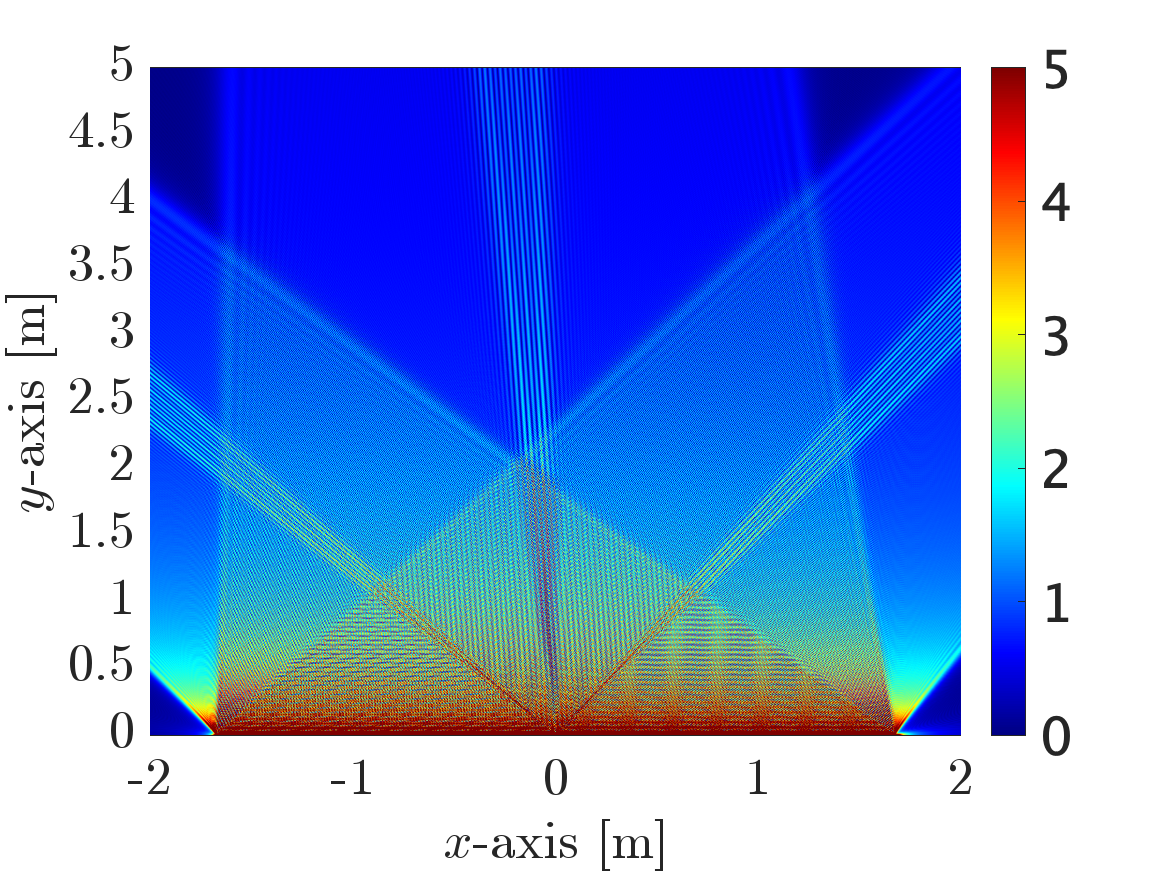}
	\label{fig:B_sampling_tw}
	}
	\end{minipage}
	\begin{minipage}{0.93\columnwidth}
	\subfigure[Along the desired direction]
	{
    \includegraphics[width=\linewidth]{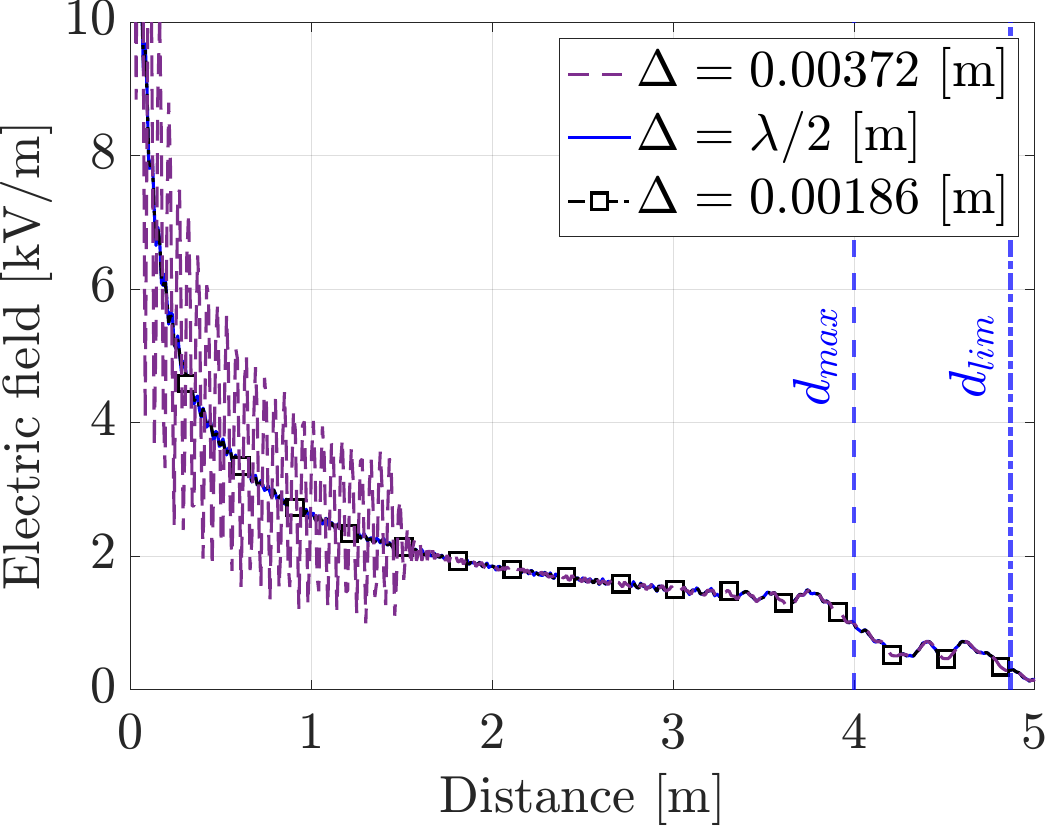}
	\label{fig:B_sampling_field}
	}
	\end{minipage}
	% \caption{Received power and phase distributions}
        \caption{Amplitude of the electric field [kV/m] of the Bessel beams toward the angle $\theta_\mathrm{A}=15$ [deg] with various \acp{ULA} in the $xy$-plane}
	\label{fig:B_sampling_15}
\end{figure*}

In this subsection, the analyses for the maximum propagation distance and the sampling theorem are verified through the simulations of the Bessel beams to achieve the desired distance $d_\mathrm{max,d}=4$ [m].
The angle $\theta_\mathrm{A}$ and the parameter $\alpha$ are fixed as $\theta_\mathrm{A}=15$ [deg] and $\alpha=20$ [deg] to simplify the simulations.
From \eqref{eq:sampling}, under those variables, the antenna spacing $\Delta$ should be shorter than $\Delta<\tfrac{\lambda}{2}\tfrac{1}{\sin(\alpha+|\theta_\mathrm{A}|)}=0.0018666$ [m] to satisfy the sampling theorem.
In Figs. \ref{fig:B_sampling_half} through \ref{fig:B_sampling_tw}, the Bessel beams with the \acp{ULA} with antenna spacing $\Delta=\tfrac{\lambda}{2}$, $\Delta=0.00186$ [m], and $\Delta=0.00186\times2=0.00372$ [m] are evaluated in terms of the amplitude of the electric field in the $xy$-plane, respectively.
Based on \eqref{eq:max_dist_num_antenna}, the number of antenna elements is set to $N=3121$, $N=1797$, $N=899$ under the spacing $\Delta=\tfrac{\lambda}{2}$, $\Delta=0.00186$, and $\Delta=0.00372$, respectively, to achieve the desired distance $d_\mathrm{max,d}=4$ [m], where the total transmit power is normalized for fair comparisons.
In Fig. \ref{fig:B_sampling_field}, the amplitude of the electric field of the Bessel beams is assessed along the desired propagation direction.

In the figures, all the Bessel beams can maintain their intensity by $d=4$ [m], which confirms the effectiveness of the analysis of the maximum propagation distance in Section \ref{sec:prop_dist}.
The blue dashed and chained lines denote the distances $d_\mathrm{max}$ and $d_\mathrm{lim}$ achieved by the \ac{ULA} with $N=3121$ antenna elements and $\Delta=\tfrac{\lambda}{2}$ antenna spacing.
Lower intensity in the region between $d_\mathrm{max}$ and $d_\mathrm{lim}$ indicates that only the beam generated by the antenna elements located in the positive region of the $x$-axis contributes to signal propagation along the desired direction in this region.

In turn, in Figs. \ref{fig:B_sampling_1} and \ref{fig:B_sampling_tw}, the grating lobes are generated due to the antenna spacing longer than half-wavelength \cite{1Balanis2005}.
However, if the antenna spacing is shorter than $\Delta<\tfrac{\lambda}{2}\tfrac{1}{\sin(\alpha+|\theta_\mathrm{A}|)}$, the unintended beam does not interfere with the desired beam, which is confirmed by Fig. \ref{fig:B_sampling_1}.
% the generation of the grating lobe does note affect the propagation direction.
%
% the grating lobe does not affect the propagation direction, which is confirmed by Fig. \ref{fig:B_sampling_1}.
%
Therefore, quasi-non-diffraction propagation is achieved even by the antenna spacing $\Delta=0.00186$ [m], as shown by the black chained line with the square markers in Fig. \ref{fig:B_sampling_field}.
In contrast, in the case of $\Delta>\tfrac{\lambda}{2}\tfrac{1}{\sin(\alpha+|\theta_\mathrm{A}|)}$, the unintended beams interfere with the desired beam, leading to unexpected attenuation of the intensity, as shown by the purple dashed line in Fig. \ref{fig:B_sampling_field}.
Those results confirm the effectiveness of the analysis for the sampling theorem in Section \ref{sec:sampling}.

%
%-----
%

\subsection{Self-Healing Capability of Bessel Beams}

\begin{figure*}[t]
\centering
	\begin{minipage}{1.0\columnwidth}
	% \subfigure[{$\theta_\mathrm{A}\approx-5.7$, $\alpha=25.7$ [deg]}]
        \subfigure[{Cuboid}]
	{
    \includegraphics[width=\linewidth]{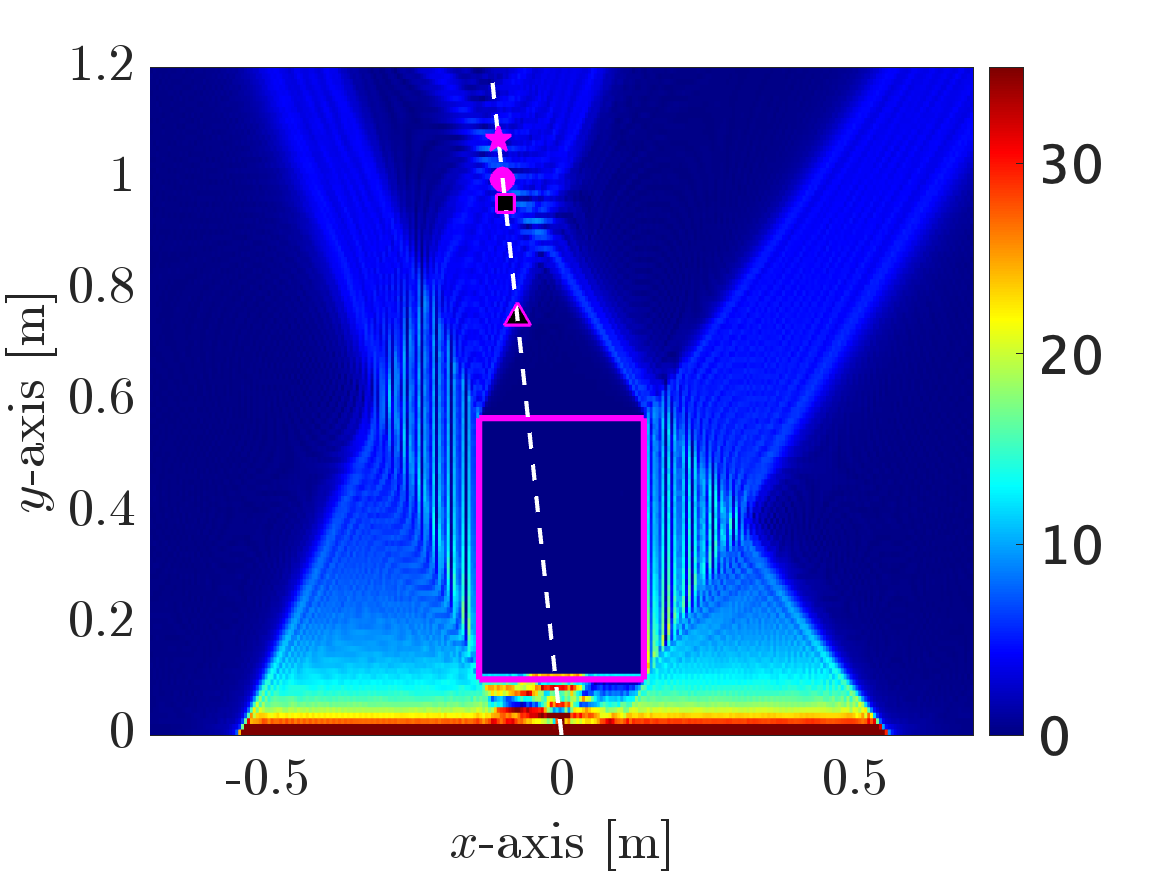}
	\label{fig:B_wobs_cu}
	}
	\end{minipage}
	\begin{minipage}{1.0\columnwidth}
        % \subfigure[{$\theta_\mathrm{A}\approx-5.7$, $\alpha=25.7$ [deg]}]
        \subfigure[{Cylinder}]
	{
  \includegraphics[width=\linewidth]{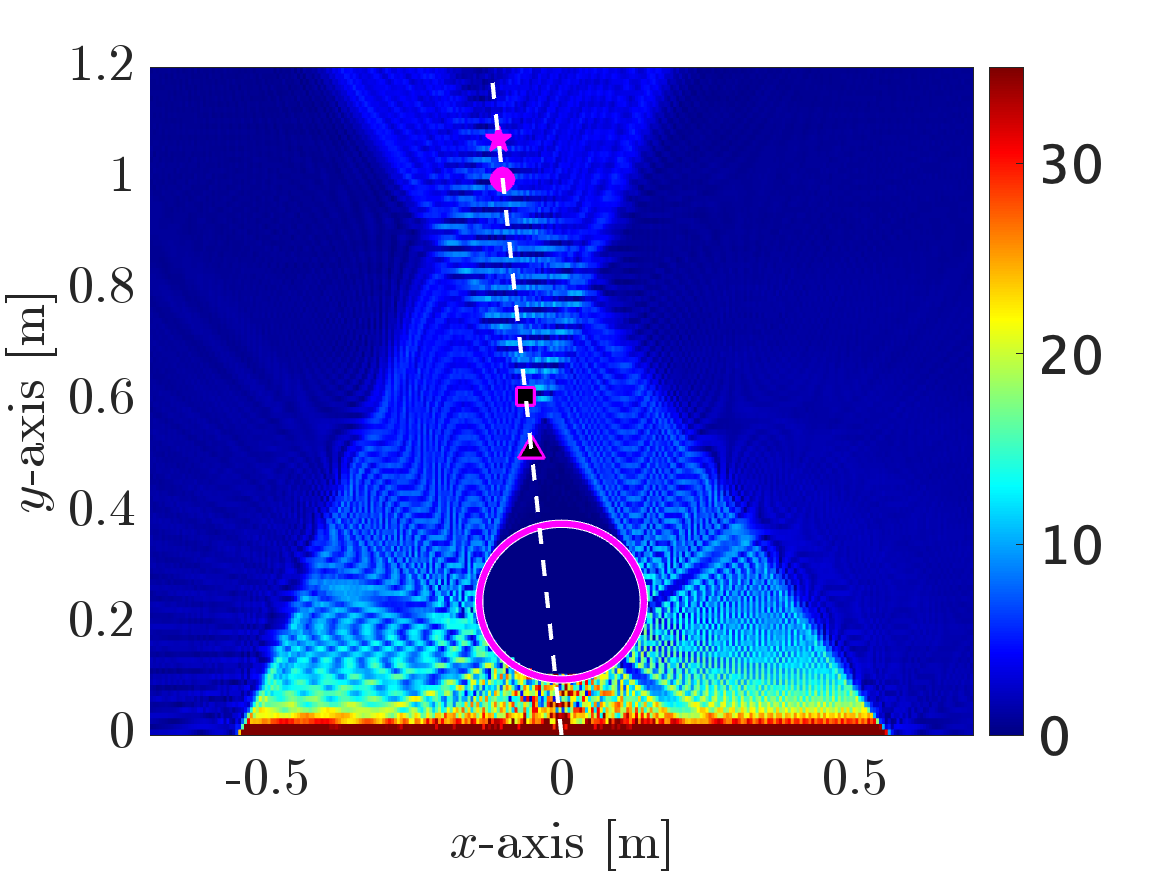}
	\label{fig:B_wobs_cy}
	}
	\end{minipage}
 \\
 \begin{minipage}{1.0\columnwidth}
	\subfigure[{$\theta_\mathrm{A}=0$, $\alpha=30$ [deg]}]
	{
    \includegraphics[width=\linewidth]{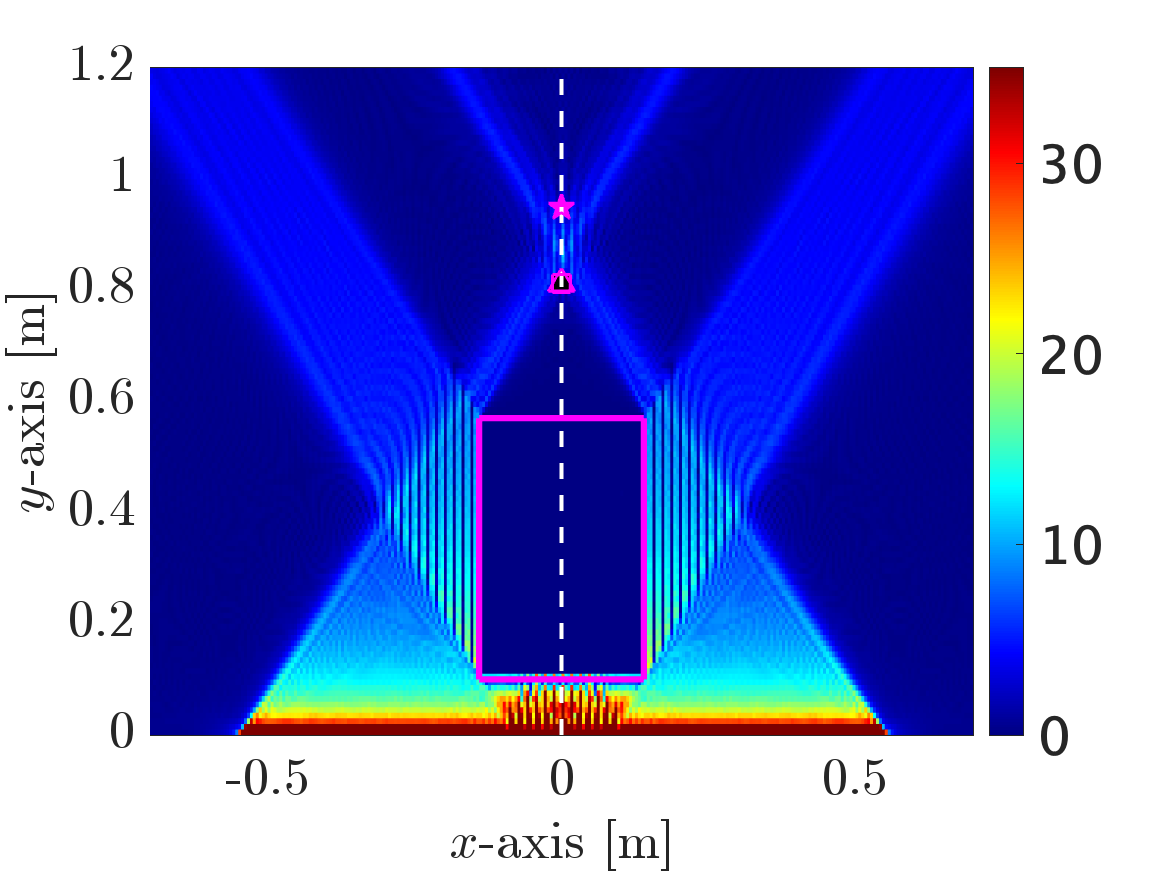}
	\label{fig:B_wobs_alpha=20}
	}
	\end{minipage}
	\begin{minipage}{1.0\columnwidth}
	\subfigure[{$\theta_\mathrm{A}=0$, $\alpha=20$ [deg]}]
	{
    \includegraphics[width=\linewidth]{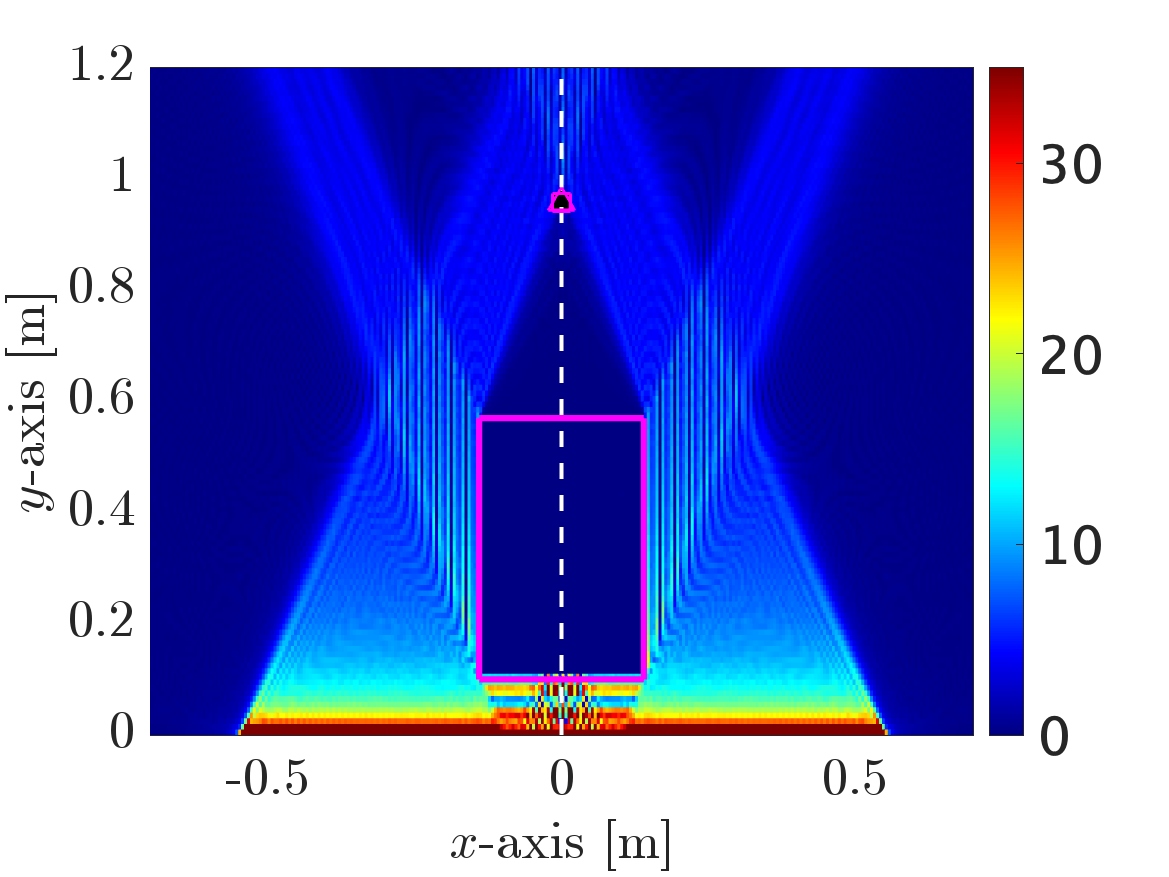}
	\label{fig:B_wobs_alpha=25}
	}
	\end{minipage}
	% \caption{Received power and phase distributions}
        \caption{Amplitude of the electric field [kV/m] of the Bessel beams in the $xy$-plane in the presence of the obstacle under various setup}
	\label{fig:B_wobs}
\end{figure*}

In this subsection, the self-healing capability of the Bessel beams is evaluated, where the number of antenna elements and the antenna spacing are set to $N=1024$ and $\Delta=\tfrac{\lambda}{2}$, respectively.
In Figs. \ref{fig:B_wobs_cu}, \ref{fig:B_wobs_alpha=20}, and \ref{fig:B_wobs_alpha=25}, the amplitude of the electric field of the Bessel beams in the $xy$-plane is evaluated in the presence of the cuboid obstacle, denoted by the pink solid line.
In Fig. \ref{fig:B_wobs_cy}, that is evaluated in the presence of the cylinder obstacle, denoted by the pink solid line.
The white dashed line and pentagram marker denote the desired propagation direction and maximum propagation distance $d_\mathrm{max}$, respectively.

In Figs. \ref{fig:B_wobs_cu} and \ref{fig:B_wobs_cy}, the user is assumed to be in the position $(x_\mathrm{u},y_\mathrm{u})=(-0.1,1)$, denoted by the circular marker, where the Bessel beam is generated based on the angle $\theta_\mathrm{A}\approx -5.7$ [deg] and the parameter $\alpha=20+|\theta_\mathrm{A}|$ [deg].
From \eqref{eq:obs_min_dist}, in the presence of the cuboid obstacle, the distances that are references for the beginning of self-healing are calculated as $d_\mathrm{h,p} = 0.9593$ and $d_\mathrm{h,m} = 0.7549$, denoted by the square and triangle markers in Fig. \ref{fig:B_wobs_cu}, respectively.
In such a situation, the relationship $d_\mathrm{h,m}(d_\mathrm{h,p})<d_\mathrm{UE}\approx 1.005 < d_\mathrm{max}\approx 1.0773$ [m] holds, and the Bessel beam is reconstructed by itself before the position of the user.
%
% Fig. \ref{fig:B_wobs_cu} also confirms that 
%
In Fig. \ref{fig:B_wobs_cy}, the relationship $d_\mathrm{h,m}\approx 0.5137 < d_\mathrm{h,p}\approx 0.6118 < d_\mathrm{UE}\approx 1.005 < d_\mathrm{max}\approx 1.0773$ [m] holds, and the reconstruction of the Bessel beam is completed before the position of the user.
Thus, those results confirm the effectiveness of the analysis for the self-healing capability described in Section \ref{sec:healing}.

In turn, in Figs. \ref{fig:B_wobs_alpha=20} and \ref{fig:B_wobs_alpha=25}, signal propagation along the $y$-axis is considered, where the Bessel beams are designed with the parameters $\alpha=30$ [deg] and $\alpha=20$ [deg], respectively.
From \eqref{eq:obs_min_dist}, in the non-steering cases, the distances $d_\mathrm{h,p}$ and $d_\mathrm{h,m}$ are the same, where the values are calculated as $0.8132$ [m] and $0.9575$ [m] with the parameters $\alpha=30$ [deg] and $\alpha=20$ [deg], respectively.
The maximum propagation distances achieved by the parameters $\alpha=30$ [deg] and $\alpha=20$ [deg] are given by $d_\mathrm{max}=0.9486$ [m] and $d_\mathrm{max}=1.5047$ [m], respectively
% , which is given by $d_\mathrm{max}=0.9486$ [m] is shorter than that achieved by the parameter $\alpha=20$ [deg], which is given by
%
To elaborate further, in the presence of the obstacle, although the Bessel beam with $\alpha=20$ [deg] is not effective in the region given by $y_\mathrm{u}\in[0.8132,0.9486)$ [m], a larger parameter $\alpha=30$ [deg] enables signal propagation in that region, which is confirmed by Fig. \ref{fig:B_wobs_alpha=25}.
In contrast, only the Bessel beam with $\alpha=20$ [deg] is effective in the region given by $y_\mathrm{u}\in[0.9575,1.5047)$ [m].
%
% However, the maximum propagation distance achieved by the parameters $\alpha=30$ [deg], which is given by $d_\mathrm{max}=0.9486$ [m] is shorter than that achieved by the parameter $\alpha=20$ [deg], which is given by $d_\mathrm{max}=1.5047$ [m].
%
From the above, transmission schemes based on the Bessel beam should use the parameters $\alpha$ smaller than 30 [deg] in signal propagation beyond the distance $d_\mathrm{max}=0.9486$ [m], which, however, implicitly decrease robustness against blockages, as described in Sections \ref{sec:sampling} and \ref{sec:healing}.
%
% Therefore, the user located in the region $y_\mathrm{u}\in(1.1745,1.5047]$ cannot receive signals by the Bessel beams designed with $\alpha=25$ [deg].

%
%---
%

\subsection{Blockage Avoidance by Curving Beams} 

\begin{figure*}[t]
\centering
	\begin{minipage}{1.0\columnwidth}
	\subfigure[{Cuboid}]
	{
    \includegraphics[width=\linewidth]{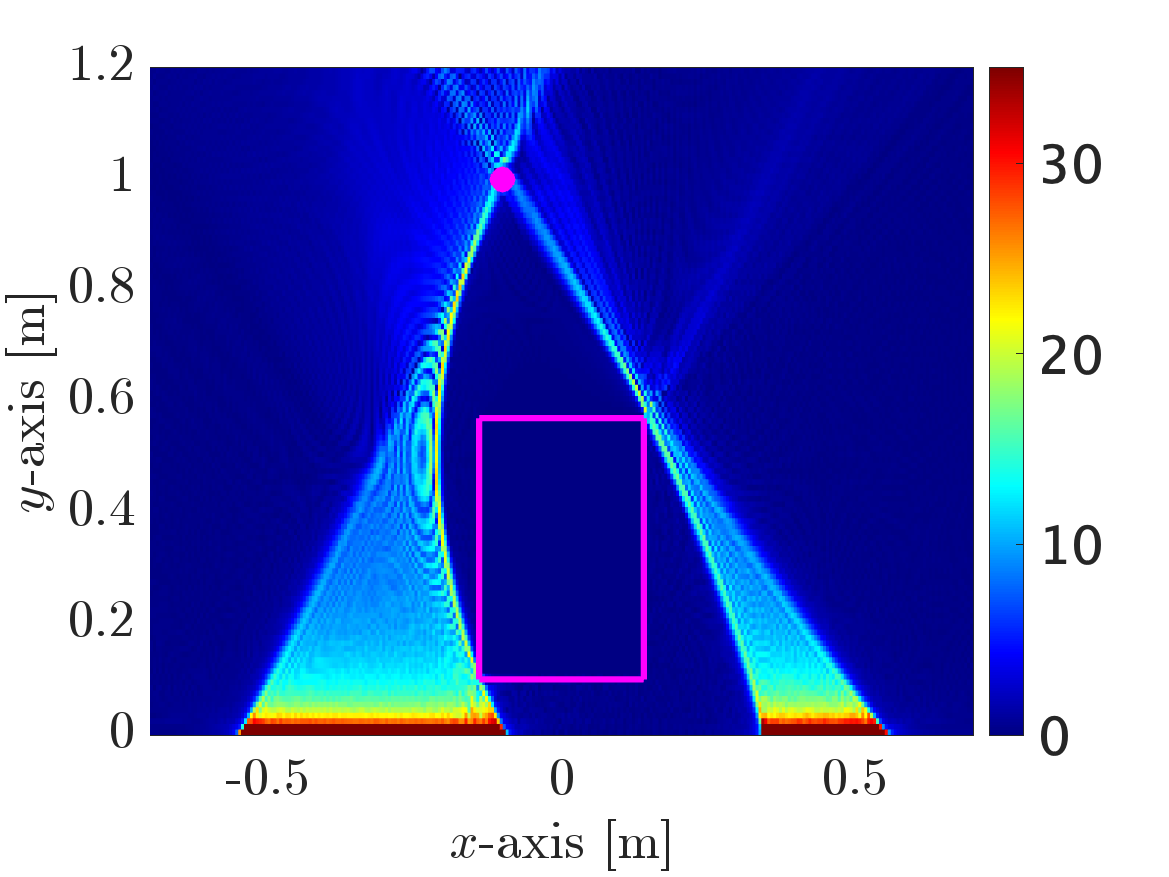}
	\label{fig:C_wobs_cu}
	}
	\end{minipage}
	\begin{minipage}{1.0\columnwidth}
        \subfigure[{Cylinder}]
	{
  \includegraphics[width=\linewidth]{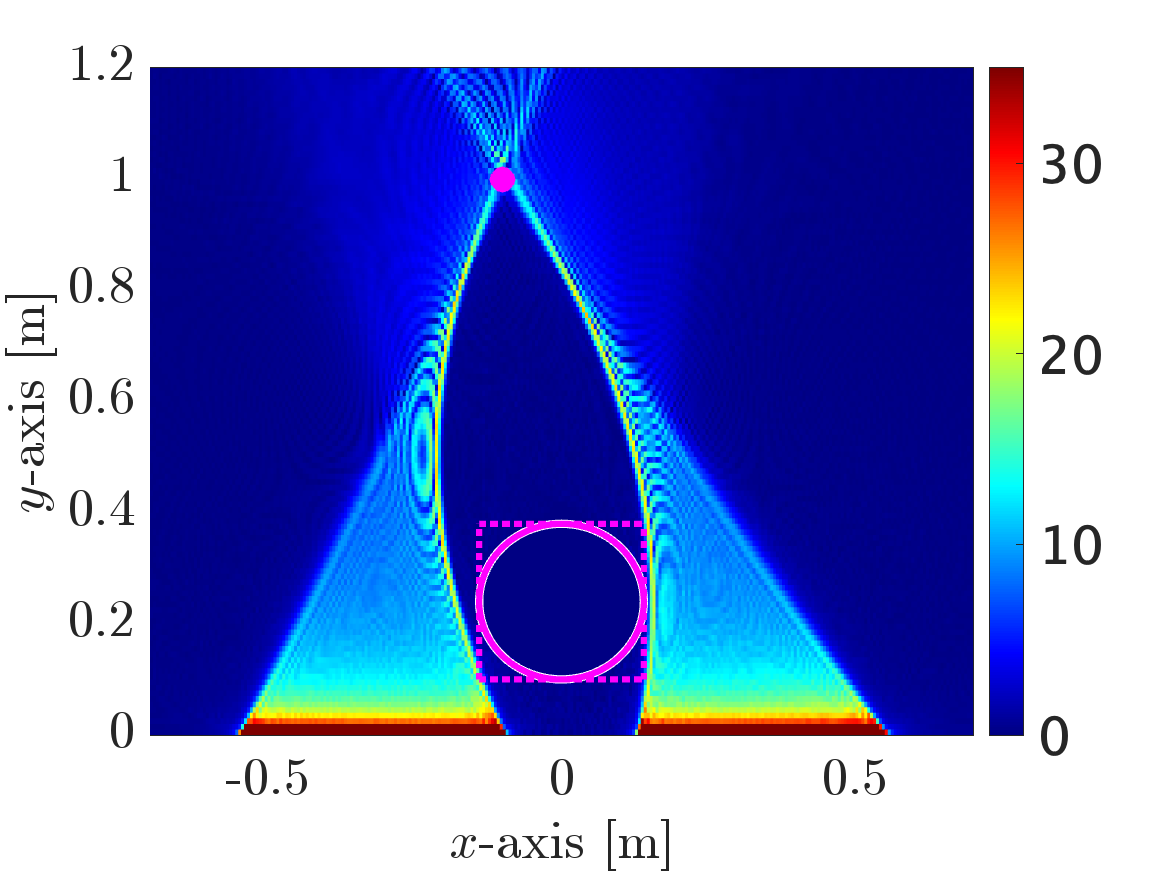}
	\label{fig:C_wobs_cy}
	}
	\end{minipage}
 \\
 \begin{minipage}{0.95\columnwidth}
	\subfigure[{At the user position}]
	{
    \includegraphics[width=\linewidth]{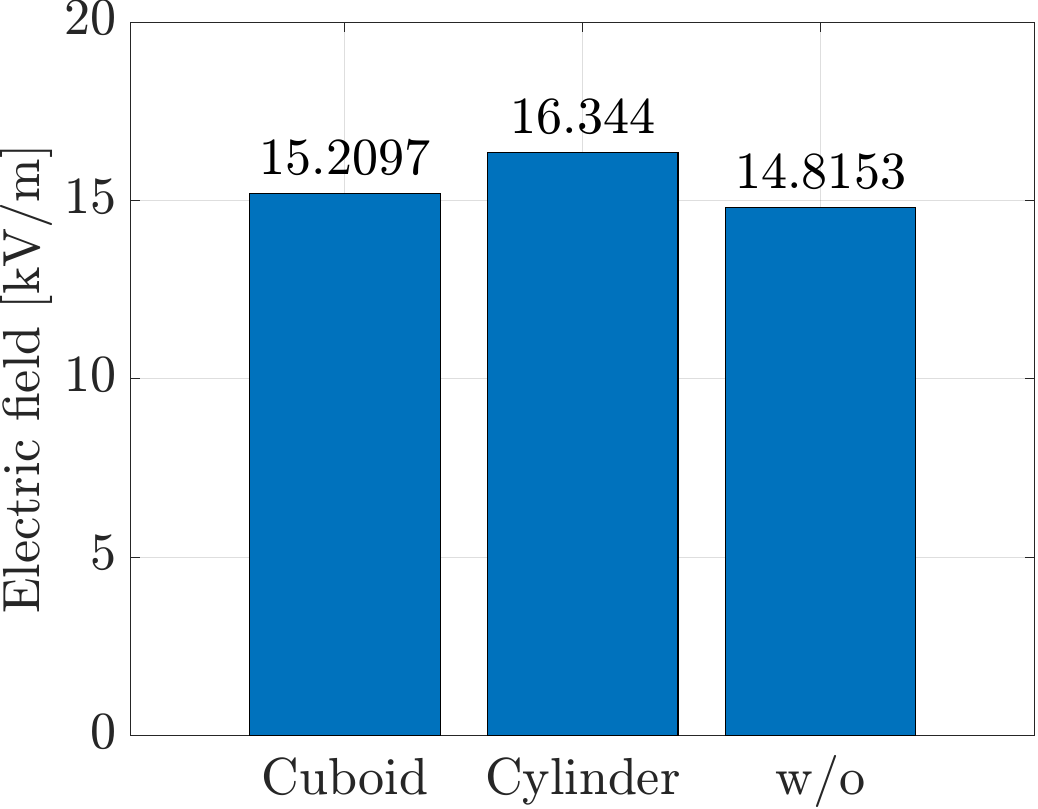}
	\label{fig:user_point}
	}
	\end{minipage}
	\begin{minipage}{1.0\columnwidth}
	\subfigure[Full exploitation case]
	{
    \includegraphics[width=\linewidth]{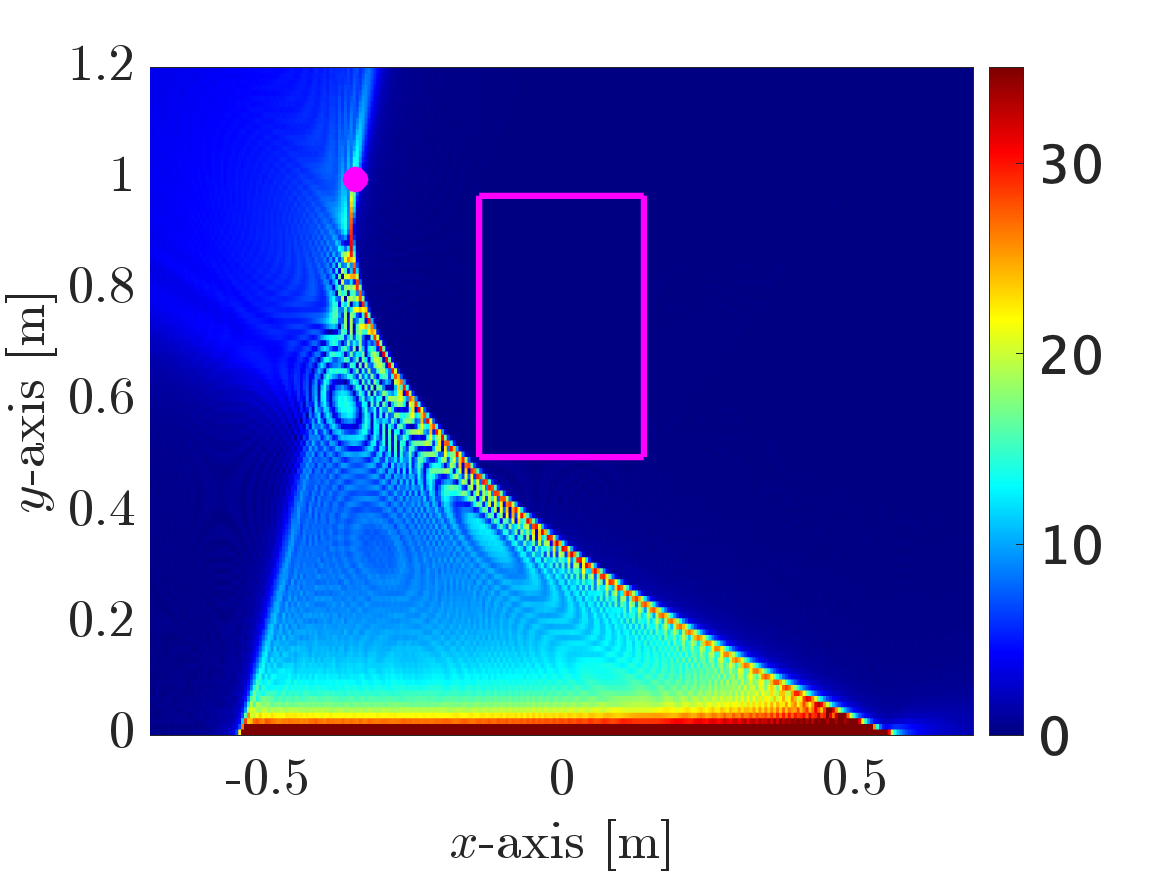}
	\label{fig:C_blkp=3}
	}
	\end{minipage}
	% \caption{Received power and phase distributions}
        \caption{Amplitude of the electric field [kV/m] of the curving beams in the $xy$-plane in the presence of the obstacles under various setup}
	\label{fig:C_wobs}
\end{figure*}

% \begin{figure}[t]
% \centering
% 	\begin{minipage}{0.45\columnwidth}
% 	\subfigure[{Cuboid}]
% 	{
%     % \includegraphics[width=\linewidth]{fig/Curving/Curving_UE2_delta=0p25_blkp=3.pdf}
%     \includegraphics[width=\linewidth]{fig/Curving/matlab/NearXY_UE2_blkp=3.pdf}
% 	\label{fig:C_wobs_cu}
% 	}
% 	\end{minipage}
% 	%
% 	\begin{minipage}{0.45\columnwidth}
%         \subfigure[{Cylinder}]
% 	{
%   % \includegraphics[width=\linewidth]{fig/Curving/Curving_UE2_delta=0p25_blkp=3_cy.pdf}
%   \includegraphics[width=\linewidth]{fig/Curving/matlab/NearXY_UE2_blkp=3_Cy.pdf}
% 	\label{fig:C_wobs_cy}
% 	}
% 	\end{minipage}
%         \caption{Curving Beams}
% 	\label{fig:C_wobs}
% \end{figure}

In this subsection, the curving beams based on the proposed trajectory designs in Section \ref{sec:curving} are evaluated under the assumption of the \ac{ULA} with $N=1024$ antenna elements and $\Delta=\tfrac{\lambda}{2}$ antenna spacing.
In Figs. \ref{fig:C_wobs_cu} and \ref{fig:C_wobs_cy}, the amplitude of the electric field of the curving beams in the $xy$-plane is evaluated in the presence of the cuboid and cylinder obstacles, respectively, where the position of the user, denoted by the circular marker, is set to the same as in Figs. \ref{fig:B_wobs_cu} and \ref{fig:B_wobs_cy}.

As shown in Figs. \ref{fig:C_wobs_cu} and \ref{fig:C_wobs_cy}, the proposed approach can design the parabolic trajectories to avoid one obstacle and to exploit as many antenna elements as possible.
Especially in Fig. \ref{fig:C_wobs_cy}, the curving beams are designed by approximating the circular shape as the square shape, denoted by the pink dotted line, in the parameter design algorithm.
This result demonstrates that the proposed design can improve robustness against blockages and positioning errors of obstacles by assuming a larger obstacle than the practice in the algorithm. %problems in \eqref{opt:final_r_p} and \eqref{opt:final_r_m}.

In Fig. \ref{fig:user_point}, the amplitude of the electric field is assessed at the position of the user, where “Cuboid” and “Cylinder” mean the curving beams shown in Figs. \ref{fig:C_wobs_cu} and \ref{fig:C_wobs_cy}, respectively.
In addition, “w/o” shows the amplitude of the electric field at the position of the user in free-space propagation, where the beam is the same as in Fig. \ref{fig:C_wobs_cu}.
The performance gap between “Cuboid” and “Cylinder” is mainly due to the difference in the number of antenna elements to exploit for beam generation.
On the one hand, lower amplitude of “w/o” than that of “Cuboid” indicates the drawback of the proposed design, which does not take the \ac{SNR} into account.
On the other hand, the results in Fig. \ref{fig:user_point} confirm that the proposed design can maintain the intensity of the curving beams even in the presence of the obstacle.

In Fig. \ref{fig:C_blkp=3}, the amplitude of the electric field of the curving beam is evaluated in the different situations from Figs. \ref{fig:C_wobs_cu} and \ref{fig:C_wobs_cy}.
As shown in Fig. \ref{fig:C_blkp=3}, the trajectory and the number of antenna elements to exploit for beam generation are adjusted based on the relationship between the positions of the user, obstacle, and \ac{ULA}, which verifies that the closed-form solutions in \eqref{eq:closed_curving} are valid.

% the proposed design leads to the proper trajectory based on the positions of the user and obstacle to avoid one obstacle and to full exploit the \ac{ULA}, as shown in the closed-form solutions in \eqref{eq:closed_curving}.

%
%----
%

\subsection{Comparisons with Different Beams}

Finally, in this subsection, Gaussian beams, beamfocusing, Bessel beams, and curving beams are compared in terms of their intensity and information used to generate beams to clarify the advantages and disadvantages of each beam.
The phase of the $n$-th antenna element to generate Gaussian beams and beamfocusing are given by $\phi_n=-k\sin\theta_\mathrm{A}x_{\mathrm{t},n}$ and $\phi_n=kr_{n}$, respectively \cite{Headland2018}.
Thus, the information used to the beam designs can be summarized as in Table \ref{table:info}, where the check mark “${\checkmark}$” denotes the information that can be used to generate the corresponding beam, and the marker “$-$” means that beam designs based on user positions implicitly use azimuth angles.
%
% The triangle marker “$\triangle$” in the column of the Bessel beam means that the positions of obstacles only enable us to estimate the reference points for the beginning of the self-healing.
%
The triangle marker “$\triangle$” in the column of the Bessel beam means that while the exploitation of the positions of obstacles cannot lead to Bessel beams that can avoid obstacles, the beginning of self-healing can be estimated based on the information.

\begin{figure*}[t]
  \centering
    \begin{minipage}{0.95\columnwidth}
    \subfigure[{Evaluation situations}]
    {
      \includegraphics[width=\linewidth]{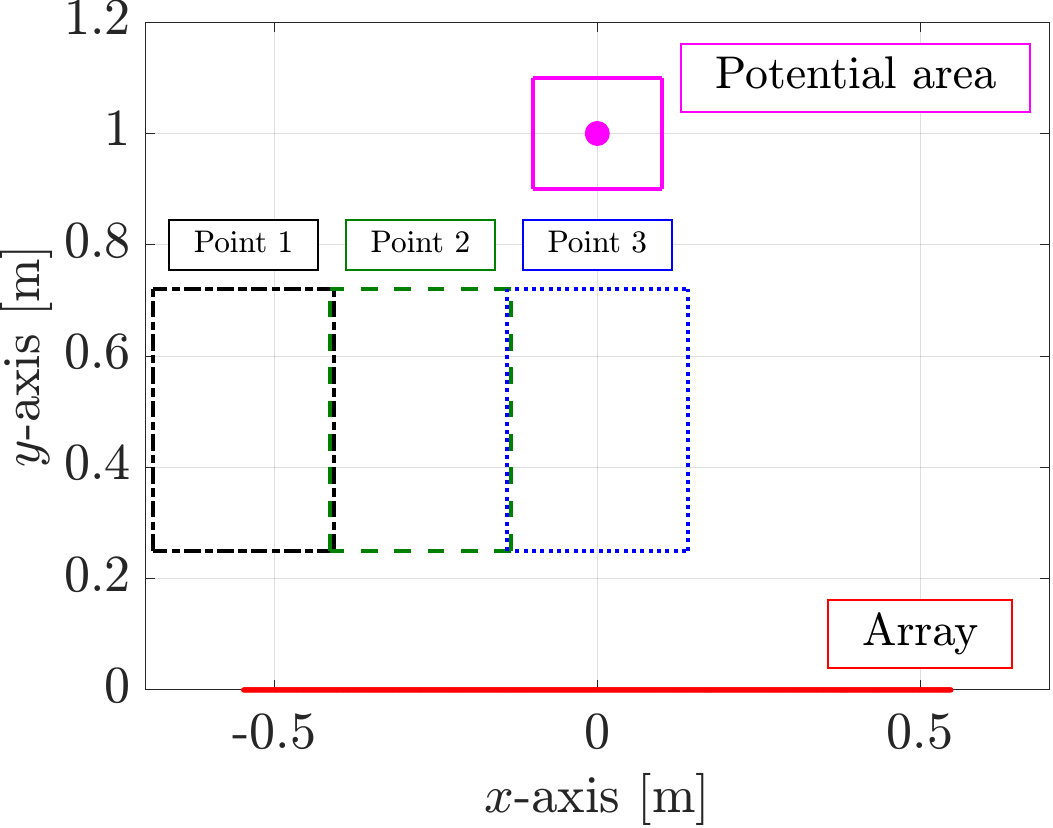}
    \label{fig:comp_system}
    }
    \end{minipage}
    \begin{minipage}{0.95\columnwidth}
          \subfigure[{Amplitude at the point}]
    {
    \includegraphics[width=\linewidth]{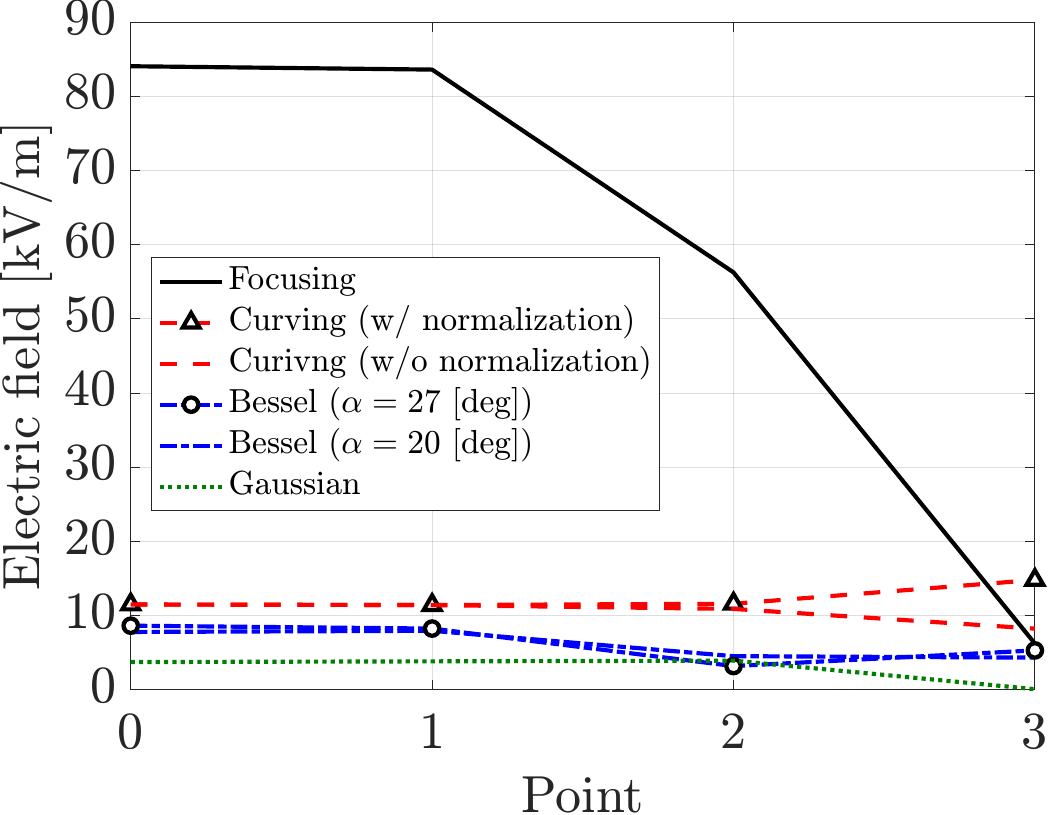}
    \label{fig:comp_point}
    }
    \end{minipage}
   \\
   \begin{minipage}{0.95\columnwidth}
    \subfigure[{Average of the amplitude}]
    {
      \includegraphics[width=\linewidth]{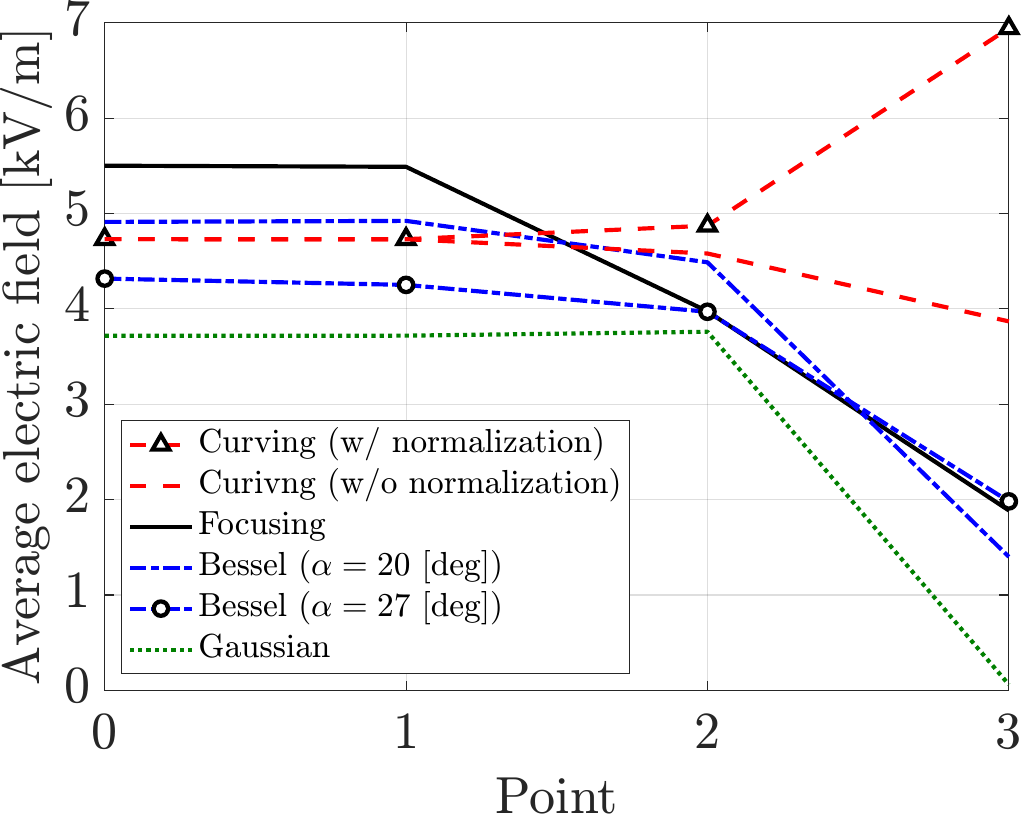}
    \label{fig:comp_ave}
    }
    \end{minipage}
    \begin{minipage}{0.95\columnwidth}
    \subfigure[CDF of the amplitude]
    {
      \includegraphics[width=\linewidth]{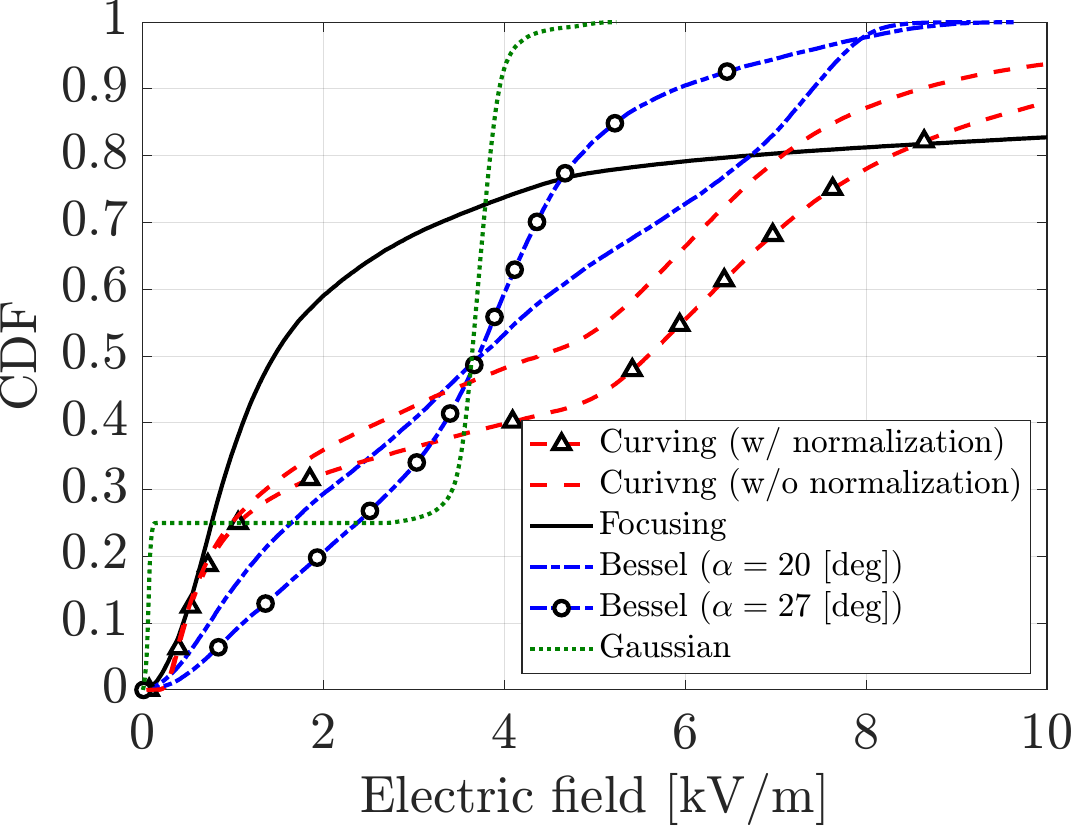}
    \label{fig:comp_CDF}
    }
    \end{minipage}
    % \caption{Received power and phase distributions}
          \caption{Amplitude of the electric field of various beams in the $xy$-plane}
    \label{fig:comp}
  \end{figure*}

In Fig. \ref{fig:comp}, the statistical behavior of the amplitude of the electric field of each beam in the $xy$-plane is evaluated.
The evaluation situations are shown in Fig. \ref{fig:comp_system}, where the position of the user is fixed at the point $(x_\mathrm{u},y_\mathrm{u})=(0,1)$, denoted by the circular marker, and the \ac{ULA} with $N=1024$ antenna elements and $\Delta=\tfrac{\lambda}{2}$ spacing is considered.
The four patterns of positions for the cuboid obstacle are considered, where the second, third, and fourth patterns are denoted by the black chained, green dashed, and blue dotted lines, respectively, in Fig. \ref{fig:comp_system}.
The first pattern is that the obstacle is far from both the user and \ac{ULA}, which, thus, is equivalent to free-space propagation.
From the above, the evaluation situation can be interpreted as the situation in which the human body crosses the communication link.
Based on the previous discussions, only the curving beam is optimized for each situation based on \eqref{opt:final_r_p} and \eqref{opt:final_r_m}.
In free-space propagation, the curving beam is designed under the assumption that the obstacle is in “Point 1” in Fig. \ref{fig:comp_system} in the parameter design algorithm.

% \begin{figure}[t]
% \centering
% 	\begin{minipage}{0.45\columnwidth}
% 	\subfigure[{Evaluation situations}]
% 	{
%     \includegraphics[width=\linewidth]{fig/Comp/matlab/System.pdf}
% 	\label{fig:comp_system}
% 	}
% 	\end{minipage}
% 	%
% 	\begin{minipage}{0.45\columnwidth}
%         \subfigure[{Amplitude at the point}]
% 	{
%   % \includegraphics[width=\linewidth]{fig/Comp/Efield_vary_Average_Alpha=20.pdf}
%   \includegraphics[width=\linewidth]{fig/Comp/matlab/Efield_point.pdf}
% 	\label{fig:comp_point}
% 	}
% 	\end{minipage}
%  \\
%  \begin{minipage}{0.45\columnwidth}
% 	\subfigure[{Average of the amplitude}]
% 	{
%     % \includegraphics[width=\linewidth]{fig/Comp/Efield_CDF_Alpha=20.pdf}
%     \includegraphics[width=\linewidth]{fig/Comp/matlab/Efield_ave.pdf}
% 	\label{fig:comp_ave}
% 	}
% 	\end{minipage}
% 	%
% 	\begin{minipage}{0.45\columnwidth}
% 	\subfigure[CDF of the amplitude]
% 	{
% 	% \includegraphics[width=\linewidth]{fig/Comp/Comp_Bessel_Efield_vary_Point.pdf}
%     \includegraphics[width=\linewidth]{fig/Comp/matlab/Efield_CDF.pdf}
% 	\label{fig:comp_CDF}
% 	}
% 	\end{minipage}
% 	% \caption{Received power and phase distributions}
%         \caption{Amplitude of the electric field of various beams in the $xy$-plane}
% 	\label{fig:comp}
% \end{figure}

In Figs. \ref{fig:comp_point} through \ref{fig:comp_CDF}, the block solid, green dotted, blue chained, and red dashed lines denote the Gaussian beams, beamfocusing, Bessel beams, and curving beams, respectively.
As mentioned in Section \ref{sec:curving} and confirmed in Fig. \ref{fig:C_wobs}, some antenna elements are not exploited to generate curving beams (\emph{i.e.}, $\gamma_n=0$).
In those cases, the transmit power is normalized, such that the sum of the square magnitude of the currents $\sum_{n=1}^{N}\gamma_n^2$ becomes the same value as the other beams, which is denoted by the red dashed line with the triangle markers.

%%%%%%%% table for references %%%%%%%%%%%%%%
\begin{table}[t]
  \caption{Information used to generate beams with a \ac{ULA}}
  \normalsize
  % \vspace{-3ex}
  \label{table:info}
  \begin{center}
  \begin{tabular}[\linewidth]{|l|c|c|c|c|c}
  \hline
  & Gaussian  & Focus & Bessel & Curving \\
  \hline
  % 
  % $\mathbf{H}[d]$ & Channel at the $d$-th delay tap\\
  % \hline
  % $\mathbf{H}[k]$ & Channel at the $k$-th subcarrier\\
  % \hline
  User position &	 & {\small \checkmark} &  & {\small \checkmark} \\
  \hline
  {Azimuth angle}  &{\small \checkmark} & $-$ & {\small \checkmark} & $-$ \\
  \hline
  Obs. position &  &  & $\triangle$  & {\small \checkmark} \\
  \hline
  \end{tabular}
  \end{center}
  % \vspace{-5ex}
  \end{table}
  %%%%%%%%%%%%%%%%%%%%%%%%

Fig. \ref{fig:comp_point} assesses the amplitude of the electric field at the position of the user as a function of the positions of the obstacle.
Beamfocusing exhibits the highest intensity, which confirms that \ac{MRT} transmissions are achieved by it in near-field communications modeled by spherical wave channels without obstacles \cite{Headland2018, Zhang2022}.
In contrast, in situations where the obstacle is near the user and the \ac{ULA}, the intensity of beamfocusing decreases severely.
In such situations, the \ac{CSI} exploited to design beamfocusing does not exactly characterize the communication environments in the presence of the obstacle.
Moreover, the Gaussian beam, which can achieve \ac{MRT} transmissions in far-field communications, exhibits the lowest intensity in all the situations.
Therefore, the results for beamfocusing and the Gaussian beam confirm that inaccurate \ac{CSI} in the beam designs leads to severe performance degradation.
Unlike those beams, the curving beams can maintain their intensity in the presence of the obstacle since the beams are designed based on both the positions of the user and the obstacle (\emph{i.e.}, accurate \ac{CSI}). 
However, the proposed curving beam designs do not consider the \ac{SNR}, which leads to lower intensity compared to beamfocusing in the situations where partial blockage effects are not significant.
The Bessel beams can also mitigate the intensity degradation caused by the obstacle, thanks to their self-healing property.
Even in the situations where partial blockage effects are negligible, the intensity of the Bessel beams is higher than that of the Gaussian beam, thanks to the additional degree of freedom brought by adjusting the parameter $\alpha$.
%
% In situations where , compared to the Gaussian beam, the additional degree of freedom in the Bessel beam design, which is brought , leads to a higher intensity. % than the Gaussian beam.
%
Nevertheless, the intensity of the Bessel beams is lower than that of the curving beams due to the lack of the exact position of the user in the beam design.
Moreover, the degradation of the intensity of the Bessel beams is more significant than that of the curving beams, since adjusting the parameter $\alpha$ leads to higher self-healing capability but does not lead to beams that avoid the obstacle.

Fig. \ref{fig:comp_ave} assesses the average of the amplitude of the electric field in the area in the $xy$-plane, which is denoted by the pink solid line in Fig. \ref{fig:comp_system} and can be interpreted as the potential area of the user in the presence of positioning errors following the uniform distribution with the closed interval $[-0.1, 0.1]$ [m] in both the $x$ and $y$-axes. 
%
% Thus, the results show robustness against both the positioning errors and blockages, where the statistical behavior of intensity in “Point 0” and “Point 1” is characterized by the positioning errors rather than the blockages.
Thus, the results show robustness against both positioning errors and blockages.
Note that the statistical behavior of intensity in “Point 0” is characterized only by the positioning errors.
The average intensity of beamfocusing is significantly lower than its maximum intensity shown in Fig. \ref{fig:comp_point}, which confirms that beamfocusing is impractical in the presence of \ac{CSI} errors \cite{Headland2018,Zhang2022}.
The Bessel beam with $\alpha=20$ [deg], denoted by the blue chained line without markers, achieves higher average intensity compared to the curving beams, thanks to the beam design independent of the exact position of the user.
In contrast, in the cases of $\alpha=27$ [deg], denoted by the blue chained line with the square markers, the average intensity of the Bessel beam is lower than that of the curving beams.
Those results indicate that a smaller $\alpha$ leads to a longer maximum propagation distance $d_\mathrm{max}$ and greater robustness against positioning errors in the distance domain.
However, compared to a smaller $\alpha$, a larger $\alpha$ mitigates the intensity degradation caused by the blockages more effectively, which is confirmed by the analysis in Section \ref{sec:healing} and the performance gap between the Bessel beams in “Point 3” in Fig. \ref{fig:comp_ave}.

Finally, in Fig. \ref{fig:comp_CDF}, the \ac{CDF} of the amplitude of the electric field in the potential area in the $xy$-plane in all the situations is assessed.
From the above discussions, Bessel beams have robustness against both positioning errors and blockages, which leads to the highest minimum intensity among all the beams.
Specifically, increasing the parameter $\alpha$ improves robustness against blockages and increases the minimum intensity, while a larger $\alpha$ leads to less robustness in the distance domain and lower average intensity.
Consequently, the Bessel beams with $\alpha=27$ [deg] and $\alpha=20$ [deg] achieve higher intensity compared to beamfocusing with probabilities greater than $75\%$ and $80\%$, respectively.
In contrast, beamfocusing achieves the maximum intensity thanks to beam designs based on the maximization of the \ac{SNR}, which, however, is rarely achieved due to less robustness against \ac{CSI} errors.
In turn, the design of the curving beams can exploit both the positions of the user and the obstacle, leading to higher minimum intensity compared to beamfocusing and higher maximum intensity compared to the Bessel beams.
The Gaussian beam do not have robustness against blockages, such that signal propagation is impossible with probability greater than $25\%$ in Fig. \ref{fig:comp_CDF}.
%
% From all the above, the analyses and results clarify the effectiveness of Gaussian beams, beamfocusing, Bessel beams, and curving beams in the \ac{ULA} systems.
From all the above, the analyses and results clarify the effectiveness of the near-field beams in wireless communications with the \ac{ULA}.
% wireless communciations systems with the \ac{ULA}.

%
%---
%

\section{Conclusion}

We derive limitations of Bessel beams and propose curving beam designs for wireless communications systems with a \ac{ULA}.
The analyses for Bessel beams reveal the maximum steering angle, the maximum propagation distance, the maximum antenna spacing, and the limitation of self-healing against blockages.
In curving beam designs, closed-form solutions for parameters of parabolic trajectories are obtained to avoid one obstacle and to fully exploit antenna elements.
The numerical results confirm the effectiveness of the analyses and the proposed curving beam designs.
Moreover, through electromagnetic wave simulations, the effectiveness of Gaussian beams, beamforming, Bessel beams, and curving beams is clarified in terms of the statistical behavior of their intensity and the information used for the beam generation.

%
%---
%

% \appendix
\setcounter{section}{0}
\renewcommand*{\thesection}{}
\section*{Appendix A}\label{appendix}

The solutions of \eqref{eq:wavefront_steering} depend on the value $x\cos\theta_\mathrm{A}-y\sin\theta_\mathrm{A}$.
% The wavefront function is obtained from the solutions of the equation in \eqref{eq:wavefront_steering} with respect to $y$, where 
%
Specifically, the condition $x\cos\theta_\mathrm{A}-y\sin\theta_\mathrm{A}\ge0$ leads to the following equation
\begin{equation}
    x\sin\theta_\mathrm{A}+y\cos\theta_\mathrm{A}
    =
    \tan\alpha{(x\cos\theta_\mathrm{A}-y\sin\theta_\mathrm{A})}, \label{eq:wave_y}
\end{equation}
where the solution with respect to $y$ is given by
\begin{equation}
    y = \tfrac{\tan\alpha\cos\theta_\mathrm{A}-\sin\theta_\mathrm{A}}{\cos\theta_\mathrm{A}+\tan\alpha\sin\theta_\mathrm{A}}x 
    = 
    {\tan(\alpha-\theta_\mathrm{A})}x,\label{eq:wavefront_eq_y}
\end{equation}
where $\tan z=\tfrac{\sin z}{\cos z}$ with an arbitrary real number $z$, $\cos\alpha\cos\theta_\mathrm{A}=\tfrac{1}{2}\{\cos(\alpha+{\theta}_\mathrm{A})+\cos(\alpha-{\theta}_\mathrm{A})\}$, $\sin\alpha\sin{\theta}_\mathrm{A} = -\tfrac{1}{2}\{\cos(\alpha+{\theta}_\mathrm{A})-\cos(\alpha-{\theta}_\mathrm{A})\}$, $\sin\alpha\cos\theta_\mathrm{A}=\tfrac{1}{2}\{\sin(\alpha+\theta_\mathrm{A})+\sin(\alpha-\theta_\mathrm{A})\}$, $\cos\alpha\sin\theta_\mathrm{A}=\tfrac{1}{2}\{\sin(\theta_\mathrm{A}+\alpha)+\sin(\theta_\mathrm{A}-\alpha)\}$, $\sin\alpha\sin{\theta}_\mathrm{A} = -\tfrac{1}{2}\{\cos(\alpha+{\theta}_\mathrm{A})-\cos(\alpha-{\theta}_\mathrm{A})\}$, $\cos\alpha\cos\theta_\mathrm{A}=\tfrac{1}{2}\{\cos(\alpha+{\theta}_\mathrm{A})+\cos(\alpha-{\theta}_\mathrm{A})\}$, and $\sin(\theta_\mathrm{A}-\alpha)=-\sin(\alpha-\theta_\mathrm{A})$ were used.

The solution given by \eqref{eq:wavefront_eq_y} rewrites the condition $x\cos\theta_\mathrm{A}-y\sin\theta_\mathrm{A}\ge0$ as 
\begin{equation}
    x\{\cos\theta_\mathrm{A}-\tan(\alpha-\theta_\mathrm{A})\sin\theta_\mathrm{A}\} = x\tfrac{\cos\alpha}{\cos(\alpha-\theta_\mathrm{A})}\ge0, 
\end{equation}
which is equivalent to $\tfrac{x}{\cos(\alpha-\theta_\mathrm{A})}\ge0$ under the definition $\alpha\in(0,\tfrac{\pi}{2})$ (\emph{i.e.}, $\cos\alpha>0$).

Similarly, under the condition of $x\cos\theta_\mathrm{A}-y\sin\theta_\mathrm{A}<0$, the solution of \eqref{eq:wavefront_steering} is given by $y =-\tan(\alpha+\theta_\mathrm{A})$, which rewrites the condition $x\cos\theta_\mathrm{A}-y\sin\theta_\mathrm{A}<0$ as $\tfrac{x}{\cos(\alpha+\theta_\mathrm{A})}<0$.

From the principle of signal propagation, to convey signals to the positive region of the $y$-axis while forming the desired conical wavefront of the \ac{ULA} located on both the positive and negative regions of the $x$-axis, the wavefront function should be defined across the first and second quadrants \cite{Headland2018}.
Thus, the wavefront function to generate the Bessel beam toward the desired angle $\theta_\mathrm{A}$ can be defined only under the condition $\cos(\alpha-\theta_\mathrm{A})\cos(\alpha+\theta_\mathrm{A})>0$.
This condition, the definition $\alpha\in(0,\tfrac{\pi}{2})$, and the assumption $\theta_\mathrm{A}\in(-\tfrac{\pi}{2},\tfrac{\pi}{2})$ lead to the relationship $0<\alpha+|\theta_\mathrm{A}|<\tfrac{\pi}{2}$.
Since the functions $\cos(\alpha-\theta_\mathrm{A})$ and $\cos(\alpha+\theta_\mathrm{A})$ take only positive values in the domain given by $0<\alpha+|\theta_\mathrm{A}|<\tfrac{\pi}{2}$, the wavefront function in \eqref{eq:wavefront_Bessel} is obtained.
Note that the wavefront function cannot be defined under the condition $\cos(\alpha-\theta_\mathrm{A})\cos(\alpha+\theta_\mathrm{A})=0$.

%
%----
%

% \appendix
\setcounter{section}{0}
\renewcommand*{\thesection}{}
\section*{Appendix B}\label{appendixB}

Let $\mathbf{n}\in\mathbb{R}^{2}$ denote the perpendicular vector to the wavefront function $f_\mathrm{w}(x)$ in \eqref{eq:wavefront_Bessel}.
Then, the position $\mathbf{p}_{\mathrm{o},n}=[x_{\mathrm{o},n},f_\mathrm{w}(x_{\mathrm{o},n})]^\mathrm{T}\in\mathbb{R}^2$ on the wavefront function to calculate the minimum distance $d_{n}=\|\mathbf{p}_{\mathrm{o},n}-\mathbf{p}_{\mathrm{t},n}\|_2$ is given by $ \mathbf{p}_{\mathrm{o},n} = \mathbf{p}_{\mathrm{t},n} - t_n\mathbf{n}$,
%
% \begin{equation}
%     \mathbf{p}_{\mathrm{o},n} = \mathbf{p}_{\mathrm{t},n} - t_n\cdot\mathbf{n}, \label{eq:min_points}
% \end{equation}
%
where $t_n\in\mathbb{R}$ is the constant to establish the equation.
From the definition of orthogonality, the perpendicular vector $\mathbf{n}$ is given by $ \mathbf{n}=\nabla \{f_\mathrm{w}(x)-y\}=\big[\tfrac{d f_\mathrm{w}(x)}{d x},-1\big]^\mathrm{T}$, where $\nabla$ denotes the nabla operator.
%
% \begin{equation}
%     \mathbf{n}=\nabla \{f_\mathrm{w}(x)-y\}=\Big[\tfrac{d f_\mathrm{w}(x)}{d x},-1\Big]^\mathrm{T}, \label{eq:perp}
% \end{equation}
%
% Under the necessary condition $\alpha + |\theta_\mathrm{A}|<\tfrac{\pi}{2}$, definition $\alpha\in(0,\tfrac{\pi}{2})$, and assumption $\theta_\mathrm{A}\in[-\tfrac{\pi}{2},\tfrac{\pi}{2}]$, the wavefront function $f_\mathrm{w}(x_{\mathrm{t},n})$ descritized by the antenna position $x_{\mathrm{t},n}\in\mathcal{X}_\mathrm{arr}$ can be rewritten as 
% %
% \begin{equation}
%     f_\mathrm{w}^\mathrm{d}(x_{\mathrm{t},n})
%     =
%     \begin{cases}
%     \tfrac{\tan\alpha-\tan\theta_\mathrm{A}}{1+\tan\alpha\tan\theta_\mathrm{A}}x_{\mathrm{t},n},~\mathrm{if}~x_{\mathrm{t},n}\ge 0
%     \\
%     \tfrac{\tan\theta_\mathrm{A}+\tan\alpha}{\tan\alpha\tan\theta_\mathrm{A}-1}x_{\mathrm{t},n},~\mathrm{otherwise}
%     \end{cases},
%     \label{eq:wavefront_re}
% \end{equation}
%
Then, the $x$ and $y$-coordinates of the position $\mathbf{p}_{\mathrm{o},n}$ are calculated by
% which yields the $x$ and $y$-coordinates of the positions $\mathbf{p}_{\mathrm{o},n}$, which are given by
%
\begin{align}
    x_{\mathrm{o},n}&\!=\!\!
    \begin{cases}
        \!x_{\mathrm{t},n}\!-\!t_n\tan(\alpha-\theta_\mathrm{A}),~\mathrm{if}~x_{\mathrm{t},n}\ge 0
        \\
        \!x_{\mathrm{t},n}\!+\!t_n\tan(\alpha+\theta_\mathrm{A}),~\mathrm{otherwise}
    \end{cases}\!\!\!\!, \label{eq:x_point}
    \\
    f_\mathrm{w}(x_{\mathrm{o},n})&=t_n. \label{eq:y_point}
\end{align}

% Since the point $\mathbf{p}_{\mathrm{o},n}$ is a part of the wavefornt function, the point in the $y$-coordinate $y_{\mathrm{o},n}$ is given by 
%
% From \eqref{eq:y_point} and the fact that the relationship between $x_{\mathrm{o},n}$ and $y_{\mathrm{o},n}$ is described as the wavefront function $f_\mathrm{wave}(x_{\mathrm{o},n})=y_{\mathrm{o},n}$, the constant $t_n$ is given by

By substituting the constant $t_n=f_\mathrm{w}(x_{\mathrm{o},n})$ into \eqref{eq:x_point}, the $x$-coordinate of the position $\mathbf{p}_{\mathrm{o},n}$ is rewritten as
%
% \begin{equation}
%     x_{\mathrm{o},n}\!\!=\!\!
%     \begin{cases}
%         \!\tfrac{x_{\mathrm{t},n}}{1+\tan^2(\alpha-\theta_\mathrm{A})}\!=x_{\mathrm{t},n}\!\cos^2(\alpha\!-\!\theta_\mathrm{A}),\mathrm{if}~x_{\mathrm{t},n}\ge 0
%         \\
%         \!\tfrac{x_{\mathrm{t},n}}{1+\tan^2(\alpha+\theta_\mathrm{A})}\!=x_{\mathrm{t},n}\!\cos^2(\alpha\!+\!\theta_\mathrm{A}),\mathrm{otherwise}
%     \end{cases}\!\!\!\!, 
%     \label{eq:min_x}
% \end{equation}
%
\begin{equation}
    x_{\mathrm{o},n}=
    \begin{cases}
        x_{\mathrm{t},n}\tfrac{1}{1+\tan^2(\alpha-\theta_\mathrm{A})},\mathrm{if}~x_{\mathrm{t},n}\ge 0
        \\
        x_{\mathrm{t},n}\tfrac{1}{1+\tan^2(\alpha+\theta_\mathrm{A})},\mathrm{otherwise}
    \end{cases}, 
    \label{eq:min_x}
\end{equation}
%
% where $\cos^2\theta_\mathrm{A}=\tfrac{1}{{\tan^2\theta_\mathrm{A}+1}}$ was used.

Based on \eqref{eq:min_x}, the $y$-coordinate of the position $\mathbf{p}_{\mathrm{o},n}$ can be described by a function of the position of the antenna element $x_{\mathrm{t},n}$ as follows
%
% \begin{equation}
%     f_\mathrm{w}(x_{\mathrm{o},n})\!\!=\!\!
%     \begin{cases}
%         \!x_{\mathrm{t},n}\!\cos(\alpha\!-\!\theta_\mathrm{A})\sin(\alpha\!-\!\theta_\mathrm{A}),\mathrm{if}~x_{\mathrm{t},n}\ge 0
%         \\
%         \!-x_{\mathrm{t},n}\!\cos(\alpha\!+\!\theta_\mathrm{A})\sin(\alpha\!+\!\theta_\mathrm{A}),\mathrm{otherwise}
%     \end{cases}\!\!\!\!, 
%     \label{eq:min_y}
% \end{equation}
%
\begin{equation}
    f_\mathrm{w}(x_{\mathrm{o},n})=
    \begin{cases}
        x_{\mathrm{t},n}\tfrac{\tan(\alpha-\theta_\mathrm{A})}{1+\tan^2(\alpha-\theta_\mathrm{A})},~~~\mathrm{if}~x_{\mathrm{t},n}\ge 0
        \\
        -x_{\mathrm{t},n}\tfrac{\tan(\alpha+\theta_\mathrm{A})}{1+\tan^2(\alpha+\theta_\mathrm{A})},~\mathrm{otherwise}
    \end{cases}, 
    \label{eq:min_y}
\end{equation}

% Thus, the minimum distance $d_{n}=\|\mathbf{p}_{\mathrm{o},n}-\mathbf{p}_{\mathrm{t},n}\|_2$ is calculated as
Thus, the minimum distance $d_{n}$ is calculated as
\begin{equation}
    d_n
    \!\!=\!\!
    \begin{cases}
        \!\sqrt{{\tfrac{x_{\mathrm{t},n}^2A^4}{(1+A^2)^2}}+\!\tfrac{x_{\mathrm{t},n}^2A^2}{(1+A^2)^2}}
        \!=\!
        \sqrt{{\tfrac{x_{\mathrm{t},n}^2A^2}{(1+A^2)}}},~\mathrm{if}~x_{\mathrm{t},n}\ge 0
        \\
        \!\sqrt{{\tfrac{x_{\mathrm{t},n}^2B^4}{(1+B^2)^2}}+\!\tfrac{x_{\mathrm{t},n}^2B^2}{(1+B^2)^2}}
        \!=\!
        \sqrt{{\tfrac{x_{\mathrm{t},n}^2B^2}{(1+B^2)}}},~\mathrm{otherwise}
    \end{cases}\!\!,
\end{equation}
where $A=\tan(\alpha-\theta_\mathrm{A})$ and $B=\tan(\alpha+\theta_\mathrm{A})$.
The properties $\sqrt{a^2}=|a|$, $\tfrac{1}{1+\tan^2 z}=\cos^2 z$, and $\tan^2 z =\tfrac{\sin^2 z}{\cos^2 z}$ lead to the phase $\phi_n=kd_n$ given by \eqref{eq:closed_Bessel}.

%
%----
%

\section*{Appendix C}\label{appendixC}

From \eqref{eq:wavefront_Bessel}, the condition $f_\mathrm{w}(x)\ge 0$ means that both the inequalities given by $\tan(\alpha-\theta_\mathrm{A})\ge0$ and $\tan(\alpha+\theta_\mathrm{A})\ge0$ should be satisfied.
Under the definition $\alpha\in(0,\tfrac{\pi}{2})$, assumption $\theta_\mathrm{A}\in(-\tfrac{\pi}{2},\tfrac{\pi}{2})$, and necessary condition to define the wavefront function $0\!<\!\alpha\!+\!|\theta_\mathrm{A}|\!<\!\tfrac{\pi}{2}$, %the functions $\cos(\alpha+\theta_\mathrm{A})$ and $\cos(\alpha-\theta_\mathrm{A})$ take positive values only.
those inequalities are equivalent to $\sin(\alpha\!-\!\theta_\mathrm{A})\!\ge\!0$ and $\sin(\alpha\!+\!\theta_\mathrm{A})\!\ge\!0$, respectively, where $\cos(\alpha\!+\!\theta_\mathrm{A})\!>\!0$ and $\cos(\alpha\!-\!\theta_\mathrm{A})\!>\!0$ were used.

Both the functions $\sin(\alpha-\theta_\mathrm{A})$ and $\sin(\alpha+\theta_\mathrm{A})$ return non-negative values in the domain given by $|\theta_\mathrm{A}|\le\alpha$ under the definition $\alpha\in(0,\tfrac{\pi}{2})$, and assumption $\theta_\mathrm{A}\in(-\tfrac{\pi}{2},\tfrac{\pi}{2})$.
Thus, from the two inequalities $0<\alpha+|\theta_\mathrm{A}|<\tfrac{\pi}{2}$ and $|\theta_\mathrm{A}|\le\alpha$, the necessary and sufficient condition in \eqref{eq:max_steering} is obtained.

\section*{Appendix D}\label{appendixD}

% The distance between the origin of the coordinate system and the reference points can 

The $\ell_2$-norms of $\mathbf{p}_{\mathrm{max,p}}$ and $\mathbf{p}_{\mathrm{max,m}}$ are calculated by
\begin{align}
     \|\mathbf{p}_{\mathrm{max,p}}\|_2 &= \tfrac{|R|}{|\tan\theta_\mathrm{A}+\tan(\alpha-\theta_\mathrm{A})|}\sqrt{\tan^2\theta_\mathrm{A}+1} \nonumber
    \\
    &=\tfrac{R}{|\sin\theta_\mathrm{A}+\cos\theta_\mathrm{A}\tan(\alpha-\theta_\mathrm{A})|},
    \\
    \|\mathbf{p}_{\mathrm{max,p}}\|_2 &= \tfrac{|R|}{|\tan\theta_\mathrm{A}-\tan(\alpha+\theta_\mathrm{A})|}\sqrt{\tan^2\theta_\mathrm{A}+1} \nonumber
    \\
    &=\tfrac{R}{|\sin\theta_\mathrm{A}-\cos\theta_\mathrm{A}\tan(\alpha+\theta_\mathrm{A})|},
\end{align}
where $\cos\theta_\mathrm{A}=\tfrac{1}{\sqrt{\tan^2\theta_\mathrm{A}+1}}$, $|R|=R$, and $|\cos\theta_\mathrm{A}|=\cos\theta_\mathrm{A}$ were used under the assumption $\theta_\mathrm{A}\in(-\tfrac{\pi}{2},\tfrac{\pi}{2})$.

Under the necessary and sufficient condition in \eqref{eq:max_steering}, the functions $\cos(\alpha-\theta_\mathrm{A})$ and $\cos(\alpha+\theta_\mathrm{A})$ take only positive values.
Thus, from the property $\tan z=\tfrac{\sin z}{\cos z}$, the $\ell_2$-norms can be rewritten as
\begin{align}
     \|\mathbf{p}_{\mathrm{max,p}}\|_2 &\!=\!\tfrac{R}{|\sin\theta_\mathrm{A}\cos(\alpha-\theta_\mathrm{A})+\cos\theta_\mathrm{A}\sin(\alpha-\theta_\mathrm{A})|}\cos(\alpha-\theta_\mathrm{A}) \nonumber 
    \\
    &\!=\!\tfrac{R}{|\sin\alpha|}\cos(\alpha-\theta_\mathrm{A}) = \tfrac{R}{\sin\alpha}\cos(\alpha-\theta_\mathrm{A}),
    \\
    \|\mathbf{p}_{\mathrm{max,p}}\|_2  &\!=\! \tfrac{R}{|\sin\theta_\mathrm{A}\cos(\alpha+\theta_\mathrm{A}) -\cos\theta_\mathrm{A}\sin(\alpha+\theta_\mathrm{A})|}\cos(\alpha+\theta_\mathrm{A}), \nonumber
    \\
     &\!=\! \tfrac{R}{|\!\sin(-\alpha)|}\cos(\alpha+\theta_\mathrm{A})\!=\!\tfrac{R}{\sin\alpha}\cos(\alpha+\theta_\mathrm{A}),
\end{align}
where $\sin z_1\cos z_2=\tfrac{1}{2}\{\sin(z_1+z_2)+\sin(z_1-z_2)\}$, $\cos z_1\sin z_2=\tfrac{1}{2}\{\sin(z_1+z_2)+\sin(z_1-z_2)\}$, and $\sin(-\alpha)=-\sin(\alpha)$ were used, and $|\sin\alpha|=\sin\alpha$ was used under the definition $\alpha\in(0,\tfrac{\pi}{2})$.

Under the necessary and sufficient condition in \eqref{eq:max_steering}, in the cases of $\theta_\mathrm{A}\in(0,\tfrac{\pi}{2})$, the relationship $\cos(\alpha+\theta_\mathrm{A}) < \cos\alpha < \cos(\alpha-\theta_\mathrm{A})$ yields the inequality $\tfrac{\cos(\alpha+\theta_\mathrm{A})}{\cos\alpha}< 1 < \tfrac{\cos(\alpha-\theta_\mathrm{A})}{\cos\alpha}$.
In contrast, in the cases of $\theta_\mathrm{A}\in(-\tfrac{\pi}{2},0)$, the inequality $\tfrac{\cos(\alpha-\theta_\mathrm{A})}{\cos\alpha}< 1 < \tfrac{\cos(\alpha+\theta_\mathrm{A})}{\cos\alpha},$ is satisfied.
%
% \begin{equation}
%     \tfrac{\cos(\alpha-\theta_\mathrm{A})}{\cos\alpha}\le 1 \le \tfrac{\cos(\alpha+\theta_\mathrm{A})}{\cos\alpha},
% \end{equation}
%
% which means that the distances $d_\mathrm{max,1}$ and $d_\mathrm{max,2}$ are determined by $\|\mathbf{p}_{\mathrm{max,m}}\|_2$ and $\|\mathbf{p}_{\mathrm{max,p}}\|_2$, respectively.
%
Thus, the distances $d_{\mathrm{max}}$ and $d_{\mathrm{lim}}$ are calculated as \eqref{eq:dist1} and \eqref{eq:dist2}, respectively. 
% %
% \begin{align}
%     d_{\mathrm{max}}&=
%     \begin{cases}
%          \tfrac{R}{\tan\alpha}\tfrac{\cos(\alpha+\theta_\mathrm{A})}{\cos\alpha},~\mathrm{if}~\theta_\mathrm{A}\in[0,\tfrac{\lambda}{2}]
%         \\
%         \tfrac{R}{\tan\alpha}\tfrac{\cos(\alpha-\theta_\mathrm{A})}{\cos\alpha},~\mathrm{if}~\theta_\mathrm{A}\in[-\tfrac{\lambda}{2},0)
%     \end{cases},
%     \\
%     d_{\mathrm{lim}}
%     &=
%     \begin{cases}
%          \tfrac{R}{\tan\alpha}\tfrac{\cos(\alpha-\theta_\mathrm{A})}{\cos\alpha},~\mathrm{if}~\theta_\mathrm{A}\in[0,\tfrac{\lambda}{2}]
%         \\
%         \tfrac{R}{\tan\alpha}\tfrac{\cos(\alpha+\theta_\mathrm{A})}{\cos\alpha},~\mathrm{if}~\theta_\mathrm{A}\in[-\tfrac{\lambda}{2},0)
%     \end{cases},
% \end{align}
% %
% which are equivalent to \eqref{eq:dist1} and \eqref{eq:dist2}, respectively.

%
%----
%

\section*{Appendix E}\label{appendixE}

As mentioned in Section \ref{sec:Bessel_closed}, the Bessel beam comprises the two beams generated by the antenna elements located in the positive and negative regions of the $x$-axis, receptively.
%
% Thus, the magnitude of the $x$-components of the wavenumber $k_\mathrm{x}$ are calculated for each beam.
Thus, the magnitude of the transverse component of the wave vector $k_\mathrm{x}$ are calculated for each beam.
% %
% This structure of the wavefront function means that the $x$-components of the wavenumber $k_\mathrm{x}$ should be calculated individually in the regions $x\ge0$ and $x<0$.

% In the origin and positive region of the $x$-axis $x\ge0$, the angle between the $x$-axis and the rays perpendicular to the wavefront function in \eqref{eq:wavefront_Bessel} is given by $\tfrac{\pi}{2}-(\alpha-\theta_\mathrm{A})$.
In the origin and positive region of the $x$-axis $x\ge0$, the angle between the $x$-axis and the propagation direction of the wave of each antenna element is given by $\tfrac{\pi}{2}-(\alpha-\theta_\mathrm{A})$.
%
% Thus, the magnitude of the $x$-component $k_\mathrm{x}$ in the positive region of the $x$-axis is given by $k_\mathrm{x}=k\cos\{\tfrac{\pi}{2}-(\alpha-\theta_\mathrm{A})\}=k\sin(\alpha-\theta_\mathrm{A})$.
Thus, the magnitude of the transverse component $k_\mathrm{x}$ in the positive region of the $x$-axis is given by $k_\mathrm{x}=k\cos\{\tfrac{\pi}{2}-(\alpha-\theta_\mathrm{A})\}=k\sin(\alpha-\theta_\mathrm{A})$.
Similarly, that in the negative region of the $x$-axis is given by $k_\mathrm{x}=k\sin(\alpha+\theta_\mathrm{A})$.
%
% In contrast, the negative region of the $x$-axis, the angle between those is given by $\tfrac{\pi}{2}+(\alpha+\theta_\mathrm{A})$.
%
% Therefore, the $x$-components $k_\mathrm{x}$ are given by $k_\mathrm{x}=k\cos\{\tfrac{\pi}{2}-(\alpha-\theta_\mathrm{A})\}=k\sin(\alpha-\theta_\mathrm{A})$ and $k_\mathrm{x}=k\cos\{\tfrac{\pi}{2}-(\alpha+\theta_\mathrm{A})\}=k\sin(\alpha+\theta_\mathrm{A})$ in the positive and negative region in the $x$-axis, respectively.
%
% \begin{equation}
%     k_\mathrm{x}
%     =
%     \begin{cases}
%         k\sin(\alpha-\theta_\mathrm{A}),~\mathrm{if}~x_{\mathrm{t},n}\ge0
%         \\
%         k\sin(\alpha+\theta_\mathrm{A}),~\mathrm{otherwise}
%     \end{cases},
% \end{equation}
%
% Thus, in signal propagation with the Bessel beam, the condition to satisfy the sampling theorem $\tfrac{\pi}{k_\mathrm{x}}$ can be rewritten as
Thus, the condition to satisfy the sampling theorem $\tfrac{\pi}{k_\mathrm{x}}$ can be rewritten as
\begin{equation}
    \Delta<
    \begin{cases}
        \tfrac{\lambda}{2}\tfrac{1}{\sin(\alpha-\theta_\mathrm{A})},~\mathrm{if}~x_{\mathrm{t},n}\ge0
        \\
        \tfrac{\lambda}{2}\tfrac{1}{\sin(\alpha+\theta_\mathrm{A})},~\mathrm{otherwise}
    \end{cases}.
\end{equation}

% Under the necessity and sufficient condition in \eqref{eq:max_steering}, the value $\tfrac{1}{\sin(\alpha+\theta_\mathrm{A})}$ is less than $\tfrac{1}{\sin(\alpha-\theta_\mathrm{A})}$ in the cases of $\theta_\mathrm{A}\in(0,\tfrac{\pi}{2})$.
% %
% In contrast, the value $\tfrac{1}{\sin(\alpha-\theta_\mathrm{A})}$ is less than $\tfrac{1}{\sin(\alpha+\theta_\mathrm{A})}$ in the cases of $\theta_\mathrm{A}\in(-\tfrac{\pi}{2},0)$.
Under the necessary and sufficient condition in \eqref{eq:max_steering}, the following inequalities $\tfrac{1}{\sin(\alpha+\theta_\mathrm{A})} <\tfrac{1}{\sin(\alpha-\theta_\mathrm{A})}$ and $\tfrac{1}{\sin(\alpha+\theta_\mathrm{A})} > \tfrac{1}{\sin(\alpha-\theta_\mathrm{A})}$ holds in the cases of $\theta_\mathrm{A}\in(0,\tfrac{\pi}{2})$ and $\theta_\mathrm{A}\in(-\tfrac{\pi}{2},0)$, respectively.
%
% the value $\tfrac{1}{\sin(\alpha+\theta_\mathrm{A})}$ is less than $\tfrac{1}{\sin(\alpha-\theta_\mathrm{A})}$ in the cases of $\theta_\mathrm{A}\in(0,\tfrac{\pi}{2})$.
% %
% In contrast, the value $\tfrac{1}{\sin(\alpha-\theta_\mathrm{A})}$ is less than $\tfrac{1}{\sin(\alpha+\theta_\mathrm{A})}$ in the cases of $\theta_\mathrm{A}\in(-\tfrac{\pi}{2},0)$.
%
Thus, to satisfy the sampling theorem in the whole \ac{ULA}, the antenna spacing should be less than $\Delta < \tfrac{\lambda}{2}\tfrac{1}{\sin(\alpha+|\theta_\mathrm{A}|)}$, which is the same condition in \eqref{eq:sampling}.

% the relationship between the terms $\sin(\alpha-\theta_\mathrm{A})$ and $\sin(\alpha+\theta_\mathrm{A})$ is described by
%
% \begin{equation}
%     \begin{cases}
%         \tfrac{1}{\sin(\alpha-\theta_\mathrm{A})} \ge \tfrac{1}{\sin(\alpha+\theta_\mathrm{A})},~\mathrm{if}~\theta_\mathrm{A}\ge0
%         \\
%         \tfrac{1}{\sin(\alpha-\theta_\mathrm{A})} < \tfrac{1}{\sin(\alpha+\theta_\mathrm{A})},~\mathrm{otherwise}
%     \end{cases}.
% \end{equation}
%

% Thus, in the cases of $\theta_\mathrm{A}\ge0$, the antenna spacing should be less than $\Delta < \tfrac{\lambda}{2}\cdot\tfrac{1}{\sin(\alpha+\theta_\mathrm{A})}$ to satisfy the sampling theorem for all the antenna elements.
%
% In contrast, in the cases of $\theta_\mathrm{A}<0$, the antenna spacing should be less than $\Delta < \tfrac{\lambda}{2}\cdot\tfrac{1}{\sin(\alpha-\theta_\mathrm{A})}$ to satisfy the sampling theorem for all the antenna elements.
%
% In conclusion, the necessity and sufficient conditions to satisfy the sampling theorem is given by \eqref{eq:sampling}.

%
%----
%

\section*{Appendix F}\label{appendixF}

The Lagrangian for the problem in \eqref{opt:final_r_p} can be defined by
% \begin{align}
%     L(\beta,\tilde{p},x_\mathrm{adj},\boldsymbol{\mu})
%     &\triangleq f_\mathrm{para} \!-\!\mu_1\beta \!-\! \mu_2(R\!+\!x_\mathrm{adj})+ \mu_3(x_\mathrm{adj}\!-\!R) \nonumber
%     \\
%     &+\mu_4\big\{\beta(y_\mathrm{n}^2-y_\mathrm{u}^2)-2\tilde{p}(y_\mathrm{n}-y_\mathrm{u})+x_\mathrm{u}- x_\mathrm{r,2}\big\} \nonumber
%     \\
%     &+\mu_5\big\{\beta(y_\mathrm{f}^2-y_\mathrm{u}^2)-2\tilde{p}(y_\mathrm{f}-y_\mathrm{u})+x_\mathrm{u}- x_\mathrm{r,2}\big\} \nonumber
%     \\
%     &+\mu_6({-2\beta y_\mathrm{u}^2+2\tilde{p} y_\mathrm{u}+x_\mathrm{u}- x_{\mathrm{adj}}}) \nonumber
%     \\
%     &-\mu_7({-2\beta y_\mathrm{u}^2+2\tilde{p} y_\mathrm{u}+x_\mathrm{u}+R}) \nonumber
%     \\
%     &-\mu_8(-\beta y_\mathrm{u}^2+2\tilde{p} y_\mathrm{u}+x_\mathrm{u}- x_{\mathrm{adj}}),
% \end{align}
%
\begin{align}
    &L(\beta,\tilde{p},x_\mathrm{adj},\boldsymbol{\mu})
    \triangleq f_\mathrm{para} \!-\!\mu_1\beta \!-\! \mu_2(R\!+\!x_\mathrm{adj})+ \mu_3(x_\mathrm{adj}\!-\!R) \nonumber
    \\
    &\!+\!\mu_4\big\{\beta(y_\mathrm{n}^2-y_\mathrm{u}^2)-2\tilde{p}(y_\mathrm{n}-y_\mathrm{u})+x_\mathrm{u}- x_\mathrm{r,2}\big\} \nonumber
    \\
    &\!+\!\mu_5\big\{\beta(y_\mathrm{f}^2-y_\mathrm{u}^2)-2\tilde{p}(y_\mathrm{f}-y_\mathrm{u})+x_\mathrm{u}- x_\mathrm{r,2}\big\} \nonumber
    \\
    &\!+\!\mu_6({-2\beta y_\mathrm{u}^2\!+\!2\tilde{p} y_\mathrm{u}\!+\!x_\mathrm{u}\!-\!x_{\mathrm{adj}}})\!-\!\mu_7({-2\beta y_\mathrm{u}^2\!+\!2\tilde{p} y_\mathrm{u}\!+\!x_\mathrm{u}\!+\!R}) \nonumber
    \\
    &\!-\!\mu_8(-\beta y_\mathrm{u}^2+2\tilde{p} y_\mathrm{u}+x_\mathrm{u}- x_{\mathrm{adj}}),
\end{align}
where $\boldsymbol{\mu}\triangleq [\mu_1,\ldots,\mu_8]^\mathrm{T}\in\mathbb{R}^8$ denote the non-negative Lagrangian multipliers.

The \ac{KKT} conditions for the problem in \eqref{opt:final_r_p} are described by the constraints in \eqref{const:array_final} through \eqref{const:sqrt_final}, the non-negative constraints for the Lagrangian multipliers $\mu_i \ge 0,(i=1,\ldots,8)$, and the following simultaneous equations
\begin{align}
    &\tfrac{\partial \mathcal{L}}{\partial \beta}=(y_\mathrm{n}^2+y_\mathrm{f}^2-2y_\mathrm{u}^2)-\mu_{1}
    +\mu_4(y_\mathrm{n}^2-y_\mathrm{u}^2)+\mu_5(y_\mathrm{f}^2-y_\mathrm{u}^2) \nonumber
    \\
    &\hspace{20ex}-2\mu_6y_\mathrm{u}^2+\mu_7y_\mathrm{u}^2+\mu_8y_\mathrm{u}^2=0, \nonumber
    \\
    &\tfrac{\partial \mathcal{L}}{\partial \tilde{p}}=2(2y_\mathrm{u}-y_\mathrm{n}-y_\mathrm{f})-2\mu_4(y_\mathrm{n}-y_\mathrm{u})-2\mu_5(y_\mathrm{f}-y_\mathrm{u}) \nonumber
    \\
    &\hspace{20ex}-2\mu_6y_\mathrm{u}-2\mu_7y_\mathrm{u}-2\mu_8y_\mathrm{u}=0,\nonumber
    \\
    &\tfrac{\partial \mathcal{L}}{\partial x_\mathrm{adj}}=-w-\mu_2+\mu_3-\mu_6+\mu_8=0, \nonumber
    \\
    &\mu_1\beta = 0, ~\mu_2(-R-x_\mathrm{adj})=0, ~\mu_3(x_\mathrm{adj}-R)=0, \nonumber
    \\
    &\mu_4\big\{\beta(y_\mathrm{n}^2-y_\mathrm{u}^2)-2\tilde{p}(y_\mathrm{n}-y_\mathrm{u})+x_\mathrm{u}- x_\mathrm{r,2}\big\}=0, \nonumber
    \\
    &\mu_5\big\{\beta(y_\mathrm{f}^2-y_\mathrm{u}^2)-2\tilde{p}(y_\mathrm{f}-y_\mathrm{u})+x_\mathrm{u}- x_\mathrm{r,2}\big\}=0, \nonumber
    \\
    &\mu_6({-2\beta y_\mathrm{u}^2+2\tilde{p} y_\mathrm{u}+x_\mathrm{u}- x_{\mathrm{adj}}})=0, \nonumber
    \\
    &\mu_7({-2\beta y_\mathrm{u}^2+2\tilde{p} y_\mathrm{u}+x_\mathrm{u}+R}) =0, \nonumber
    \\
    &\mu_8(-\beta y_\mathrm{u}^2+2\tilde{p} y_\mathrm{u}+x_\mathrm{u}- x_{\mathrm{adj}})=0, \nonumber
\end{align}
where $\mathcal{L}\triangleq \mathcal{L}(\beta,\tilde{p},x_\mathrm{adj},\boldsymbol{\mu})$, and the solutions are candidates for the points that satisfy the \ac{KKT} conditions.

Obviously, the solutions of the above simultaneous equations that include the negative Lagrangian multipliers that are independent of the positions $x_\mathrm{u}$, $y_\mathrm{u}$, $y_\mathrm{n}$, $y_\mathrm{f}$, $x_\mathrm{r,2}$, and $R$ do not satisfy the \ac{KKT} conditions.
In addition, as described in Section \ref{sec:closed_curving}, the solutions that include $\beta=0$ or $x_\mathrm{adj}=-R$, should be dropped.
% In addition, the solutions that include $\beta=0$ or $x_\mathrm{adj}=-R$, should be dropped.
%
Thus, the possible combinations of the optimal solutions for the problem in \eqref{opt:final_r_p} are given by \eqref{eq:closed_curving}.

%
%----
%

\bibliographystyle{IEEEtran}
\bibliography{References}

\begin{thebibliography}{10}
\providecommand{\url}[1]{#1}
\csname url@samestyle\endcsname
\providecommand{\newblock}{\relax}
\providecommand{\bibinfo}[2]{#2}
\providecommand{\BIBentrySTDinterwordspacing}{\spaceskip=0pt\relax}
\providecommand{\BIBentryALTinterwordstretchfactor}{4}
\providecommand{\BIBentryALTinterwordspacing}{\spaceskip=\fontdimen2\font plus
\BIBentryALTinterwordstretchfactor\fontdimen3\font minus
  \fontdimen4\font\relax}
\providecommand{\BIBforeignlanguage}[2]{{%
\expandafter\ifx\csname l@#1\endcsname\relax
\typeout{** WARNING: IEEEtran.bst: No hyphenation pattern has been}%
\typeout{** loaded for the language `#1'. Using the pattern for}%
\typeout{** the default language instead.}%
\else
\language=\csname l@#1\endcsname
\fi
#2}}
\providecommand{\BIBdecl}{\relax}
\BIBdecl

\bibitem{Wang2023}
Wang \emph{et~al.}, ``On the road to {6G}: Visions, requirements, key
  technologies, and testbeds,'' \emph{IEEE Commun. Surv. Tut.}, vol.~25, no.~2,
  pp. 905--974, Feb. 2023.

\bibitem{Jiang2024_Suv}
W.~Jiang \emph{et~al.}, ``Terahertz communications and sensing for 6g and
  beyond: A comprehensive review,'' \emph{IEEE Communications Surveys \&
  Tutorials}, vol.~26, no.~4, pp. 2326--2381, Apr. 2024.

\bibitem{Jornet2011}
J.~M. Jornet and I.~F. Akyildiz, ``Channel modeling and capacity analysis for
  electromagnetic wireless nanonetworks in the terahertz band,'' \emph{IEEE
  Trans. Wireless Commun.}, vol.~10, no.~10, pp. 3211--3221, Oct. 2011.

\bibitem{Andrews2017}
J.~G. Andrews \emph{et~al.}, ``Modeling and analyzing millimeter wave cellular
  systems,'' \emph{IEEE Trans. Commun.}, vol.~65, no.~1, pp. 403--430, Jan.
  2017.

\bibitem{Liu2021}
Y.~Liu, X.~Liu, X.~Mu, T.~Hou, J.~Xu, M.~Di~Renzo, and N.~Al-Dhahir,
  ``Reconfigurable intelligent surfaces: Principles and opportunities,''
  \emph{IEEE Commun. Surv. \& Tut.}, vol.~23, no.~3, pp. 1546--1577, May 2021.

\bibitem{Bodet2023_OJ}
D.~M. Bodet \emph{et~al.}, ``Directional antennas for sub-{THz} and {THz}
  {MIMO} systems: Bridging the gap between theory and implementation,''
  \emph{IEEE Open J. Commun. Society}, vol.~4, pp. 2261--2273, Sep. 2023.

\bibitem{Petrov2017}
V.~Petrov, M.~Komarov, D.~Moltchanov, J.~M. Jornet, and Y.~Koucheryavy,
  ``Interference and {SINR} in millimeter wave and terahertz communication
  systems with blocking and directional antennas,'' \emph{IEEE Trans. Wireless
  Commun.}, vol.~16, no.~3, pp. 1791--1808, Mar. 2017.

\bibitem{Uchimura2023}
S.~Uchimura \emph{et~al.}, ``Outage-minimization coordinated multi-point for
  millimeter-wave {OFDM} with random blockages,'' \emph{IEEE Trans. Veh.
  Technol.}, vol.~72, no.~7, pp. 8783--8796, Jul. 2023.

\bibitem{Ayach2014}
O.~E. Ayach, S.~Rajagopal, S.~Abu-Surra, Z.~Pi, and R.~W. Heath, ``Spatially
  sparse precoding in millimeter wave {MIMO} systems,'' \emph{IEEE Trans.
  Wireless Commun.}, vol.~13, no.~3, pp. 1499--1513, Mar. 2014.

\bibitem{Uchimura2024_TWC}
S.~Uchimura \emph{et~al.}, ``Blockage-robust hybrid beamforming enabling high
  sum rate for millimeter-wave {OFDM} systems,'' \emph{IEEE Trans. Wireless
  Commun.}, vol.~23, no.~7, pp. 7095--7110, Jul. 2024.

\bibitem{Uchimura2025_TWC}
S.~Uchimura, K.~Ando, G.~T.~F. de~Abreu, and K.~Ishibashi, ``Joint design of
  equalization and beamforming for single-carrier mimo transmission over
  millimeter-wave and sub-terahertz channels,'' \emph{IEEE Trans. Wireless
  Commun.}, vol.~24, no.~3, pp. 1978--1991, Mar. 2025.

\bibitem{Uchimura2024_TVT}
S.~Uchimura \emph{et~al.}, ``Efficient channel tracking based on compressive
  sensing for {OFDM} millimeter-wave systems,'' \emph{IEEE Trans. Veh.
  Technol.}, vol.~73, no.~8, pp. 11\,411--11\,426, Aug. 2024.

\bibitem{Heath2016}
R.~W. Heath \emph{et~al.}, ``An overview of signal processing techniques for
  millimeter wave {MIMO} systems,'' \emph{IEEE J. Sel. Topics Sig. Process.},
  vol.~10, no.~3, pp. 436--453, 2016.

\bibitem{Sarieddeen2021}
H.~Sarieddeen, M.-S. Alouini, and T.~Y. Al-Naffouri, ``An overview of signal
  processing techniques for terahertz communications,'' \emph{Proc. IEEE}, vol.
  109, no.~10, pp. 1628--1665, Aug. 2021.

\bibitem{MacCartney2016}
G.~R. MacCartney, S.~Deng, S.~Sun, and T.~S. Rappaport, ``Millimeter-wave human
  blockage at 73 {GHz} with a simple double knife-edge diffraction model and
  extension for directional antennas,'' in \emph{Proc. IEEE 84th Veh. Technol.
  Conf. (VTC-Fall)}, Sep. 2016, pp. 1--6.

\bibitem{Mukherjee2022}
S.~Mukherjee \emph{et~al.}, ``Scalable modeling of human blockage at
  millimeter-wave: A comparative analysis of knife-edge diffraction, the
  uniform theory of diffraction, and physical optics against 60 {GHz} channel
  measurements,'' \emph{IEEE Access}, vol.~10, pp. 133\,643--133\,654, Dec.
  2022.

\bibitem{Fieramosca2024}
F.~Fieramosca, V.~Rampa, S.~Savazzi, and M.~D'Amico, ``On the impact of the
  antenna radiation patterns in passive radio sensing,'' \emph{IEEE Antennas
  Wireless Propag. Lett.}, vol.~23, no.~2, pp. 503--507, Oct. 2024.

\bibitem{Poddar2024}
H.~Poddar \emph{et~al.}, ``A tutorial on {NYUSIM}: Sub-terahertz and
  millimeter-wave channel simulator for {5G}, {6G}, and beyond,'' \emph{IEEE
  Commun. Surv. \& Tut.}, vol.~26, no.~2, pp. 824--857, Dec. 2024.

\bibitem{1Balanis2005}
C.~A. Balanis, \emph{Antenna Theory: Analysis and Design}.\hskip 1em plus 0.5em
  minus 0.4em\relax USA: Wiley-Interscience, 2005.

\bibitem{Headland2018}
\BIBentryALTinterwordspacing
D.~Headland \emph{et~al.}, ``Tutorial: Terahertz beamforming, from concepts to
  realizations,'' \emph{APL Photonics}, vol.~3, no.~5, p. 051101, 02 2018.
  [Online]. Available: \url{https://doi.org/10.1063/1.5011063}
\BIBentrySTDinterwordspacing

\bibitem{Petrov2023}
V.~Petrov \emph{et~al.}, ``Near-field {6G} networks: Why mobile terahertz
  communications {MUST} operate in the near field,'' in \emph{Proc. IEEE Global
  Commun. Conf. (GLOBECOM)}, Feb. 2023, pp. 3983--3989.

\bibitem{Liu2023}
Y.~Liu \emph{et~al.}, ``Near-field communications: A tutorial review,''
  \emph{IEEE Open J. Commun. Soc.}, vol.~4, pp. 1999--2049, Aug. 2023.

\bibitem{Bodet2024}
\BIBentryALTinterwordspacing
D.~Bodet, V.~Petrov, S.~Petrushkevich, and J.~M. Jornet, ``Sub-terahertz near
  field channel measurements and analysis with beamforming and bessel beams,''
  \emph{Scientific Reports}, vol.~14, no.~1, p. 19675, 2024. [Online].
  Available: \url{https://doi.org/10.1038/s41598-024-70542-z}
\BIBentrySTDinterwordspacing

\bibitem{Chen2024}
A.~Chen, L.~Chen, Y.~Chen, N.~Zhao, and C.~You, ``Near-field positioning and
  attitude sensing based on electromagnetic propagation modeling,'' \emph{IEEE
  J. Sel. Areas Commun.}, vol.~42, no.~9, pp. 2179--2195, Sep. 2024.

\bibitem{Castellanos2024}
M.~R. Castellanos and R.~W. Heath, ``Electromagnetic manifold characterization
  of antenna arrays,'' \emph{IEEE Trans. Wireless Commun.}, vol.~24, no.~3, pp.
  1772--1785, Mar. 2025.

\bibitem{Monemi2024}
M.~Monemi, S.~Bahrami, M.~Rasti, and M.~Latva-aho, ``A study on
  characterization of near-field sub-regions for phased-array antennas,''
  \emph{IEEE Trans. Commun.}, pp. 1--16, 2024, {Early Access}.

\bibitem{You2025}
C.~You \emph{et~al.}, ``Next generation advanced transceiver technologies for
  6g and beyond,'' \emph{IEEE J. Sel. Areas Commun.}, vol.~43, no.~3, pp.
  582--627, Mar. 2025.

\bibitem{Durnin1987}
\BIBentryALTinterwordspacing
J.~Durnin, J.~J. Miceli, and J.~H. Eberly, ``Diffraction-free beams,''
  \emph{Phys. Rev. Lett.}, vol.~58, pp. 1499--1501, Apr 1987. [Online].
  Available: \url{https://link.aps.org/doi/10.1103/PhysRevLett.58.1499}
\BIBentrySTDinterwordspacing

\bibitem{Siviloglou2007}
\BIBentryALTinterwordspacing
G.~A. Siviloglou \emph{et~al.}, ``Observation of accelerating airy beams,''
  \emph{Phys. Rev. Lett.}, vol.~99, p. 213901, Nov. 2007. [Online]. Available:
  \url{https://link.aps.org/doi/10.1103/PhysRevLett.99.213901}
\BIBentrySTDinterwordspacing

\bibitem{Zhang2022}
H.~Zhang \emph{et~al.}, ``Beam focusing for near-field multiuser {MIMO}
  communications,'' \emph{IEEE Trans. Wireless Commun.}, vol.~21, no.~9, pp.
  7476--7490, Mar. 2022.

\bibitem{Reddy2023}
I.~Reddy, D.~Bodet, A.~Singh, V.~Petrov, C.~Liberale, and J.~Jornet,
  ``Ultrabroadband terahertz-band communications with self-healing bessel
  beams,'' \emph{Commun. Eng.}, vol.~2, no.~1, p.~70, Oct. 2023.

\bibitem{Guerboukha2024}
\BIBentryALTinterwordspacing
H.~Guerboukha \emph{et~al.}, ``Curving {THz} wireless data links around
  obstacles,'' \emph{Commun. Eng.}, vol.~3, no.~1, p.~58, 2024. [Online].
  Available: \url{https://doi.org/10.1038/s44172-024-00206-3}
\BIBentrySTDinterwordspacing

\bibitem{Yang2023}
C.~Yang \emph{et~al.}, ``Terahertz bessel beam scanning enabled by
  dispersion-engineered metasurface,'' \emph{IEEE Trans. Microw. Theory
  Techn.}, vol.~71, no.~8, pp. 3303--3311, Aug. 2023.

\bibitem{Zhongsheng2024}
\BIBentryALTinterwordspacing
Z.~Zhai, J.~Huang, X.~Yu, Q.~Lv, N.~Offiong, and D.~Liu, ``High uniformity
  bessel beams with angle-controllable steering,'' \emph{Opt. Express},
  vol.~32, no.~19, pp. 33\,811--33\,829, Sep. 2024. [Online]. Available:
  \url{https://opg.optica.org/oe/abstract.cfm?URI=oe-32-19-33811}
\BIBentrySTDinterwordspacing

\bibitem{Lee2025}
D.~Lee, Y.~Yagi, K.~Suzuoki, and R.~Kudo, ``Experimental demonstration of
  wireless transmission using airy beams in sub-thz band,'' \emph{IEEE Open J.
  Commun. Soc.}, vol.~6, pp. 1091--1102, 2025.

\bibitem{Gabriel2022}
\BIBentryALTinterwordspacing
G.~Lasry, Y.~Brick, and T.~Melamed, ``Manipulation of curved beams using
  beam-domain optimization,'' \emph{Opt. Express}, vol.~30, no.~4, pp.
  6061--6075, Feb. 2022. [Online]. Available:
  \url{https://opg.optica.org/oe/abstract.cfm?URI=oe-30-4-6061}
\BIBentrySTDinterwordspacing

\bibitem{Petrov2024}
V.~Petrov \emph{et~al.}, ``Wavefront hopping: An enabler for reliable and
  secure near field terahertz communications in {6G} and beyond,'' \emph{IEEE
  Wireless Commun.}, vol.~31, no.~1, pp. 48--55, Feb. 2024.

\bibitem{Petrov2024_TC}
------, ``Wavefront hopping for physical layer security in 6g and beyond
  near-field thz communications,'' \emph{IEEE Trans. Commun.}, pp. 1--17, Oct.
  2024, {Early Access}.

\bibitem{Simon2024}
A.~Simon\v{c}i\v{c}, A.~Hrovat, G.~Morano, T.~Kocevska, and T.~Javornik,
  ``Near-field beam steering with planar antenna array,'' in \emph{Proc. 7th
  Int. Balkan Conf. Commun.Netw. (BalkanCom)}, Jun. 2024, pp. 31--36.

\bibitem{Droulias2024}
S.~Droulias, G.~Stratidakis, and A.~Alexiou, ``Bending beams for {6G}
  near-field communications,'' \emph{IEEE Trans. Wireless Commun.}, vol.~24,
  no.~2, pp. 1467--1480, Feb. 2025.

\bibitem{Marchand1966}
\BIBentryALTinterwordspacing
E.~W. Marchand, ``Electromagnetic theory and geometrical optics (morris kline
  and irvin w. kay),'' \emph{SIAM Rev.}, vol.~8, no.~1, p. 119–120, Jan.
  1966. [Online]. Available: \url{https://doi.org/10.1137/1008025}
\BIBentrySTDinterwordspacing

\bibitem{Pizzo2022}
A.~Pizzo, A.~d.~J. Torres, L.~Sanguinetti, and T.~L. Marzetta, ``Nyquist
  sampling and degrees of freedom of electromagnetic fields,'' \emph{IEEE
  Trans. Sig. Process.}, vol.~70, pp. 3935--3947, Jun. 2022.

\bibitem{Paknys2016_2}
\BIBentryALTinterwordspacing
R.~Paknys, \emph{Uniform Theory of Diffraction}.\hskip 1em plus 0.5em minus
  0.4em\relax John Wiley \& Sons, Ltd, 2016, ch.~8, pp. 268--316. [Online].
  Available:
  \url{https://onlinelibrary.wiley.com/doi/abs/10.1002/9781119127444.ch8}
\BIBentrySTDinterwordspacing

\bibitem{Aguilar2022}
A.~G. Aguilar \emph{et~al.}, ``A novel faceted {UTD} solver in altair feko for
  antenna placement applications,'' in \emph{Proc. 3rd URSI Atlantic and Asia
  Pacific Radio Science Meeting (AT-AP-RASC)}, May 2022, pp. 1--3.

\end{thebibliography}

\vfill

\end{document}